\documentclass[%
 reprint,
 amsmath,amssymb,
 aps,
floatfix,
]{revtex4-2}
\usepackage{graphicx}% Include figure files
\usepackage{dcolumn}% Align table columns on decimal point
\usepackage{bm}% bold math
\usepackage[utf8]{inputenc}
\usepackage[T1]{fontenc}
\usepackage{mathptmx}
\usepackage{etoolbox}
\usepackage{adjustbox}
\usepackage[dvipsnames]{xcolor}
\usepackage{diagbox}
\usepackage{multirow}
\newcolumntype{C}[1]{>{\centering\arraybackslash}p{#1}}

\usepackage[caption=false,font=normalsize,labelfont=sf,textfont=sf]{subfig}
\begin{document}

\preprint{APS/123-QED}

\title[Point defects in CdTe and CdSeTe alloy: a first principles investigation with DFT+U]{Point defects in CdTe and CdSeTe alloy: a first principles investigation with DFT+U}
% Force line breaks with \\
\author{Xiaofeng Xiang}
% \altaffiliation[Also at ]{Molecular Engineering \& Sciences Institute, University of Washington, Seattle, Washington 98195, USA}%Lines break automatically or can be forced with \\
\affiliation{ 
Molecular Engineering \& Sciences Institute, University of Washington, Seattle, Washington 98195, USA %\\This line break forced with \textbackslash\textbackslash
}%
\author{Yijun Tong}%
% \email{Second.Author@institution.edu.}
\affiliation{ 
Department of Electrical and Computer Engineering, University of Washington, Seattle, Washington 98195, USA %\\This line break forced with \textbackslash\textbackslash
}%

\author{Aaron Gehrke}%
% \email{Second.Author@institution.edu.}
\affiliation{ 
Department of Materials Science \& Engineering, University of Washington, Seattle, Washington 98195, USA %\\This line break forced with \textbackslash\textbackslash
}%

\author{Scott T. Dunham}
% \homepage{http://www.Second.institution.edu/~Charlie.Author.}
\affiliation{%
Department of Electrical and Computer Engineering, University of Washington, Seattle, Washington 98195, USA%\\This line break forced% with \\
}%

\date{\today}% It is always \today, today,

\begin{abstract}
CdTe and its alloy CdSeTe are widely used in optoelectronic devices, such as radiation detectors and solar cells, due to their superior electrical properties. However, the formation of defects and defect complexes in these materials can significantly affect their performance. As a result, understanding the defect formation and recombination processes in CdTe and CdSeTe alloy is of great importance. In recent years, density functional theory (DFT) calculations have emerged as a powerful tool for investigating the properties of defects in semiconductors. In this paper, we use DFT+U calculations to comprehensively study the properties of intrinsic defects as well as extrinsic defects induced by commonly used dopants, such as Cu and group V elements, in CdTe and CdSeTe alloy. This work provides insights into the effects of these defects on the electrical and optical properties of the material. %(We found that AX center formation is more favorable in the alloy, which can potentially reduce the dopability of As and P in CdSeTe alloy. Our findings not only provide new insights into the mechanisms of As and P doping in CdSeTe but also have implications for the development of high-performance solar cells. )
\end{abstract}

\maketitle

% \begin{quotation}
% The ``lead paragraph'' is encapsulated with the \LaTeX\ 
% \verb+quotation+ environment and is formatted as a single paragraph before the first section heading. 
% (The \verb+quotation+ environment reverts to its usual meaning after the first sectioning command.) 
% Note that numbered references are allowed in the lead paragraph.
% %
% The lead paragraph will only be found in an article being prepared for the journal \textit{Chaos}.
% \end{quotation}

%\section{\label{sec:level1}First-level heading:\protect\\ The line
%break was forced \lowercase{via} \textbackslash\textbackslash}
\section{\label{sec:level1}Introduction\protect\\}

Cadmium telluride (CdTe) and its alloy with selenium (CdSeTe) have been extensively studied due to their high potential for use in optoelectronic devices, including solar cells and radiation detectors. Compared to CdTe, the addition of Se to the alloy has been shown to improve the material's electronic properties, such as its carrier lifetime and carrier mobility, allowing tuning of bandgap, making it a promising candidate for high-efficiency solar cell applications \cite{fiducia2019understanding}.
The performance of these devices, however, can be significantly impacted by the presence of point defects, which can affect the material's electrical and optical properties. To better understand the effects of defects on the properties of CdSeTe, it is necessary to investigate the formation and recombination of defects and defect complexes in this material.
 In this work, we use density functional theory (DFT) calculations to investigate the properties of intrinsic defects,  copper and group V dopants in CdSeTe, with a focus on the formation of point defects and their impact on the material's electronic properties. We consider intrinsic defects that have been widely reported and are considered to be significant, such as $V_{Cd}$\cite{yang2016review, krasikov2016theoretical, kavanagh2021rapid, shepidchenko2015small, biswas2012causes}, $Te_{Cd}$\cite{yang2014tuning, orellana2019self}, $V_{Te}$\cite{yang2014tuning, du2008native, pan2018spin, menendez2019symmetry}, $Cd_{Te}$\cite{yang2014tuning, du2008native, pan2018spin}, $Cd_{int}$\cite{yang2014tuning}, $Te_{int}$\cite{krasikov2016theoretical, yang2014tuning, krasikov2018beyond, du2008native}, $Se_{Cd}$\cite{selvaraj2021passivation}, $V_{Cd}+Te_{Cd}$\cite{krasikov2017defect} and $Te_{int}+Te_{Cd}$\cite{krasikov2017defect}. For copper in CdSeTe, we consider four dominant extrinsic defects reported in the literature, including $Cu_{Cd}$\cite{yang2016first, dou2021effects}, $Cu_{int}$\cite{yang2016first}, $Cu_{Cd}+Cd_{int}$\cite{yang2016first} and $Cu_{Cd}+Cu_{int}$\cite{yang2016first}. And for group V dopants, in addition to the commonly recognized AX defects and substitutional defects $As_{Te}$\cite{krasikov2018beyond, yang2015enhanced}, we also consider two types of defects complexes that has not been thoroughly studied, including $(Cd_{int}+As_{Te})^{+}$ and $(V_{Te}+As_{Te})^{+}$. The choice of these two defects is based on the potential Coulombic interaction between $As_{Te}^{-1}$ and $Cd_{int}^{2+}$ or $V_{Te}^{2+}$.
 
\section{Methods}
\subsection{\label{sec:level2}FIRST-PRINCIPLES CALCULATION METHODS}

Most DFT calculations apply the generalized gradient approximation (GGA) or local density approximation (LDA) as the exchange-correlation functional. However, it is widely known that DFT with LDA or GGA is likely to underestimate the bandgap of semiconductors. For CdTe, GGA gives us a bandgap of 0.68 eV, which is much smaller than experimental values ($\approx$ 1.5 eV). This can be explained by the overestimation of the delocalization of Cd-4$d$ electrons, which lifts the valence electron energies (the Te-5$p$ valence band) \cite{wu2015lda+}. There are many approaches to correct the exchange-correlation functional, such as self-interaction correction (SIC) calculations\cite{menendez2014electronic}, and hybrid functional of Heyd, Scuseria and Ernzerhof (HSE)\cite{kavanagh2021rapid, yang2016review}. However, these methods are computationally expensive, which makes them impractical for study of defects in alloys due to the massive number of possible configurations. DFT with coulomb self-interaction potentials (GGA+U) is another method that has been frequently used to correct the calculated bandgaps \cite{wu2015lda+}. This method combines Hubbard-like model for a portion of states in the system with Coulomb self-interaction potentials (U) to select bands for correction. Non-integer or double occupations of states are described by introducing of two parameters: (1) U, which reflects the intensity of the on-site Coulomb interaction, and (2) J, which adjusts the intensity of the exchange interaction. Typically, for simplicity, an effective parameter U\textsubscript{eff} $=$ U $-$ J is used. This effective parameter U\textsubscript{eff} is typically referred to as U.

\subsection{\label{sec:level3}Computational Details}

Our investigation reveals that setting U=12.2 eV\cite{xiang2023understanding} for Cd-4$d$ orbitals in CdTe closely matches the experimental lattice constant and bandgap, showcasing the effectiveness of the GGA+U method. In comparison, while the HSE method accurately reproduces the experimental bandgap, it tends to overestimate lattice constants, indicating a larger error margin compared to the GGA+U approach.

Applying the optimized U parameter to zinc blende and wurtzite structures of CdSe, we observe a surprisingly good agreement with experimental data, surpassing the performance of the HSE06 hybrid method. Specifically, the bandgap and lattice parameters of zinc blende CdSe align closely with experimental results, as detailed in Table~\ref{tab:comparison}. For the wurtzite structure, although GGA+U slightly underestimates lattice parameters, the bandgap predictions remain consistent.

This consistent accuracy across both CdTe and CdSe can be attributed to the similar \(d\)-\(s\) coupling inherent to Cd-Te and Cd-Se bonds, with both Te and Se being group VI elements. The GGA+U method effectively lowers the Cd 4\(d\) bands, enhancing \(d\)-\(s\) coupling. With U=12.2 eV, it provides a suitable correction for both materials. The details of U value determination are provided in Appendix~\ref{app:Uvalues}.

Leveraging this optimized U value, we extend our study to the full composition range of the CdSeTe alloy. A notable advantage of the GGA+U method is its lower computational cost compared to HSE06, facilitating the application to supercells of the CdSeTe alloy with varied Se/Te arrangements and defect calculations. This approach also allows for the extension to larger supercells at a manageable computational expense when some defect calculations yield less reliable results due to finite cell-size errors\cite{castleton2004finite}.

Structural optimizations and energy calculations were carried out using the VASP code. For the GGA+U calculations, we employed the Perdew, Burke, and Ernzerhof (PBE) exchange-correlation functional. Additionally, the HSE06 functional with default parameters was utilized for comparative analysis. The calculations were performed with a plane wave cutoff energy of 450 eV for the wave functions. The chosen energy cutoff value is validated in Appendix~\ref{app:cutoff}. Spin-polarized calculations are performed by setting ISPIN to 2. The Brillouin zones for all structures under investigation were sampled using \(\Gamma\)-centered k-point grids. Specifically, a 2\(\times\)2\(\times\)2 k-point grid was employed for both 64-atom and 216-atom supercells, while a 1\(\times\)1\(\times\)1 grid was applied to the 512-atom supercell.

In the study of CdSeTe alloy, we focus on compositions of CdSe\textsubscript{0.25}Te\textsubscript{0.75} and CdSe\textsubscript{0.50}Te\textsubscript{0.50}. This selection is informed by the fact that the cubic phase of CdSe\textsubscript{x}Te\textsubscript{1-x} is stable for $0 \leq x \leq 0.45$\cite{schenk1998lattice}. In contrast, the hexagonal phase (wurtzite structure) of CdSe\(_{x}\)Te\(_{1-x}\) emerges for $0.55 \leq x \leq 1$, which is not considered to be photoactive\cite{poplawsky2016structural, schenk1998lattice}. Moreover, in most state-of-the-art CdSeTe technologies, the Se ratio typically remains below 40\%\cite{li202220, fiducia2019understanding, kuciauskas2023increased}. Consequently, we will not extend our study to higher Se ratios.

Utilizing a supercell program\cite{okhotnikov2016supercell}, we generate 30 configurations for each alloy composition, prioritizing those with the highest frequency of occurrence. The configuration selected as the reference structure is the one whose formation energy is closest to the Boltzmann distribution average of the formation energies for all configurations sharing the same alloy ratio. This reference supercell then serves as the basis for generating defect supercells. In Appendix~\ref{app:Se_dep}, we demonstrate that the Se-Se interaction, energy difference, and lattice constant difference in CdSeTe alloys with the same Se ratio but different Se arrangements are all small. Therefore, the selection of one reference structure will not significantly affect the overall results. In addressing defects within the alloy, we examine multiple configurations that exhibit varying Se/Te arrangement environments. We found some defects, such as $Te_{Cd}$, $As_{Te}$ AX defects, show a significant dependence on the local environment. However, most defects with \(T_d\) symmetry do not exhibit strong dependence on local arrangements. For all these defects, we opt for the Boltzmann distribution average at 873 K as a representative formation energy values across at least four configurations for each point defect. 873K is a typical annealing temperature for CdTe deposition process\cite{scarpulla2023cdte, albin2002effect}.

\begin{table*}[ht]
\centering
\caption{\scriptsize LATTICE CONSTANT, BANDGAP FOR CDTE AND CDSE OBTAINED FROM GGA+U
HSE06 AND EXPERIMENTS.}
\label{tab:comparison}
\begin{adjustbox}{max width=1.0\textwidth}
\begin{tabular}{|c|c|c|c|c|c|c|}
\hline
\textbf{Method} & \multicolumn{2}{c|}{\textbf{GGA+U}}  & \multicolumn{2}{c|}{\textbf{HSE06}}  & \multicolumn{2}{c|}{\textbf{Experiment\cite{landolt1966numerical}}} \\ 
\hline
& \textbf{Bandgap (eV)} & \textbf{$a_0$ (Å)}& \textbf{Bandgap (eV)} & \textbf{$a_0$ (Å)}& \textbf{Bandgap (eV)} & \textbf{$a_0$ (Å)} \\
\hline
CdTe (Zinc Blende) & 1.50 & 6.46 & 1.50 & 6.58 & 1.50 & 6.48 \\
\hline
CdSe (Zinc Blende) & 1.72 & 5.95 & - & - & 1.71 & 5.98 \\
\hline
CdSe (Wurtzite) & 1.79 & $a=b=4.21$ $c=6.86$ & 1.68 & $a=b=4.30$ $c=7.01$ & 1.80 & $a=b=4.30$ $c=7.02$ \\ \hline
\end{tabular}
\end{adjustbox}
\end{table*}

The defect formation energy can be derived from DFT using the supercell method\cite{freysoldt2014first} by the following equation,
\begin{equation}\label{dft_form}
%\begin{align}
E^{f}_{i,q} = E_{i,q} - E_\mathrm{bulk} - \sum n \mu \nonumber + q(E_\mathrm{VBM} + \mu_\mathrm{Fermi}) + E_\mathrm{corr} \ ,
%\end{align}
\end{equation}
where $E^{f}_{i,q}$ is the formation energy of a defect $i$, $E_{i,q}$ is the total energy of one defect state with charge $q$, $E_\mathrm{bulk}$ is the total energy of a bulk unit cell with the same size of the defected cell, and $n\mu$ is the reference energy of $n$ added/removed atoms of an element at given chemical potential $\mu$. For any defects in this paper, the chemical potentials of all species are defined with reference to the elemental standard states. $E_\mathrm{VBM}$ is the valence band maximum as computed from electronic structure of bulk supercell.

$E_\mathrm{corr}$ is the charge correction energy to account for interaction between periodic images. The conventional FNV correction method\cite{freysoldt2009fully} exhibits a tendency for overestimating or underestimating the charge correction, especially when dealing with partially or fully delocalized defects\cite{komsa2012finite}. In order to address this issue and enhance the accuracy of the correction process, a strategic modification is implemented. Instead of directly applying the FNV correction to these complex defects, we choose to initiate the correction process by employing the density derived electrostatic and chemical (DDEC) method to determine the net atomic charges (NACs)\cite{tong2024pervoskite, limas2016introducing, manz2016introducing}. This preliminary step serves the purpose of precisely ascertaining the charge localization within the system. We consider the charge on the defect atom, and also compensating charges nearby.
\begin{figure*}[!t]
\centering
\captionsetup{justification=centering}
\subfloat[]{\label{fig:Corr_As}\includegraphics[width=0.4\linewidth]{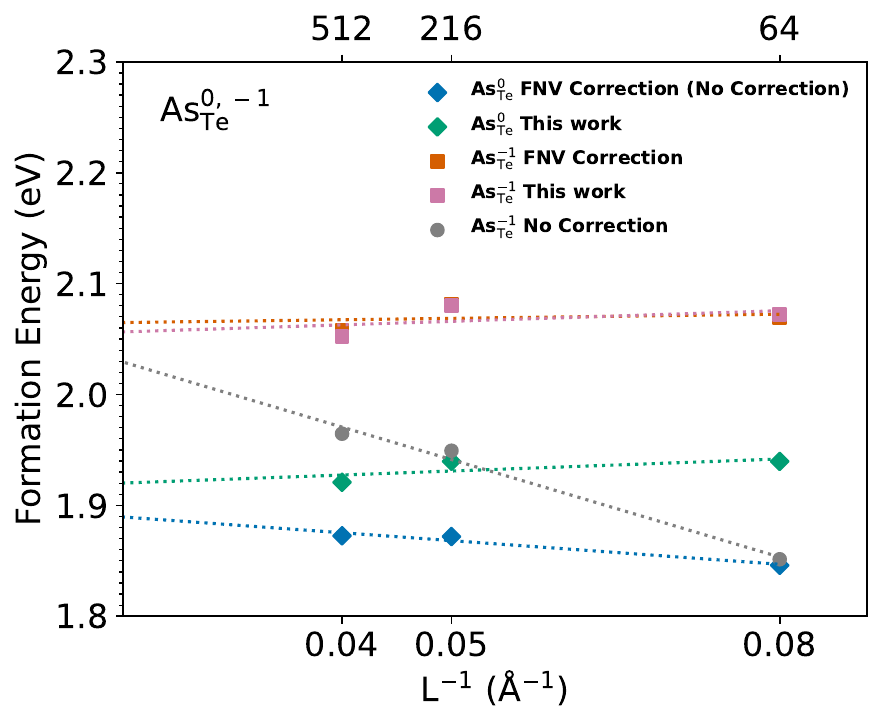}}
%\hspace{-0.5in}
\subfloat[]{\label{fig:Corr_As_comp}\includegraphics[width=0.4\linewidth]{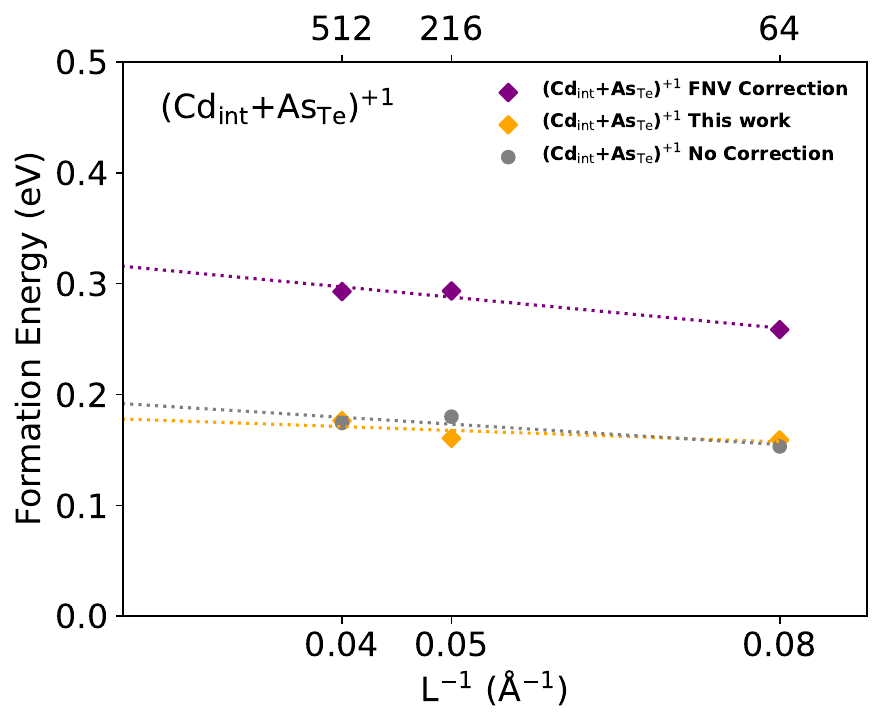}}\\
%\hspace{-0.5in}
\subfloat[]{\label{fig:Corr_Cuint}\includegraphics[width=0.4\linewidth]{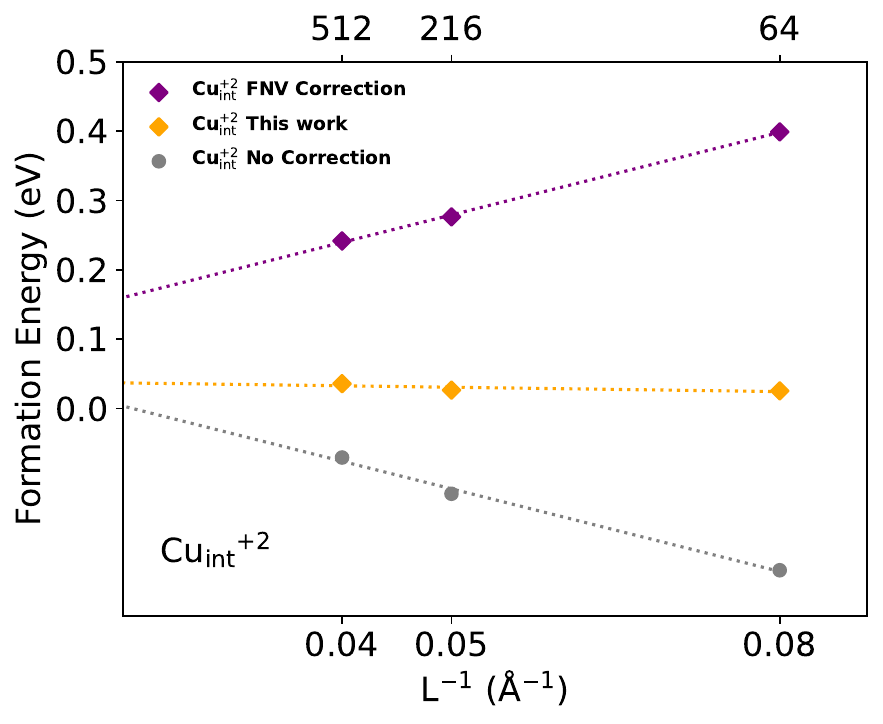}}
\caption{Defect formation energies vs inverse supercell size for (a) $As_{Te}^{0, -1}$, (b) $(Cd_{int}+As_{Te})^{+}$ and (c) $Cu_{int}^{+2}$ . The defects are relaxed with fixed lattice constants. "L" on the bottom x-axis denotes the length of corresponding CdTe supercell. The Fermi level is set to the VBM in these plots.} 
\label{fig:Charge_corr}
\end{figure*}
Following the DDEC analysis, the $q_\mathrm{NAC}$ of the defects are determined. To find out the unscreened localized charge used in FNV correction, a scaling factor $k$ is introduced. This scaling factor $k$ is carefully calibrated to reintroduce the screening effect that may be lost during the charge localization process. Subsequently, $q_\mathrm{NAC}$, scaled by the aforementioned factor $k$, is treated as the input localized charge $q_\mathrm{loc}$ for the FNV charge correction method, 
\begin{equation}\label{localize_charge}
q_\mathrm{loc} = k * q_\mathrm{NAC}.
\end{equation}

The factor $k$ can be determined using one of two approaches. The first approach involves selecting defects that are fully localized; in this scenario, the nominal charge of the defects is equal to \(q_{\mathrm{loc}}\), facilitating straightforward determination of $k$. However, this method may not always be feasible due to the prerequisite knowledge required about defect localization. Alternatively, the second approach entails adjusting the supercell size and iteratively testing $k$ values to achieve convergence in the defect formation energies across varying supercell dimensions. For both strategies, it is advisable to select defects with substantial $q_{\mathrm{NAC}}$ values, as this can lead to significant correction values. Such a selection ensures the derived $k$ values are both reasonable and effective. 

In this study, we select $Te_{Cd}^{+}$ as the reference defect due to its advantageous properties. Specifically, the Kohn-Sham level of $Te_{Cd}^{+}$ is located within the band gap, and it exhibits a large capture cross section for both the (0/+1) and (+1/+2) transitions, as discussed in Sec.~\ref{sec:Intrinsic} and corroborated by existing literature\cite{krasikov2016theoretical}. We firstly apply first approach to determine universal $k$ value of 3.5 for CdTe, CdSe\textsubscript{0.25}Te\textsubscript{0.75}, and CdSe\textsubscript{0.50}Te\textsubscript{0.50}. This $k$ value is then verified by the second approach by increasing supercell size.

Figure~\ref{fig:Charge_corr} showcases the effectiveness of our charge correction method in overcoming the limitations associated with the FNV correction approach. Our method notably addresses several critical issues:
\begin{itemize}
    \item Non-zero NACs in neutral defects, such as $As_{Te}^{0}$ illustrated in Fig.~\ref{fig:Corr_As}, result in image interactions that necessitate charge correction, a scenario not accommodated by FNV correction. This issue has been observed in other work as well \cite{oba2008defect}. It occurs when a carrier fills states with delocalized characteristics, such as states close to the CBM or VBM. Consequently, the free carrier charges do not fully screen the nuclear charge of point defects, leaving a long-range Madelung interaction between the point defects even in neutral supercells. Our method's application of charge correction to these defects is supported by faster convergence, as evident in Fig.~\ref{fig:Corr_As}. For instance, the transition level for $As_{Te}$ (0/-1) shifts from 0.22 eV without charge correction to 0.13 eV with it, more closely aligning with experimental (\(\approx 94\) meV)~\cite{nagaoka2020comparison} and hydrogenic model predictions (\(\approx 113\) meV)~\cite{seeger2013semiconductor}. Similarly, the transition level for $P_{Te}$ (0$/$-1) improves from 0.16 eV to 0.11 eV, nearing experimental values (\(\approx 87\) meV)~\cite{nagaoka2020comparison}.
    \item For charged defects that are not fully localized, our method accurately accounts for the discrepancy between localized charge $q_\mathrm{loc}$ and nominal charge, avoiding the overcorrection observed with FNV correction. This accuracy is validated by the consistent formation energy results for $Cu_{int}$ across supercells of 64, 216, and 512 atoms, as shown in Fig.~\ref{fig:Corr_Cuint}.
    \item Additionally, our approach adeptly determines the localized charge distribution of complex defects, a task for which FNV correction's charge assignment proves ambiguous and results in poor convergence. Our method's ability to assign $q_\mathrm{loc}$ to specific defect positions leads to improved convergence for complex defects such as $(Cd_{int}+As_{Te})^{+}$, as demonstrated in Fig.~\ref{fig:Corr_As_comp}.
\end{itemize}

For localized defects, such as $As_{Te}^{-1}$ in Fig.~\ref{fig:Corr_As}, $Te_{Cd}^{+1}$ and $Te_{Cd}^{+2}$, our method shows little difference compared to FNV correction. These findings highlight the comprehensive capabilities of our charge correction method in accurately modeling defect properties, surpassing the limitations of existing approaches.

\section{Results and Discussion}

\subsection{\label{sec:Intrinsic}Intrinsic Defects}

\begin{figure*}[!t]
%\centering
%\captionsetup{justification=centering}
\subfloat[]{\label{fig:defects_CdTe_Cd_rich}\includegraphics[width=0.3\linewidth]{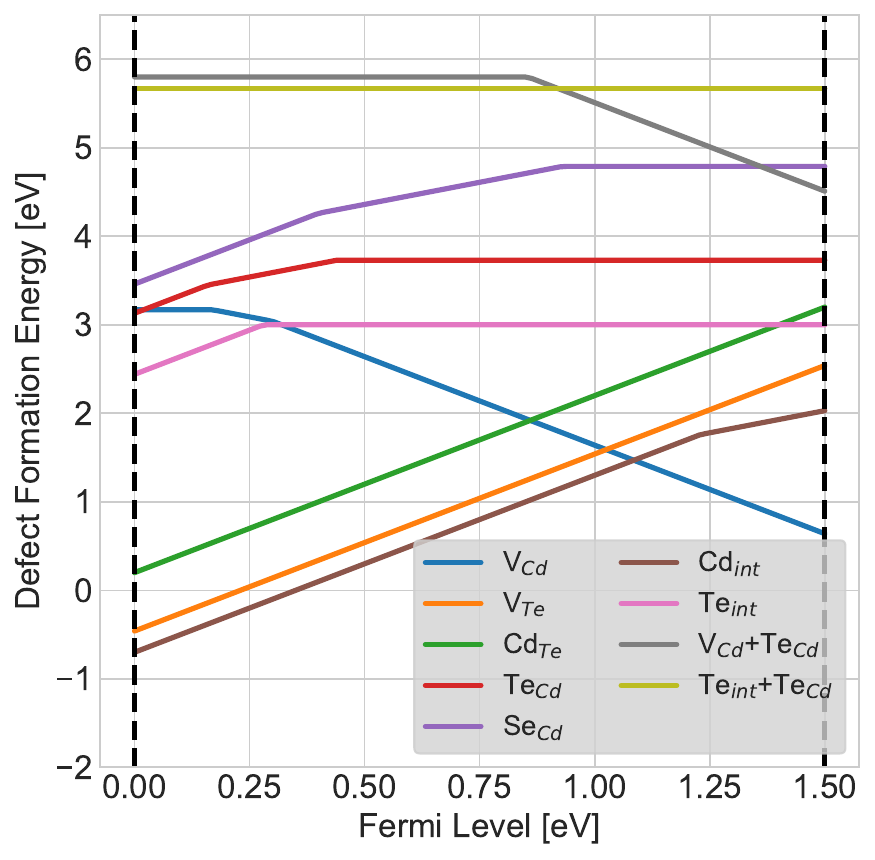}}
%\hspace{-0.5in}
\subfloat[]{\label{fig:defects_CdTe0.75_Cd_rich}\includegraphics[width=0.3\linewidth]{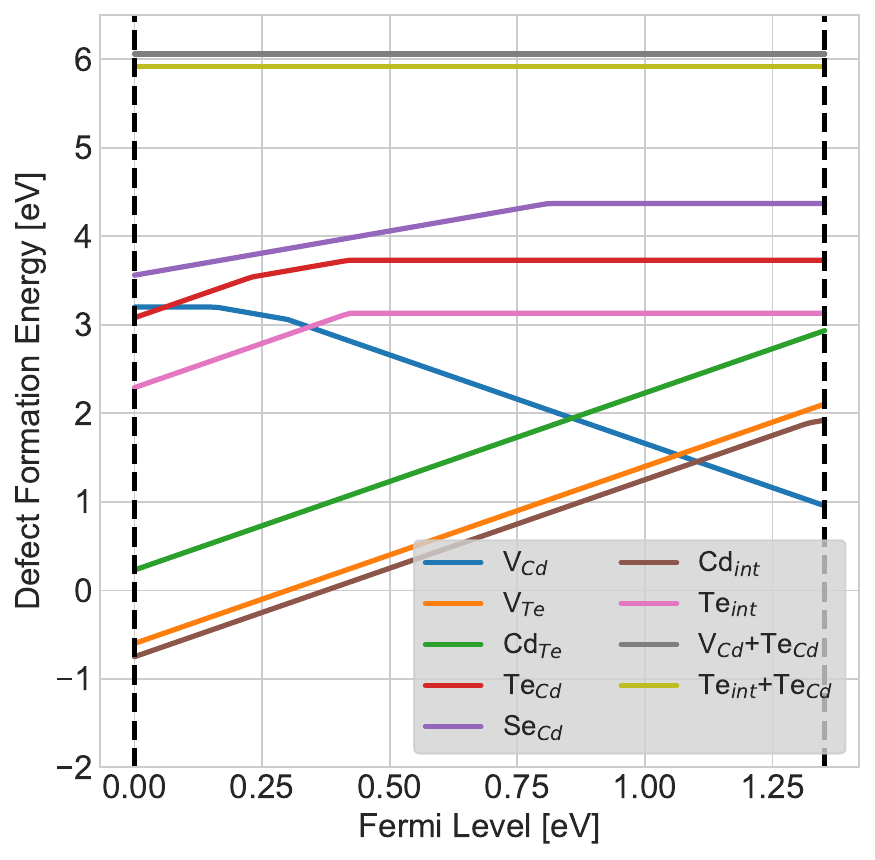}}
\subfloat[]{\label{fig:defects_CdTe0.50_Cd_rich}\includegraphics[width=0.3\linewidth]{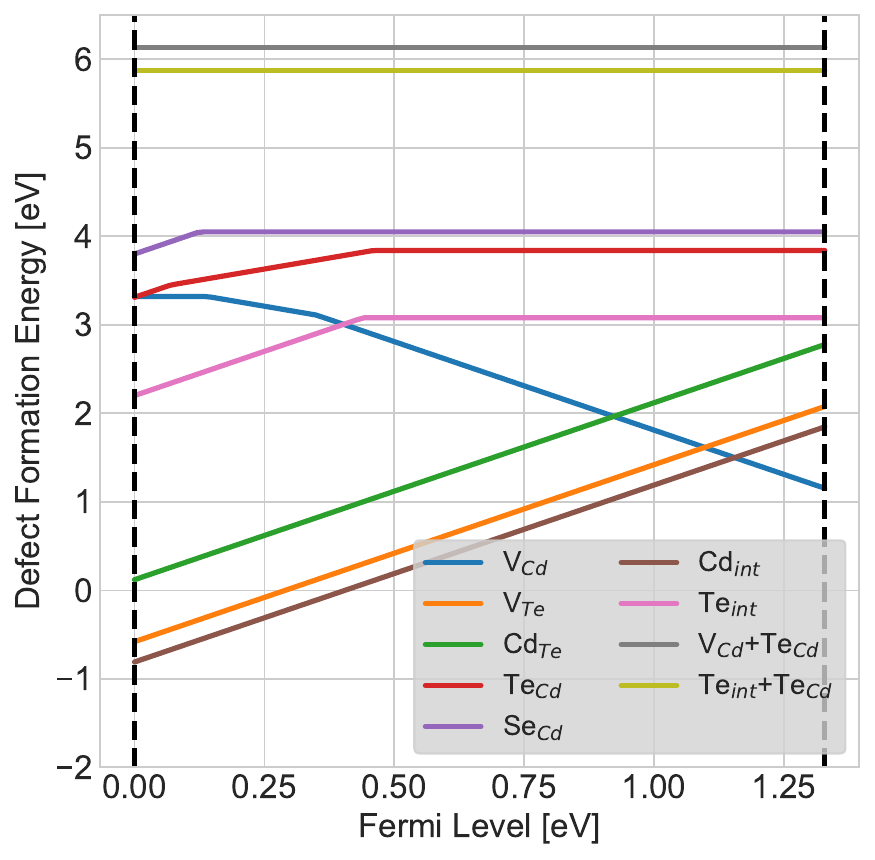}}\\
\subfloat[]{\label{fig:defects_CdTe_Te_rich}\includegraphics[width=0.3\linewidth]{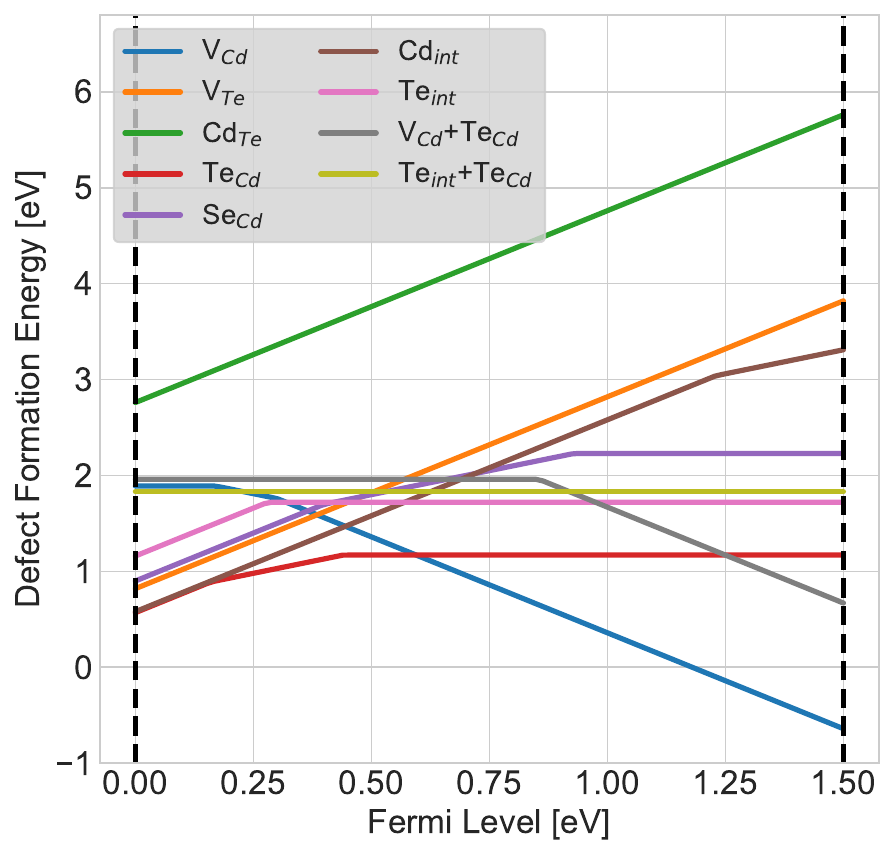}}
%\hspace{-0.5in}
\subfloat[]{\label{fig:defects_CdTe0.75_Te_rich}\includegraphics[width=0.3\linewidth]{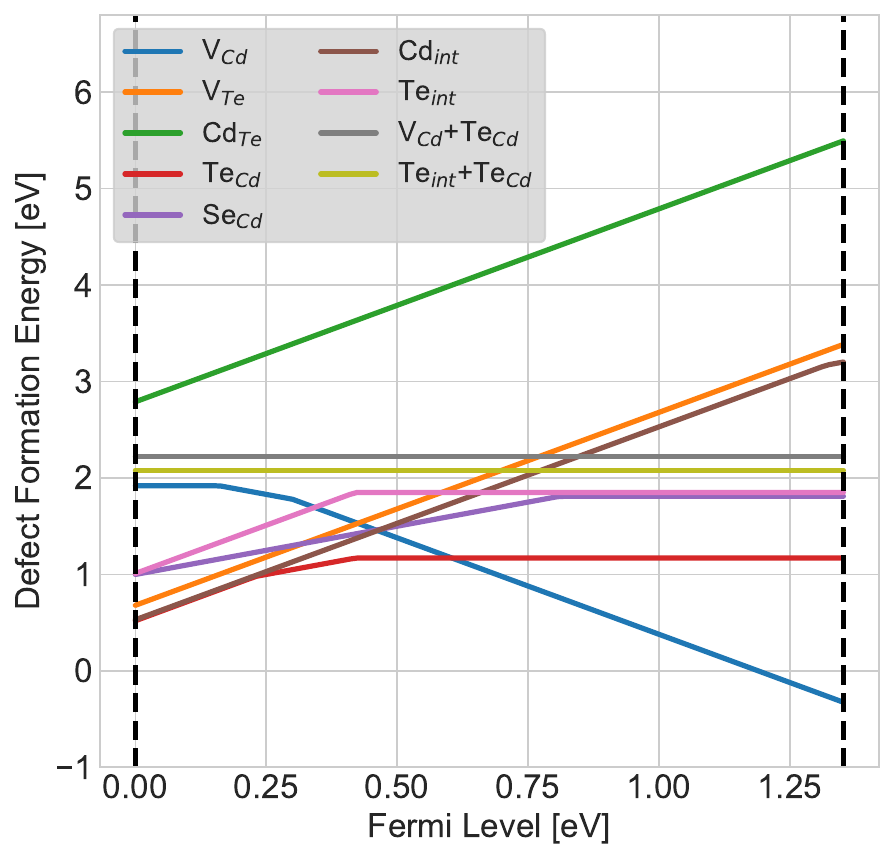}}
%\hspace{-0.5in}
\subfloat[]{\label{fig:defects_CdTe0.50_Te_rich}\includegraphics[width=0.3\linewidth]{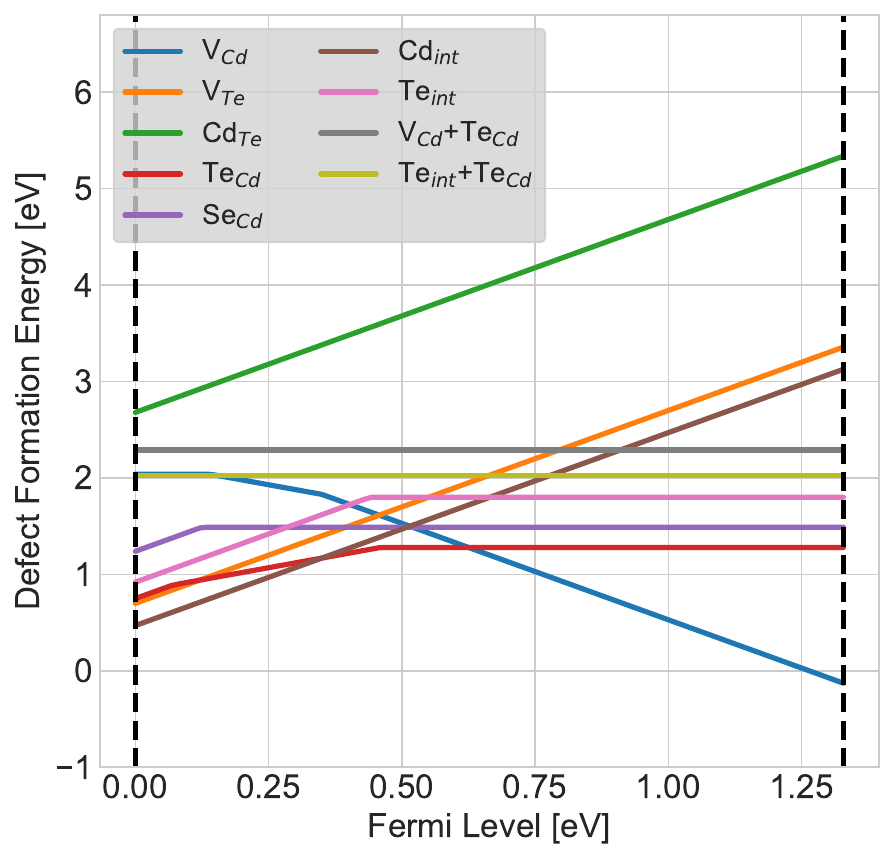}}
%\hspace{-0.5in}
\caption{Defect formation energies vs Fermi level of intrinsic defects in 
CdTe (a and d), CdSe\textsubscript{0.25}Te\textsubscript{0.75} (b and e), and CdSe\textsubscript{0.50}Te\textsubscript{0.50} (c and f) under Cd-rich (a-c) and Te-rich (d-f) conditions. The chemical potential conditions are $\mu_{Cd}+\mu_{Te} = -1.28$ eV and $\mu_{Cd}+\mu_{Se} < 0.15$ eV.}
\label{fig:Defect_trans_all}
\end{figure*}
Defect formation energies of intrinsic defects are depicted in Fig.~\ref{fig:Defect_trans_all}. It is worth noting that there are some negative values of defect formation energies in certain regions. These negative values suggest a very high concentration of defects in these regions. However, these regions are unlikely to be achieved under normal conditions due to their coupling with Fermi level. For example, in Fig.~\ref{fig:defects_CdTe_Cd_rich}, the undoped material is n-type as the dominant defects $V_{Cd}$ and $Cd_{int}$ will cause Fermi level pinning at around 1.1 eV (referenced to VBM). When the system approaches p-type, it will lower the formation energy of $Cd_{int}$, thus more donors will be generated, keeping the system n-type (or at most weakly p-type). Therefore, the negative values near VBM for $Cd_{int}$ indicates that achieving a strong p-type Fermi level in this material is extremely difficult. A similar situation occurs in Fig.~\ref{fig:Defect_trans_all_Cu}.

The cadmium vacancy $V_{Cd}$ is a pivotal intrinsic defect in CdTe and its alloys, with its configuration and transition levels being subjects of significant debate in both theoretical and experimental literature\cite{yang2016review}. Our research reveals that the neutral state of $V_{Cd}$ primarily adopts \(D_{2d}\) symmetry (dimer structure), featuring two fully occupied degenerate states and one unoccupied state. In contrast, the -1 and -2 charged states are most stable in \(T_d\) symmetry. The transition levels of $V_{Cd}$, as established in various studies, range from 0.1 to 0.8 eV above the VBM \cite{emanuelsson1993identification, berding1999native, wei2002chemical, chang2006symmetrized, du2008native, castaldini1998deep, reislohner1998band}. In this work, the transition levels for $V_{Cd}$ in CdTe are identified as 0.17 eV for the (-1/0) transition and 0.30 eV for the (-2/-1) transition. For CdSe\textsubscript{0.25}Te\textsubscript{0.75}, these levels are 0.16 eV and 0.30 eV, respectively, while for CdSe\textsubscript{0.50}Te\textsubscript{0.50} they are 0.14 eV and 0.35 eV, respectively. These findings indicate that the transition levels of $V_{Cd}$ exhibit relative consistency in alloys with a selenium ratio below 50\%. 

\begin{figure}[!t]
\centering
\captionsetup{justification=centering}
\subfloat[]{\label{fig:CCD_V_cd_dimer}\includegraphics[width=0.45\linewidth]{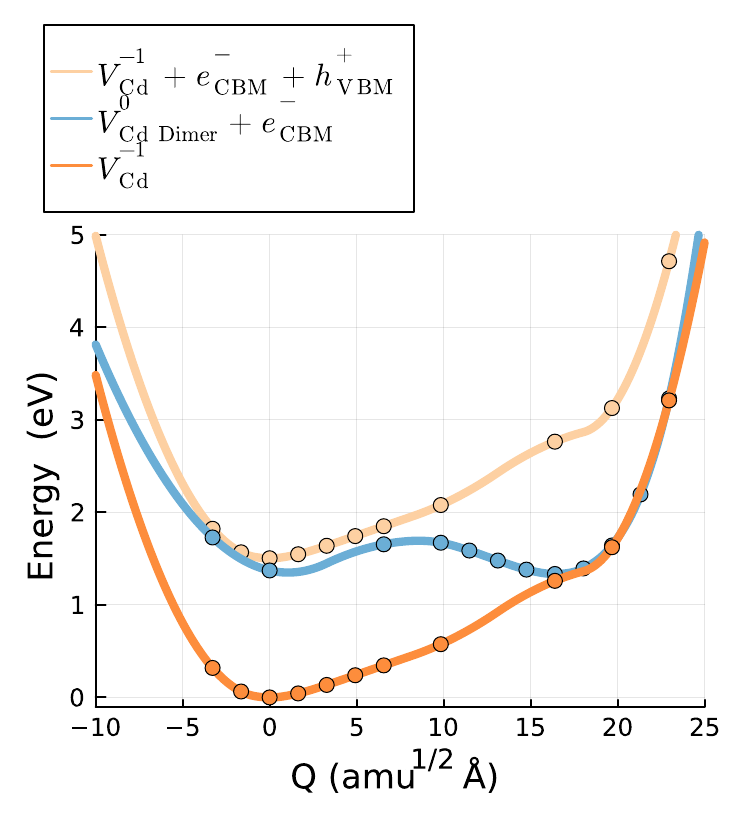}}
\subfloat[]{\label{fig:CCD_V_cd_Td}\includegraphics[width=0.45\linewidth]{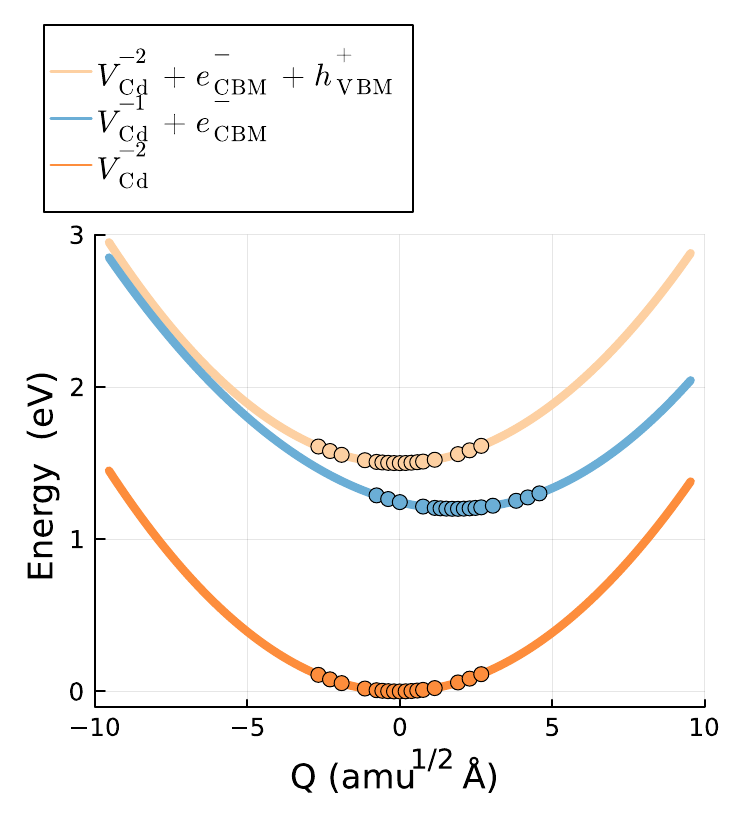}}
\caption{Configuration Coordinate Diagram of $V_{Cd}$ (-1$/$0) and (-2$/$-1). $Q$ in X-axis corresponds to the mass-weighted configuration coordinate. $Q$ indicates the configurational coordinate path between equilibrium configurations. $Q=0$ indicates defects ground state with \(T_d\) symmetry.}
\label{fig:CCD_Vcd}
\end{figure}

Furthermore, we calculated the defect capture cross section of $V_{Cd}$ in CdTe based on static approximation\cite{park2018point, freysoldt2014first, stoneham1981non, alkauskas2014first, Kim2019kesterite, kim2019anham}. The results, presented in Fig.~\ref{fig:CCD_Vcd} and Table~\ref{tab:deep_level_cap}, reveal that the potential energy surface (PES) of the (-2/-1) transition is quasi-harmonic, while the PES of the (-1/0) transition is anharmonic. This discrepancy can be attributed to the proximity of metastable \(D_{2d}\) $V_{Cd}^{-1}$ and \(T_{d}\) $V_{Cd}^{0}$ states, which merge with the most stable configurations of these respective charge states—a phenomenon also observed elsewhere in HSE06 calculation of $V_{Cd}$\cite{kavanagh2021rapid} and defects in GaAs\cite{kim2019anham}. From the carrier capture cross section calculations, we determined that the (-1/0) transition exhibits strong nonradiative recombination intensity, whereas the (-2/-1) transition demonstrates only weak nonradiative recombination. This observation aligns with previous reports using the HSE06 functional\cite{kavanagh2021rapid, krasikov2016theoretical}.

\begin{table}[htbp]
\centering
\caption{\footnotesize Capture cross section of deep level defects in CdTe at 300K}
%\begin{adjustbox}{width=0.4\textwidth}
\begin{tabular}{|c|c|c|c|}
\hline
Defect & Transition Level & Trapping Process & $\sigma(T = 300K) [cm^2]$ \\
\hline
\multirow{2}{*}{$V_{Cd}$} & \multirow{2}{*}{(-2/-1)} & hole capture & $8.0 \times 10^{-17} $ \\\cline{3-4}
 &  & electron capture & $1.9 \times 10^{-20} $ \\\cline{2-3}
 \hline
\multirow{2}{*}{$V_{Cd}$} & \multirow{2}{*}{(-1/0)} & hole capture & $2.8 \times 10^{-16} $ \\\cline{3-4}
 &  & electron capture & $2.1 \times 10^{-15} $ \\\cline{2-3}
\hline
\multirow{2}{*}{$Te_{Cd}$} & \multirow{2}{*}{(+2/+1)} & hole capture & $2.1 \times 10^{-18}$ \\\cline{3-4}
 &  & electron capture & $1.0 \times 10^{-15} $ \\\cline{2-3}
 \hline
 \multirow{2}{*}{$Te_{Cd}$} & \multirow{2}{*}{(+1/0)} & hole capture & $3.3 \times 10^{-15}$ \\\cline{3-4}
 &  & electron capture & $8.1 \times 10^{-18} $ \\\cline{2-3}
 \hline
\end{tabular}
%\end{adjustbox}
\label{tab:deep_level_cap}
\end{table}

\begin{figure}[!t]
\centering
\captionsetup{justification=centering}
{\includegraphics[width=0.4\linewidth]{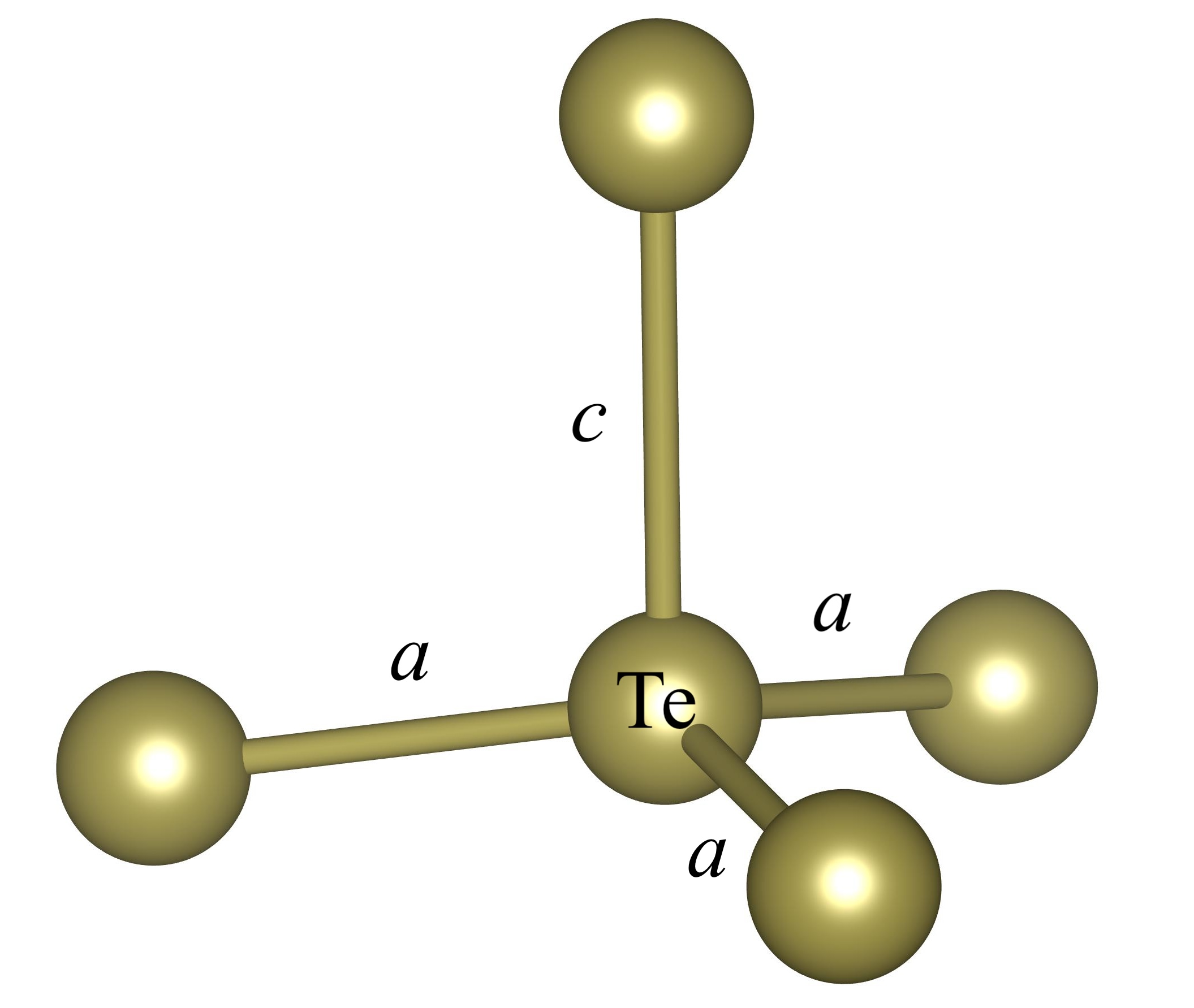}}
\caption{Schematic of $Te_{Cd}$ \(C_{3v}\) defect configuration. $c$ denotes length of the long bond. $a$ denotes length of three short bonds.}
\label{fig:Te_cd_c3v}
\end{figure}

\begin{figure}[!t]
\centering
\captionsetup{justification=centering}
{\includegraphics[width=0.7\linewidth]{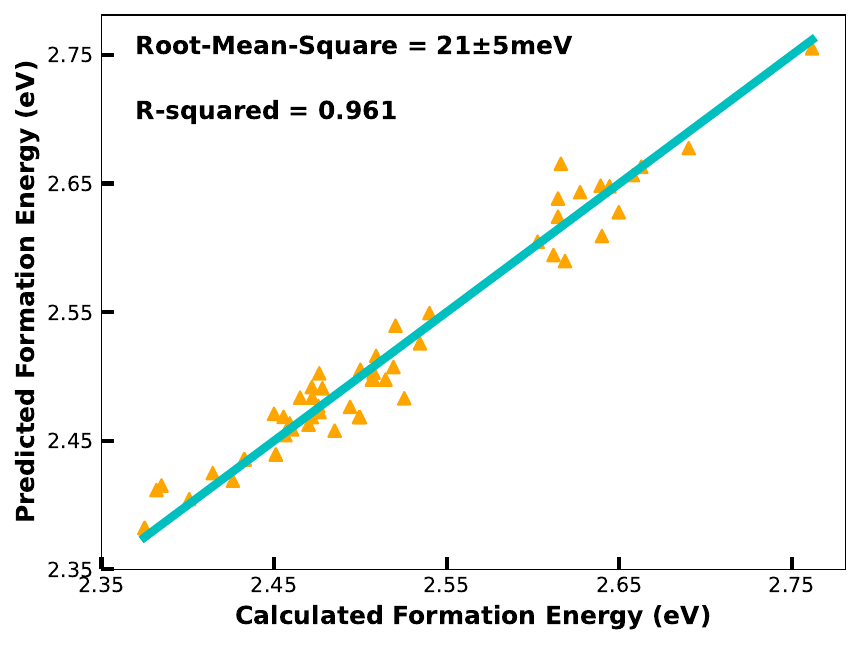}}
\caption{Comparison between Defect Neighbor Model predictions and DFT calculations for the $Te_{Cd}^{0}$ defect across Se ratios ranging from 0\% to 50\% (56 data included). The root mean square error (RMSE) for test data and correlation coefficient (\(R^2\)) demonstrate excellent agreement with the data.}
\label{fig:Regression_TeCd}
\end{figure}

The second critical intrinsic defect under investigation is $Te_{Cd}$. The charge states $Te_{Cd}^{+}$ and $Te_{Cd}^{0}$ both exhibit \(C_{3v}\) symmetry as shown in Fig.~\ref{fig:Te_cd_c3v}, while $Te_{Cd}^{2+}$ displays \(T_d\) symmetry. A significant dependence on the Se/Te arrangement is observed for $Te_{Cd}$ defects with \(C_{3v}\) symmetry. To exemplify this, we considered $Te_{Cd}^{0}$ and calculated its formation energy for 56 different local Se/Te environments. These 56 $Te_{Cd}^{0}$ defects were randomly split into an 80\% training set and a 20\% testing set. We employed a neighbor counting linear regression model, as detailed in Table~\ref{tab:neighbor model TeCd}, which effectively captures this dependence using eight features. This regression model was trained using a bootstrapping strategy, a resampling technique that involves repeatedly drawing samples from our dataset with replacement\cite{tibshirani1993introduction}. We repeat this process 10 times. The fitting results, illustrated in Fig.~\ref{fig:Regression_TeCd}, demonstrate the model's accuracy.

The eight features considered in the model include the Se ratio, the square of Se ratio, the presence of a first nearest neighbor (1NN) Se atom with a long bond, the count of 1NN Se atoms with short bonds, the square of the count of 1NN Se atoms with short bonds, the presence of second nearest neighbors (2NN) with one long and one short bond, 2NN with one long bond, and 2NN with one short bond. The positive coefficients for the Se ratio and Se ratio squared suggest that higher Se ratios make the formation of $Te_{Cd}^{0}$ less favorable. The feature 1NN Se with a long bond indicates whether the elongated bond in the \(C_{3V}\) configuration is occupied by Se, with its negative coefficient implying that the presence of Se in this bond promotes \(C_{3V}\) defect formation. The features 1NN Se with short bonds and its square capture a quadratic relationship between the formation energy of \(C_{3V}\) defects and the occupancy of short bonds by Se, with the minimum of the quadratic function occurring near a Se ratio of 0.8. This indicates a decrease in formation energy as more Se atoms occupy the 1NN short bonds with Se ratio less than 80 \%. The final three features account for the 2NN Se/Te atoms and whether they are neighbors to the 1NN with elongated or short bond, or both. The positive coefficient for 2NN (1 Long, 1 Short) and negative coefficients for both 2NN (1 Long) and 2NN (1 Short) reflect the diverse preferences for the \(C_{3V}\) defect configuration. This model can be further integrated with Lattice Monte Carlo simulations to elucidate the impact of the local environment on $Te_{Cd}$. Additionally, we performed calculations for the capture cross section of this defect, with results presented in Table~\ref{tab:deep_level_cap}. These findings validate that $Te_{Cd}$ can create deep-level trap centers, aligning with previous HSE06 calculations and experimental observations\cite{krasikov2016theoretical, ma2013dependence}.

\begin{table}[htbp]
\centering
\caption{\small $Te_{Cd}^{0}$ Defect Neighbor Model: coefficients $\Delta E$ and their standard deviations for each feature. For nearest neighbor counting features, the count is specific to Se atoms. The distinction between short bond and long bond is shown in Fig.~\ref{fig:Te_cd_c3v}.}
\begin{adjustbox}{max width=0.47\textwidth}
\begin{tabular}{|c|c|c|c|}
\hline
Feature & $\Delta E$ (meV) & Feature & $\Delta E$ (meV) \\
\hline
Se ratio & $383\pm52$ & (1NN Se with short bonds)$^2$ & $32\pm2$ \\
\hline
(Se ratio)$^2$ & $760\pm70$ & 2NN (1 Long, 1 Short) & $38\pm2$ \\
\hline
1NN Se with long bond & $-74\pm4$ & 2NN (1 Long) & $-44\pm2$ \\
\hline
1NN Se with short bonds & $-51\pm8$ & 2NN (1 Short)  & $-29\pm2$ \\
\hline
\end{tabular}
\end{adjustbox}
\label{tab:neighbor model TeCd}
\end{table}

The defects $V_{Te}$, $Cd_{int}$, and $Cd_{Te}$ are identified as shallow donors, exhibiting several common characteristics. Both $V_{Te}$ and $Cd_{Te}$ display negative-U behavior. The (+2/0) transition states of $V_{Te}$ and $Cd_{Te}$, along with the (+2/+1) transition state of $Cd_{int}$, are situated near the conduction band minimum (CBM) or embedded within the conduction band (CB). Consequently, the +2 charge state is predominant for all three defects in \(p\)-type CdTe and its alloys. Only minor variations in their formation energies are observed with Se alloying, suggesting that these defects will continue to act as compensating defects in intrinsic CdSeTe alloys. Among these, $Cd_{Te}$ exhibits the highest formation energy under both Cd-rich and Te-rich conditions, indicating that it is unlikely to be the dominant compensating defect. The neutral state of $V_{Te}$ is most stable in a Jahn-Teller distorted structure, consistent with HSE06 calculations\cite{yang2016review}. The $V_{Te}^{+2}$ state is most stable in \(T_d\) symmetry, while $Cd_{int}^{+2}$ is most stable when positioned in a tetrahedral vacant site surrounded by four group VI atoms.

%Need Aaron's input, MD simulation details, a structure figure would be better%
Interstitial tellurium $Te_{int}$ is identified as a donor. We used ab-initio molecular dynamics (MD) to run simulated annealing, a technique for finding global energy minimums of $Te_{int}$ without an initial guess structure. Essentially, the system begins at high temperature and is slowly cooled, eventually settling into a low energy configuration. MD was run in the canonical (NVT) ensemble, using the Nose-Hoover thermostat to control the system temperature. The system was advanced with a 6 femtosecond time step for 1000 steps, starting at 1000K and ending at 1K. We found $Te_{int}$ prefers to form split interstitial oriented along the [110] directions. Ref. \cite{ma2014correlation} find identical stable configurations. Some studies suggest that $Te_{int}$ could be a deep-level defect\cite{krasikov2016theoretical, ma2013dependence}, and our findings indicate that the transition level of $Te_{int}$ is indeed located near the mid-gap. However, the formation energy of $Te_{int}$ is not particularly favorable, even under Te-rich conditions. Based on its formation energy and our compositionally constrained thermodynamic (CCT) method\cite{mutter2015calculation}, we estimate that the density of $Te_{int}$ at 300K is around three orders of magnitude lower than the carrier density and the densities of two other deep-level defects, $Te_{Cd}$ and $V_{Cd}$, under Te-rich condition. Consequently, the effectiveness of $Te_{int}$ in limiting carrier lifetime is questionable. This aspect will be further explored in our future work.

\begin{figure}[!t]
\centering
\captionsetup{justification=centering}
{\includegraphics[width=0.6\linewidth]{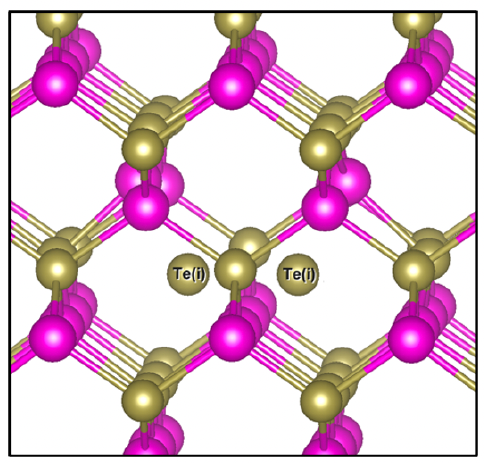}}
\caption{Schematic of $Te_{int}$ split defect configuration}
\label{fig:Te_int}
\end{figure}

The defect $Te_{Cd}$ can potentially form complexes with $Te_{int}$ and $V_{Cd}$ due to strong binding energy\cite{krasikov2017defect}. We have investigated the formation energies of the $V_{Cd}+Te_{Cd}$ and $Te_{int}+Te_{Cd}$ complexes in Fig.~\ref{fig:Defect_trans_all} and employed the CCT to estimate their densities. However, the results are similar to those for $Te_{int}$. We determined that the density of these two complexes at 300 K is approximately three orders of magnitude lower than the carrier density. Further investigation is required to ascertain whether these complexes significantly impact the defect distribution.

We have also explored dominant deep level defect formation energy in CdSeTe alloy with varied Se/Te ratio. Dominant defect formation energy in CdTe and CdSeTe for different charge states are depicted in Fig.~\ref{fig:Defect_trans_all}. It is clearly observable that deep level defects have higher formation energy in Cd-rich condition. In fact, both experimental observation and HSE06 results validate this observation that as CdTe becomes more Cd-rich, a longer minority carrier lifetime can be achieved\cite{ma2013dependence}. This conclusion remains valid in CdSeTe alloy. 
Moreover, from experiment observation\cite{balcioglu2000deep, krasikov2019selenium}, the bulk recombination in CdSeTe alloy is lowered in CdTe. Fig.~\ref{fig:Defect_trans_all} shows that the formation energy of deep level defects ($Te_{Cd}$ and $V_{Cd}$) slightly increases as we raise the Se ratio. Specifically, The $E_f$ of $V_{Cd}^{0}$ is 0.15 eV higher in 50\% alloy while $E_f$ of $Te_{Cd}^{+2}$ is 0.18 eV higher. This suggests a lower density of deep level centers in CdSeTe alloy. In addition, Se alloying could possibly deactivate the recombination centres by forming Se-complexes\cite{krasikov2019selenium}. Thus, it is expected that bulk recombination will be reduced in CdSeTe alloy. 

Examination of Fig.~\ref{fig:defects_CdTe_Te_rich}, Fig.~\ref{fig:defects_CdTe0.75_Te_rich}, and Fig.~\ref{fig:defects_CdTe0.50_Te_rich} reveals that the Fermi level $E_F$ is consistently pinned around 0.6 eV across CdTe, CdSe\textsubscript{0.25}Te\textsubscript{0.75}, CdSe\textsubscript{0.50}Te\textsubscript{0.50} compositions for Te-rich condition. Under such condition, the dominant acceptor is $V_{Cd}$ and the dominant compensating donors are $Te_{Cd}$, $Cd_{int}$, $V_{Te}$ and $Se_{Cd}$. For Cd-rich condition, $E_F$ is observed to be pinned between 1.2 eV and 1.3 eV for the same compositions. Under such condition, the dominant acceptor is $V_{Cd}$ and the dominant compensating donors are $Cd_{int}$, and $V_{Te}$. These findings align closely with Yang's first-principles calculations using the HSE06 functional\cite{yang2016review}, highlighting the intrinsic self-doping limits under equilibrium growth conditions. Interestingly, this limitation may potentially be circumvented by employing rapid cooling from high temperatures\cite{yang2016review}. Nonetheless, the current method for calculating defect formation energies in the alloy, utilizing a Boltzmann distribution average at 873 K, simplifies the analysis but might not adequately address the intricacies arising from defects significantly influenced by the Se/Te arrangement. In addition, defect interaction is omitted when the material is cooling down from high temperature. Achieving a more detailed defect profile necessitates the integration of Lattice Monte Carlo simulations and continuum simulations\cite{xiang2021coupled, gehrke2023atomistic, xiang2024coupled}. These advanced simulation techniques are planned for inclusion in our future work, promising a deeper understanding of defect dynamics within the alloy.

\subsection{Group V Dopants}
\begin{figure}[!t]
\centering
\captionsetup{justification=centering}
{\includegraphics[width=0.6\linewidth]{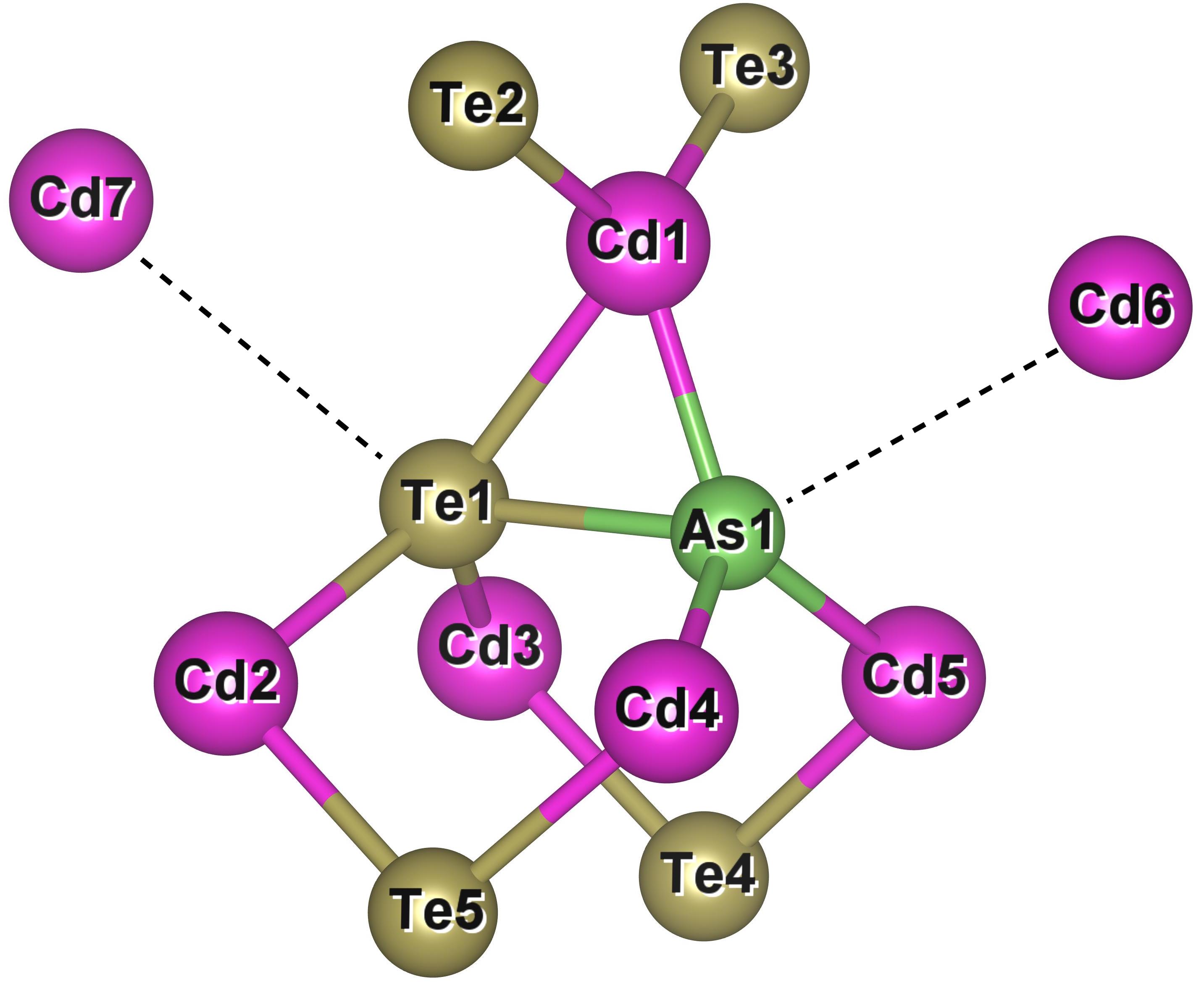}}
\caption{Schematic of AX center defect configuration}
\label{fig:AsTe}
\end{figure}

\begin{figure}[!t]
\centering
\captionsetup{justification=centering}
{\includegraphics[width=1.0\linewidth]{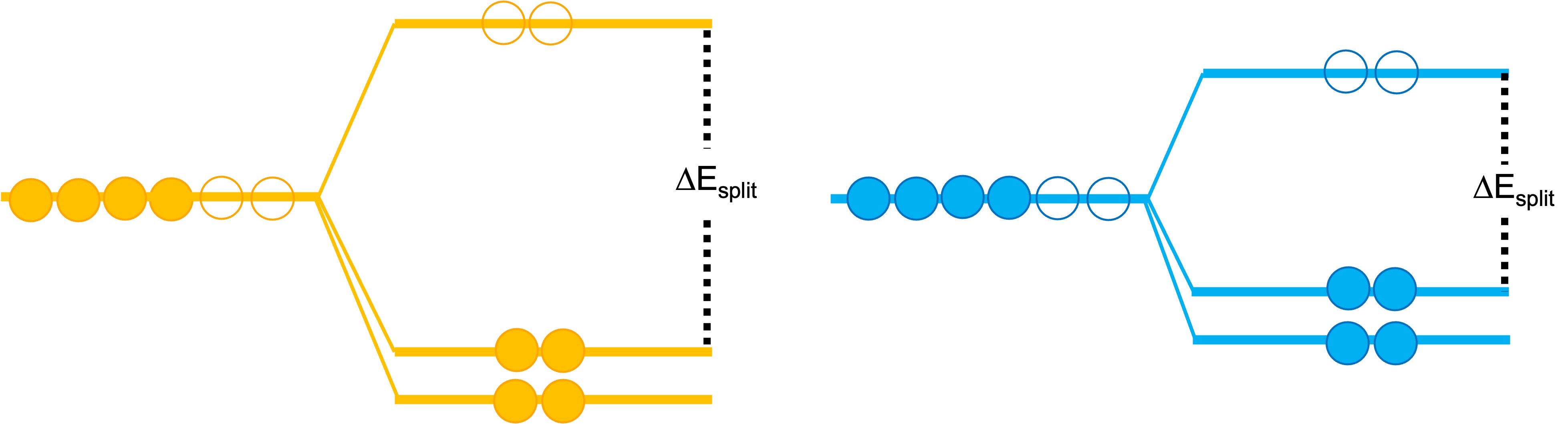}}
\caption{Schematic of AX center defect band splitting. Left one indicates AX in CdTe and right one indicates AX in CdSe\textsubscript{0.25}Te\textsubscript{0.75}. $\triangle E_{split}$ is the band splitting energy}
\label{fig:Bandsplit}
\end{figure}
Arsenic (As) and phosphorus (P) have been identified as effective dopants in CdTe-based materials, offering potential for enhancing p-type doping in solar cells\cite{yang2015enhanced}. However, the behavior of As and P in CdSeTe, such as the primary sources of compensating defects in As- and P-doped CdTe, remains less understood\cite{chatratin2023doping}.

As anticipated, As can substitute for Te or Se in CdSeTe, acting as an acceptor due to its group V nature, possessing one electron fewer than Te. In this study, we focus on the substitutional defect $As_{Te}$ and $P_{Te}$, as other related defects, like As/P interstitials, exhibit relatively higher formation energies\cite{yang2015enhanced}. Our calculations show that the neutral and -1 charged states of $As_{Te}$ and $P_{Te}$ display \(T_d\) symmetry, consistent with previous reports\cite{yang2015enhanced}. The calculated (0/-1) transition levels for \(As_{Te}\) and \(P_{Te}\) are 0.13 eV and 0.18 eV, respectively (Table~\ref{tab:As_Defect_Formation}, Table~\ref{tab:P_Defect_Formation}, Fig.~\ref{fig:defects_trans_CdTe}), indicating shallow levels suitable for $p$-type CdTe doping. However, the transition level of $As_{Te}$ shifts toward the CBM, reaching 0.31 eV above the valence band maximum (VBM) when the Se ratio is 50\% (Table~\ref{tab:As_Defect_Formation}, Table~\ref{tab:P_Defect_Formation}, Fig.~\ref{fig:defects_trans_CdSe0.25}, Fig.~\ref{fig:defects_trans_CdSe0.50}). Conversely, the transition level of $P_{Te}$ remains close to the VBM even as the Se ratio increases. This suggests that for As-doped CdSeTe, careful control of the Se ratio is necessary to maintain reasonable As doping efficiency.

Furthermore, As and P can form a +1 charged AX center defect, converting it into a donor\cite{yang2015enhanced, krasikov2018beyond}. As depicted in Fig.~\ref{fig:AsTe}, this asymmetric defect occurs when As/P shifts toward one of the neighboring Te atoms, causing the triply degenerate \(T_d\) state to split into two fully occupied states and one empty state, as illustrated in Fig.~\ref{fig:Bandsplit}. This process results in the breaking of one bond with a neighboring Cd atom for each of the two atoms. If the band splitting energy $\Delta E_{split}$ exceeds the bonding energy $E_\mathrm{bond}$ with Cd, the AX defect becomes more stable than the \(T_d\) As/P substitutional defect. However, our calculations indicate that in CdTe the formation energy of the As AX center is 1.92 eV, similar to that of \(T_d\) $As_{Te}^{+1}$ at 1.93 eV. Similarly, the P AX center has a formation energy of 2.04 eV, close to \(T_d\) $P_{Te}^{+1}$ at 2.10 eV, suggesting that the AX center does not provide additional stability.

As the Se ratio increases, the \(T_d\) +1 charge state of $As_{Te}$ exhibits a formation energy of 1.93 eV in CdTe, 1.89 eV in CdSe\textsubscript{0.25}Te\textsubscript{0.75}, and 1.84 eV in CdSe\textsubscript{0.50}Te\textsubscript{0.50}, indicating a slight preference for the \(T_d\) +1 defect with higher Se alloying. However, the formation energy of the AX defect significantly decreases with increasing Se ratio in the CdSeTe alloy. For instance, in CdSe\textsubscript{0.25}Te\textsubscript{0.75}, the formation energy of the AX defect ranges from 1.57 eV to 1.79 eV for different Se/Te arrangements, compared to 1.92 eV in CdTe. This suggests a possible interaction between Se and the AX defect that could lower the formation energy of the AX defect and compensate for p-type doping. As shown in Fig.~\ref{fig:Bandsplit}, the favorable formation energy of the AX defect is not attributed to an increase in $\Delta E_\mathrm{split}$, but rather to the nearby Se atom potentially weakening the bond between Cd and As/Te. In fact, our study of a 64-atom supercell of CdSeTe alloy with only one Se atom in the group VI lattice revealed that the lowest formation energy of the AX defect occurred when either the Te4 or Te5 atom was replaced with a Se atom. From Fig.~\ref{fig:AsTe}, it can be inferred that when a smaller Se atom is placed into positions Te4 and Te5, the nearby Cd atoms are compressed in the direction of As1-Te1, strengthening the As1-Te1 bond but weakening the As1-Cd6 and Te1-Cd7 bonds. Moreover, we observed that the As atom tends to move toward the Te atom to form a dimer structure in the alloy, while moving toward the Se atom is not energetically favorable. A similar behavior is observed for $P_{Te}$ as well.

\begin{figure}[!t]
\centering
\captionsetup{justification=centering}
\subfloat[]{\label{fig:As_CCD_As_Te}\includegraphics[width=0.50\linewidth]{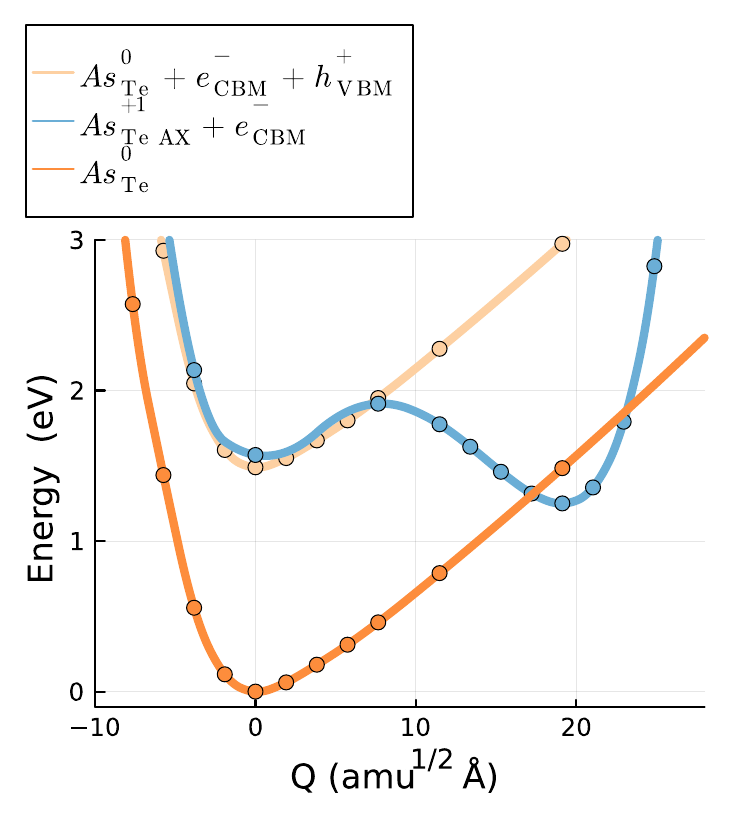}}
\subfloat[]{\label{fig:As_CCD_Cd_int}\includegraphics[width=0.50\linewidth]{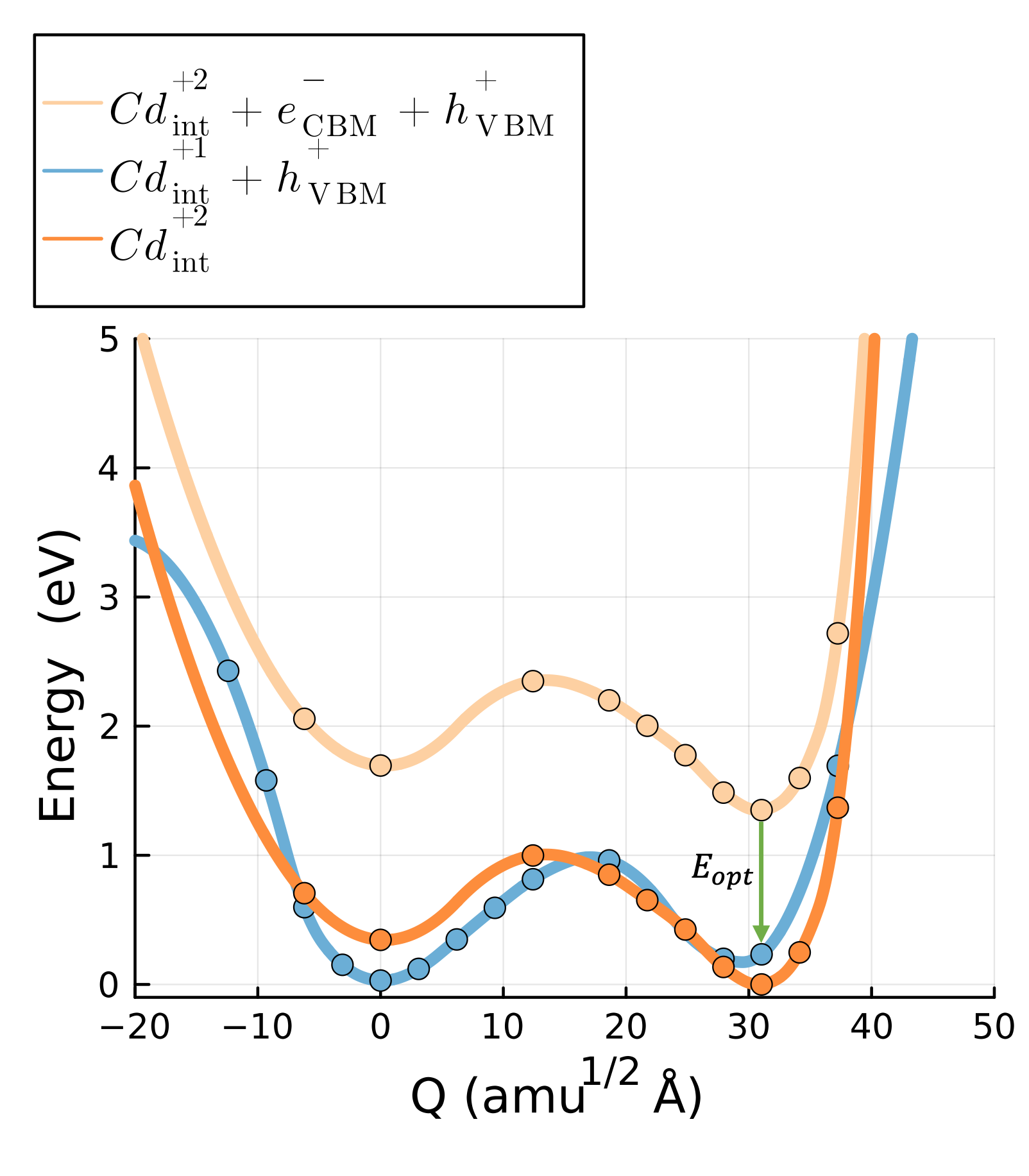}}
\caption{Configuration Coordinate Diagram of $As_{Te}$ (+1\_AX/0) and $Cd_{int}$ (+2/+1) in CdSe\textsubscript{0.25}Te\textsubscript{0.75}. $As_{Te}^{+}$ with formation energy close to the Boltzmann distribution at 300 K is selected as the excited state. $As_{Te}^{0}$ exhibits minimal dependence on the Se/Te arrangement, and the choice of $As_{Te}^{0}$ does not significantly affect the results. In (b), $E_{opt}$ corresponds to the optical transition level. $Q=0$ indicates interstitial ground state positioned in a octahedra vacant site. $Q\approx30$ indicates interstitial ground state positioned in a tetrahedral vacant site. $Cd_{int}^{+1}$ is favorable in octahedra site while $Cd_{int}^{+2}$ is favorable in tetrahedral site.}
\label{fig:As_CCD}
\end{figure}

To further unveil the impact of AX center defects on group V dopability, we explore the transition mechanism between the AX center defect and the neutral $As_{Te}$ substitutional defect within a 25\% CdSeTe alloy. We plot the configuration coordinate diagram (CCD) of $As_{Te}$ (+1\_AX/0) in Fig.~\ref{fig:As_CCD_As_Te}. The significant barrier (\(\sim 0.4\) eV) between the AX and \(T_d\) $As_{Te}^{+1}$ states likely limits the occurrence of AX defects, a finding consistent with similar transition barriers reported in CdTe\cite{krasikov2018beyond}. Although certain alloy arrangements may promote AX defect formation, the overall quantity of AX defects may still be limited, due to the notable transition barrier between the AX defect and the positively charged \(T_d\) $As_{Te}$ defect. Recent computational findings\cite{chatratin2023doping} indicate that AX centers do not constitute a bottleneck for p-type doping in CdTe and CdSeTe alloys. Therefore, the significance of AX centers on the dopability of group V elements requires further investigation, taking into account both the kinetic and thermodynamic aspects of AX defect formation.

\begin{table}[!t]
\centering
\caption{\small Arsenic Defect Formation Energy (eV) in CdSe\textsubscript{x}Te\textsubscript{1-x}.}
\begin{adjustbox}{max width=0.47\textwidth}
\begin{tabular}{|c|c|c|c|c|c|c|c|}
\hline
 & $As_{Te}^{+1}$ AX & $As_{Te}^{+1}$ $T_d$ & $As_{Te}^{0}$ & $As_{Te}^{-1}$ & (-1/0) & $Cd_{int}+As_{Te}$ & $V_{Te}+As_{Te}$ \\
\hline
CdTe & 1.92 & 1.93 & 1.94 & 2.07 & 0.13 & 0.16 & 2.34 \\
\hline
CdSe\textsubscript{0.25}Te\textsubscript{0.75} & 1.57$\sim$1.79 & 1.89 & 1.93 & 2.11 & 0.18 & 0.09 & 2.42  \\
\hline
CdSe\textsubscript{0.50}Te\textsubscript{0.50} & 1.70$\sim$1.87 & 1.84 & 1.87 & 2.19 & 0.31 & 0.06 & 2.42  \\
\hline
\end{tabular}
\end{adjustbox}
\label{tab:As_Defect_Formation}
\end{table}

\begin{table}[!t]
\centering
\caption{\small Phosphorus Defect Formation Energy (eV) in CdSe\textsubscript{x}Te\textsubscript{1-x}.}
\begin{adjustbox}{max width=0.47\textwidth}
\begin{tabular}{|c|c|c|c|c|c|c|c|}
\hline
 & $P_{Te}^{+1}$ AX & $P_{Te}^{+1}$ $T_d$ & $P_{Te}^{0}$ & $P_{Te}^{-1}$ & (-1/0) & $Cd_{int}+P_{Te}$ & $V_{Te}+P_{Te}$ \\
\hline
CdTe & 2.04 & 2.10 & 1.93 & 2.04 & 0.11 & 0.20 & 2.51 \\
\hline
CdSe\textsubscript{0.25}Te\textsubscript{0.75} & 1.72$\sim$1.94 & 2.08 & 2.05 & 2.09 & 0.04 & 0.08 & 2.46  \\
\hline
CdSe\textsubscript{0.50}Te\textsubscript{0.50} & 1.82$\sim$2.01 & 2.10 & 2.07 & 2.13 & 0.06 & 0.05 & 2.45  \\
\hline
\end{tabular}
\end{adjustbox}
\label{tab:P_Defect_Formation}
\end{table}

A remaining question is the identity of the dominant compensating donor. Chatratin 2023\cite{chatratin2023doping} suggests that $V_{Te}$ or $Cd_{Te}$ could be potential compensating donors, but they exclude $Cd_{int}$ due to its high mobility. However, as shown in Fig.~\ref{fig:Defect_trans_all}, it is questionable whether $Cd_{Te}$ can become the dominant compensating donor, as its formation energy is less favorable compared to the other two, particularly in Cd-rich conditions where As and P doping is efficient\cite{nagaoka2020comparison}. Thus, we lean towards $V_{Te}$ and $Cd_{int}$ as the likely dominating compensating defects. Although the high mobility of $Cd_{int}$ might lead to instability, we find that $Cd_{int}$ can bind with $As_{Te}^{+1}$ or $P_{Te}^{+1}$, forming $Cd_{int}+As_{Te}$ or $Cd_{int}+P_{Te}$ complexes due to electrostatic interaction, thereby increasing its stability. The formation energies of these complexes are relatively low, suggesting they could be significant in CdSeTe (Table~\ref{tab:As_Defect_Formation}, Table~\ref{tab:P_Defect_Formation}). In fact, CCT results indicate that over 80\% of $Cd_{int}$ is in complex form.

Recently, \textit{Kuciauskas et al.}\cite{kuciauskas2023increased} investigated defects in CdSeTe with As doping. Their spectral data revealed a defect with an activation energy of 0.14 eV – 0.22 eV from the CBM in undoped CdSeTe, which changes or disappears after As doping. They speculated that this defect could be $V_{Te}$. To test this hypothesis, we calculated the formation energy and transition levels of $V_{Te}$ in a 25\% CdSeTe alloy. From two typical configurations of $V_{Te}$, we found that the transition levels are 0.64\(\sim\)0.66 eV for (0/+1) and 1.38\(\sim\)1.41 eV for (+1/+2), which do not support the assumption that the defect is $V_{Te}$.

Furthermore, we investigated the defect-assisted optical transition of $Cd_{int}$ using first-principles methods\cite{dreyer2020radiative}. Intriguingly, from the CCD of $Cd_{int}$ in CdSe\textsubscript{0.25}Te\textsubscript{0.75} shown in Fig.~\ref{fig:As_CCD_Cd_int}, the optical transition level $E_{opt}$ of tetrahedral site interstitial $Cd_{int}$ is approximately 1.12 eV from the VBM, or 0.23 eV to the CBM. This level aligns well with the photoluminescence (PL) signal detected by \textit{Kuciauskas et al.}\cite{kuciauskas2023increased} \cite{kuciauskas2023increased}. The significant binding between $Cd_{int}$ and $As_{Te}^{+1}$ could explain why the PL signal disappears after As doping. These calculations suggest that the change in the PL signal is likely due to $Cd_{int}$ rather than $V_{Te}$. However, it is important to note that optical transitions may not be well described in DFT. Time-dependent density functional theory (TDDFT)\cite{ullrich2016excitons} can be used to more accurately study the properties and dynamics of optical transitions in the system.

\subsection{Copper}
\begin{figure*}[!t]
\centering
\captionsetup{justification=centering}
\subfloat[]{\label{fig:defects_CdTe_Cd_rich_Cu}\includegraphics[width=0.3\linewidth]{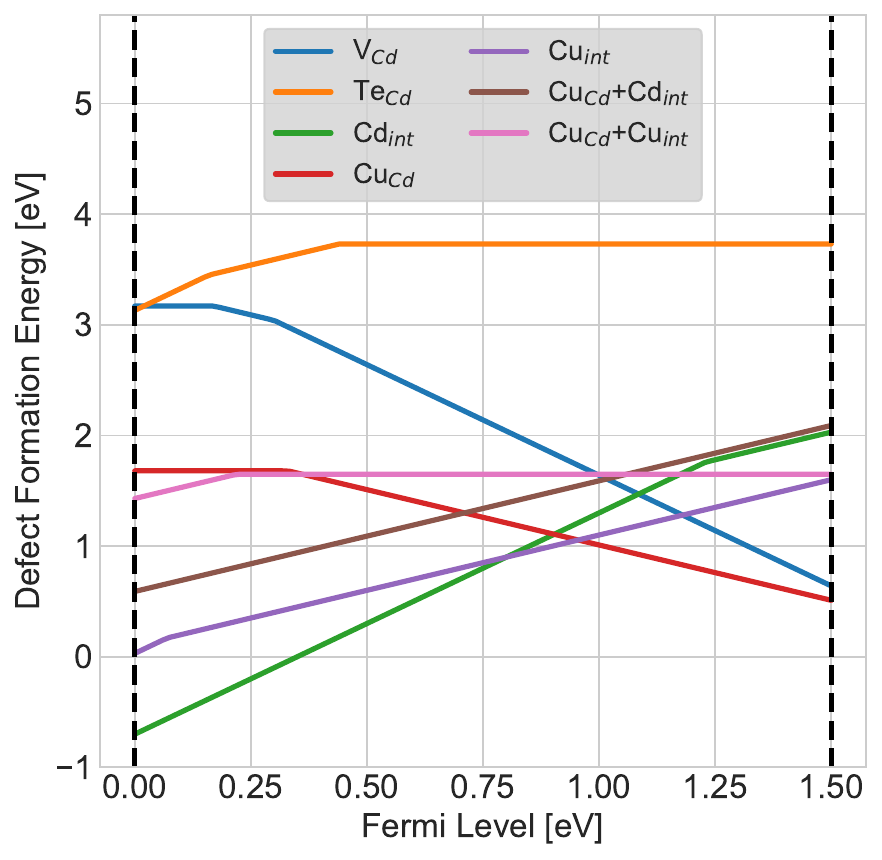}}
\subfloat[]{\label{fig:defects_CdTe0.75_Cd_rich_Cu}\includegraphics[width=0.3\linewidth]{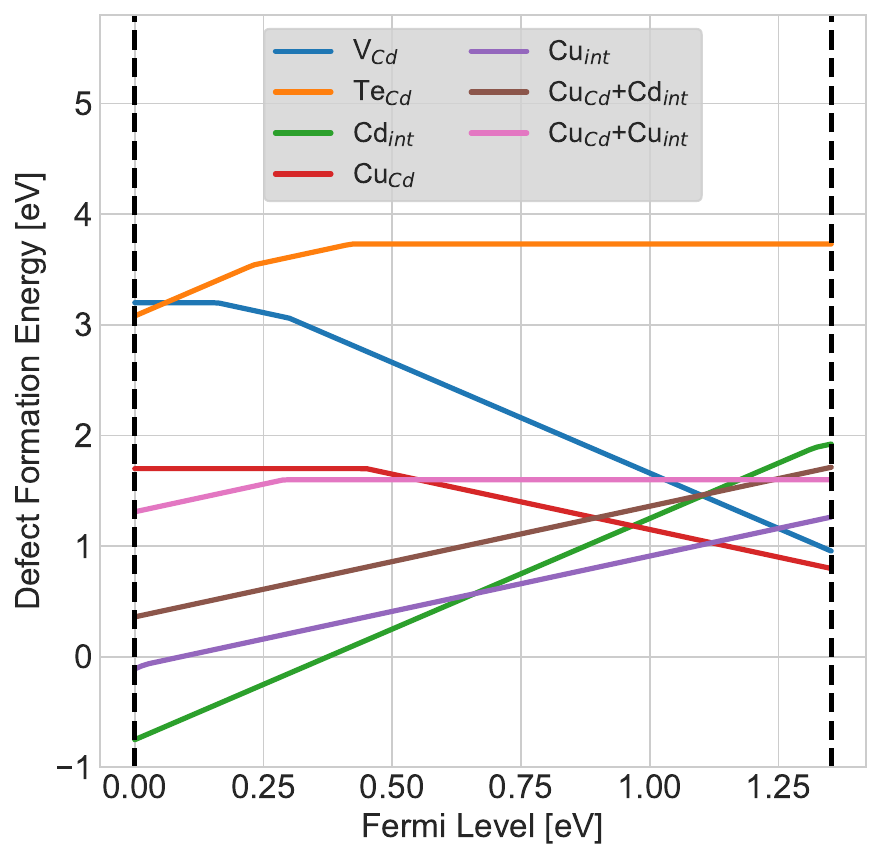}}
\subfloat[]{\label{fig:defects_CdTe0.50_Cd_rich_Cu}\includegraphics[width=0.3\linewidth]{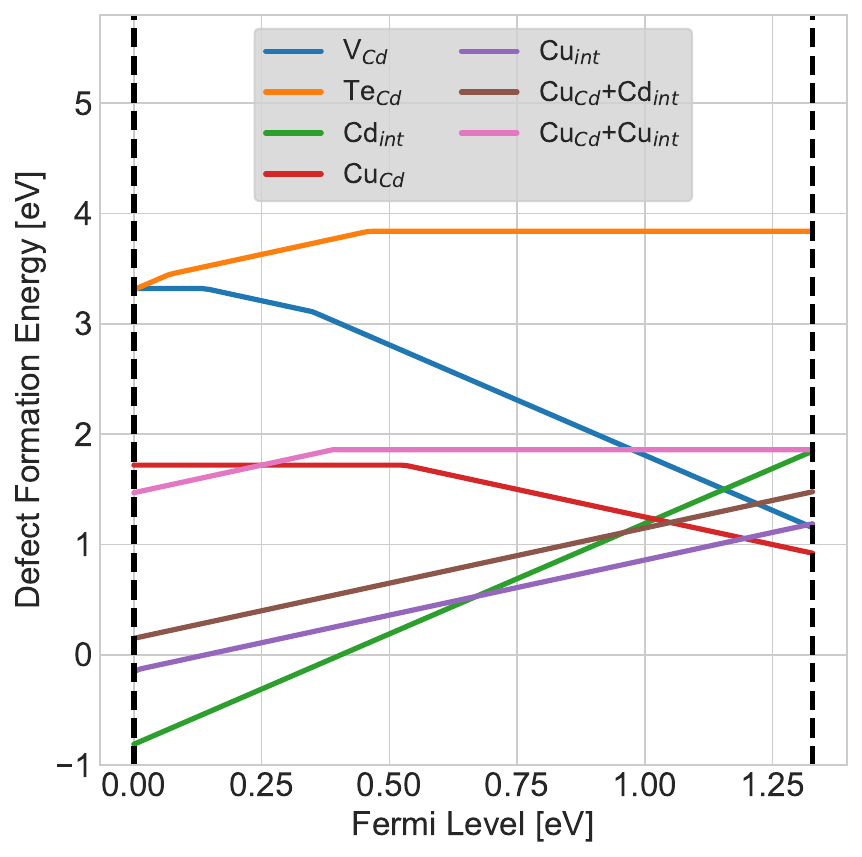}}\\
\subfloat[]{\label{fig:defects_CdTe_Te_rich_Cu}\includegraphics[width=0.3\linewidth]{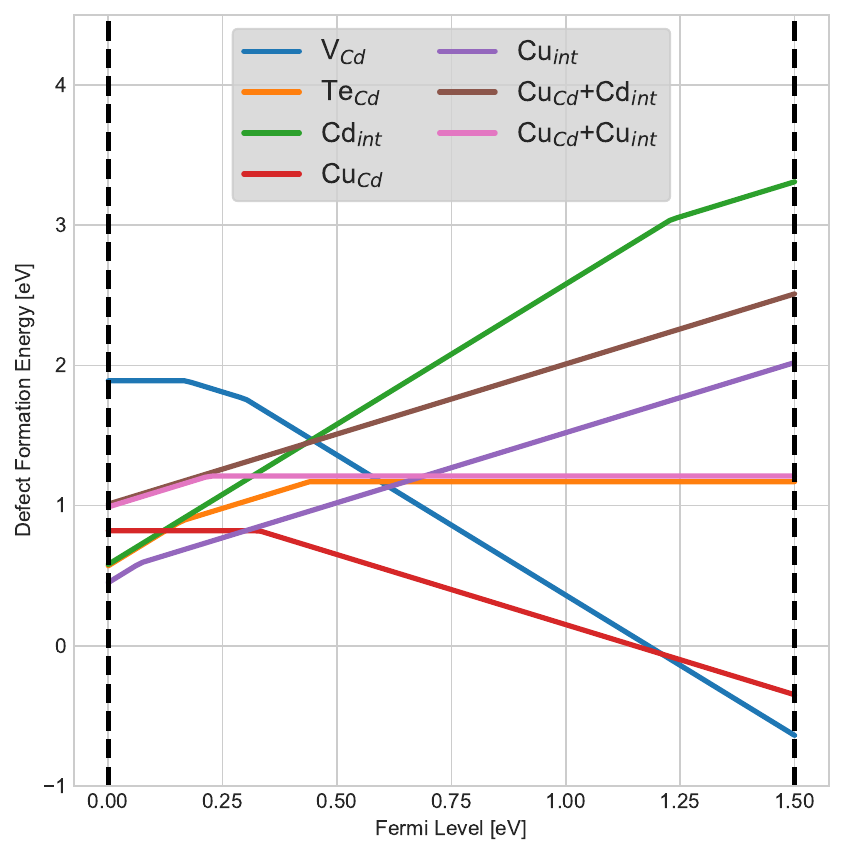}}
\subfloat[]{\label{fig:defects_CdTe0.75_Te_rich_Cu}\includegraphics[width=0.3\linewidth]{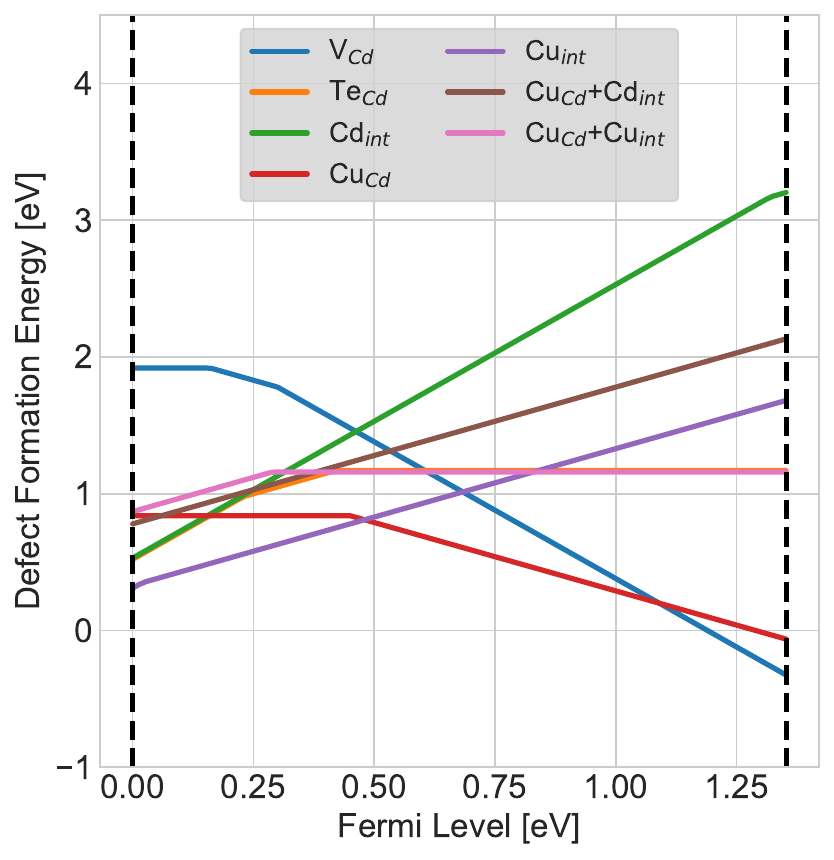}}
\subfloat[]{\label{fig:defects_CdTe0.50_Te_rich_Cu}\includegraphics[width=0.3\linewidth]{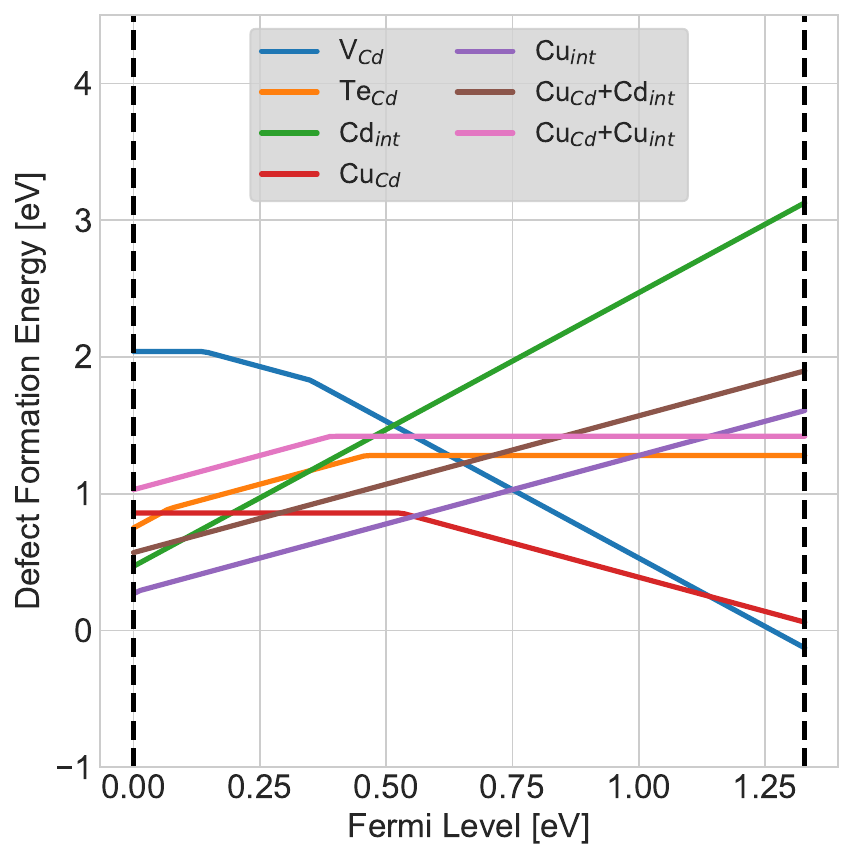}}\\
%\hspace{-0.5in}
\caption{Defect formation energies vs Fermi level of Cu related defects and dominant intrinsic defects in 
CdTe under (a) Cd-rich and (d) Te-rich condition, CdSe\textsubscript{0.25}Te\textsubscript{0.75} under (b) Cd-rich and (e) Te-rich condition, CdSe\textsubscript{0.50}Te\textsubscript{0.50} under (c) Cd-rich and (f) Te-rich condition. The chemical potential condition is $\mu_{Cd}+\mu_{Te} = -1.28$ eV, $\mu_{Cd}+\mu_{Se} < 0.15$ eV and $\mu_{Cu}+\mu_{Te} < -0.42$ eV.}
\label{fig:Defect_trans_all_Cu}
\end{figure*}

Copper (Cu) is well-known for enhancing carrier density in CdTe solar cells, leading to higher power conversion efficiency\cite{li2021low}. However, the doping instability of Cu and the presence of a large number of compensating defects\cite{artegiani2020amount, krasikov2021comparative, chou1996copper} result in inferior efficiency and stability compared to group V-doped counterparts\cite{li2021low, krasikov2021comparative}. To explore the dopability of Cu in CdSeTe, we calculated the formation energies of Cu defects under Cd-rich and Te-rich conditions (Fig.~\ref{fig:Defect_trans_all_Cu}). Four dominant defects, $Cu_{Cd}$, $Cu_{int}$, $Cu_{Cd}+Cd_{int}$ and $Cu_{Cd}+Cu_{int}$, along with the dominant intrinsic defects, are considered in this study.

Our calculations show that the defect transition level of $Cu_{Cd}$ (-1/0) in CdTe is 0.33 eV above the valence band maximum (VBM). Experimental results for this transition level vary between 0.15 and 0.37 eV\cite{balcioglu2000deep, gippius1974deep, chamonal1982identification, said1990excited}, with our result falling within the upper bound of this range, while HSE06 calculations suggest lower bound values around 0.20 eV\cite{yang2016review}. Despite appearing deep in the gap, Cu can still act as an effective acceptor due to its relatively low formation energy, as shown in Fig.~\ref{fig:defects_CdTe_Te_rich_Cu}. In CdTe, $Cu_{int}$ and $Cd_{int}$ emerge as dominant compensating defects, with the Fermi level being pinned around 0.3 eV, a significant improvement over intrinsic CdTe (Fig.~\ref{fig:defects_CdTe_Te_rich}). However, as the Se ratio increases, the pinned level shifts toward the conduction band minimum (CBM) (Fig.~\ref{fig:defects_CdTe0.75_Te_rich_Cu}, Fig.~\ref{fig:defects_CdTe0.50_Te_rich_Cu}, Fig.~\ref{fig:defects_trans_CdSe0.25}, Fig.~\ref{fig:defects_trans_CdSe0.50}), rendering the system more \(n\)-type. This shift suggests a decrease in the doping efficiency of Cu in CdSeTe, which may account for the observed degradation in net acceptor density in Cu-doped devices\cite{krasikov2021comparative}.

In conclusion, we believe that group V dopants hold more promise than Cu in CdSeTe alloys. Apart from the doping efficiency challenges associated with Cu in CdSeTe, Cu implantation in CdTe typically necessitates a Te-rich environment\cite{zhang2008effect, artegiani2020amount}, which may lead to the introduction of more deep-level traps, as illustrated in Fig.~\ref{fig:Defect_trans_all} Fig.~\ref{fig:Defect_trans_all_Cu}. In contrast, group V dopants generally favor a Cd-rich environment\cite{nagaoka2019arsenic, nagaoka2020comparison}, which can enhance carrier lifetime in solar cells.

\section{Conclusions}

In this study, we have undertaken a comprehensive investigation of intrinsic defects, copper, and group V dopants in CdSeTe alloys using DFT calculations. Our tailored localized charge correction approach demonstrates its advantage over traditional FNV correction methods, particularly when applied to defects with delocalized charge or multiple charge centers. Systematic evaluations reveal that our refined methodology not only improves convergence characteristics but also delivers results that are more consistent with experimental data. Specifically, when analyzing defects such as As\textsubscript{Te} and P\textsubscript{Te}, the transition levels calculated by our method show close agreement with experimental findings, underscoring the enhanced accuracy and reliability of our proposed approach.

Within the CdSeTe framework, the formation energy of point defects is noticeably influenced by the local spatial arrangement of Se and Te atoms. This phenomenon is notably evident in instances such as Te\textsubscript{Cd} (+1, 0) and As\textsubscript{Te} (+1, AX), where the interplay of Se and Te positions exerts a pronounced effect. Our investigation extends beyond the mere identification of point defects to a deeper exploration of their intricate relationships with the local atomic environment.

Our study provides insights into the effects of arsenic and phosphorus defects on the electrical properties of CdSeTe alloy. We found that AX formation is more favorable in the alloy due to the interaction of selenium and group V species, which could potentially deteriorate the dopability of As/P in CdSeTe alloy. Our findings could help guide the development of new doping strategies for the fabrication of more efficient solar cells based on CdSeTe alloy.

Overall, we believe that group V dopants hold more promise than Cu in CdSeTe alloys, due to the doping efficiency challenges associated with Cu in CdSeTe and the potential introduction of more deep-level traps in a Te-rich environment. In contrast, group V dopants generally favor a Cd-rich environment, which can enhance carrier lifetime in solar cells.

In light of our findings, we conclude that optimizing the doping strategy and understanding the defect dynamics in CdSeTe are crucial for further improving the performance of optoelectronic devices based on this material. Future work will focus on integrating advanced simulation techniques, such as Lattice Monte Carlo simulations, continuum simulations and device simulation, to achieve a more detailed understanding of defect behavior in CdSeTe alloys and develop possible optimization strategy that could further enhance the efficiency and stability of CdSeTe-based optoelectronic devices.

\begin{acknowledgments}
This work was supported by the U.S. Department of Energy's Office of Energy Efficiency and Renewable Energy (EERE) under the Solar Energy Technology Office Award Number DE-EE0008556, NSF MRSEC DMR-1719797 and NSF MRSEC DMR-2308979. This work was facilitated through the use of the Hyak supercomputer system at the University of Washington provided via the Clean Energy Institute, the MEM-C MRSEC, and the Student Technology Fund. We also extend our gratitude to Dr. Darius Kuciauskas at the National Renewable Energy Laboratory (NREL) for fruitful discussions at the 50th IEEE Photovoltaic Specialists Conference (PVSC).
\end{acknowledgments}

\section*{Data Availability Statement}

% AIP Publishing believes that all datasets underlying the conclusions of the paper should be available to readers. Authors are encouraged to deposit their datasets in publicly available repositories or present them in the main manuscript. All research articles must include a data availability statement stating where the data can be found. In this section, authors should add the respective statement from the chart below based on the availability of data in their paper.
The data that support the findings of this study are available from the corresponding author upon reasonable request.

\appendix

\section{\label{app:Uvalues}Hubbard U Values Determination}
\begin{figure}[!t]
\centering
\captionsetup{justification=centering}
{\includegraphics[width=0.8\linewidth]{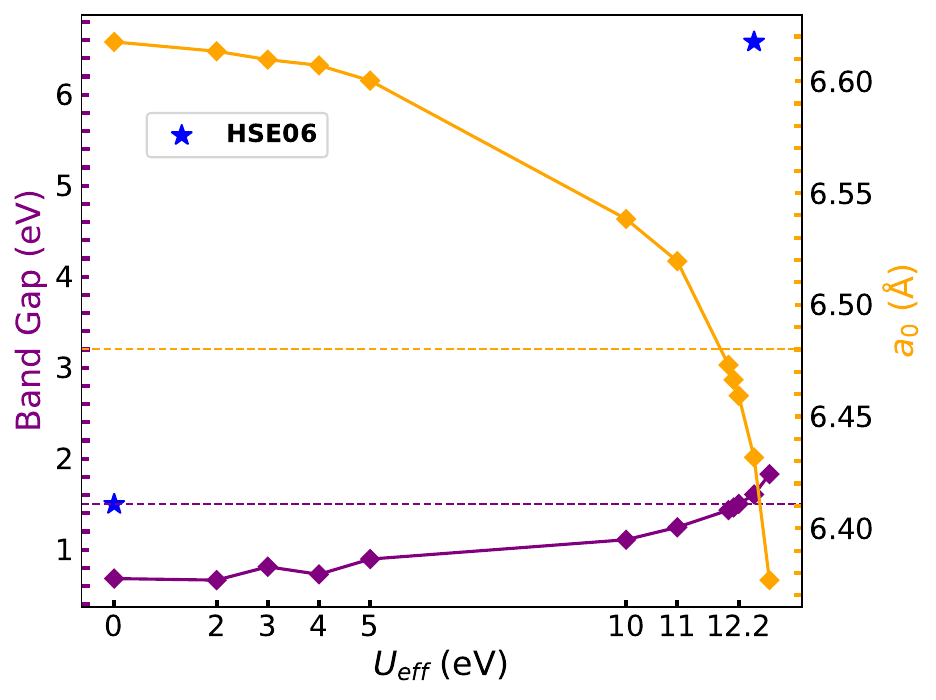}}
\caption{Dependence of lattice parameter $a_0$ and bandgap on $U$ of different methods: standard DFT (U=0 eV) and GGA+U. The stars represent the HSE06 values. The dashed lines represent the experimental values.}
\label{fig:GGA+U+corr}
\end{figure}

The determination of the optimal U value for each material remains a matter of debate because a U value is often optimal for some calculated properties, but not all of them\cite{wu2015lda+}. Here, we systematically vary the U value and compare the calculated results with experimental data, such as bandgap, lattice constant, and formation enthalpy. The comparison with other methods is shown in Fig~\ref{fig:GGA+U+corr}. The U value we chose (U=12.2 eV) show excellent agreement with the experimental bandgap and lattice constant. As a crosscheck, the experimental formation enthalpy of CdTe is -1.30 eV\cite{brebrick2010high}, which is close to -1.28 eV obtained in this work. In addition, the good agreement between lattice constants and bandgap in both zinc blende and wurtzite CdSe validates that this U value provides universal local effect correction for Cd in CdTe, CdSe, and their alloys.

\begin{figure}[!t]
\centering
\captionsetup{justification=centering}
\subfloat[]{\label{fig:GGA_DOS}\includegraphics[width=1\linewidth]{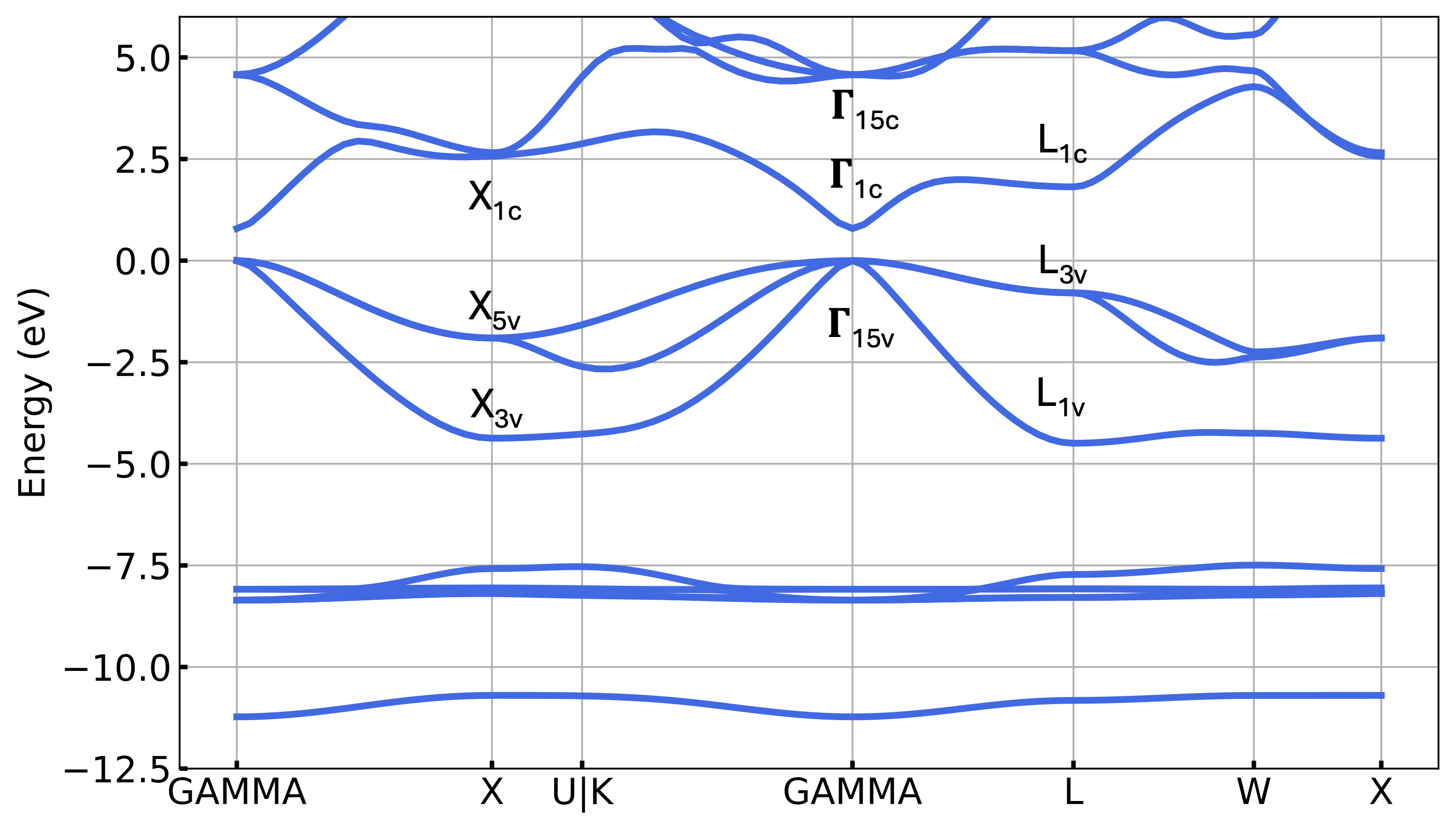}} \\
\subfloat[]{\label{fig:GGAplusU_DOS}\includegraphics[width=1\linewidth]{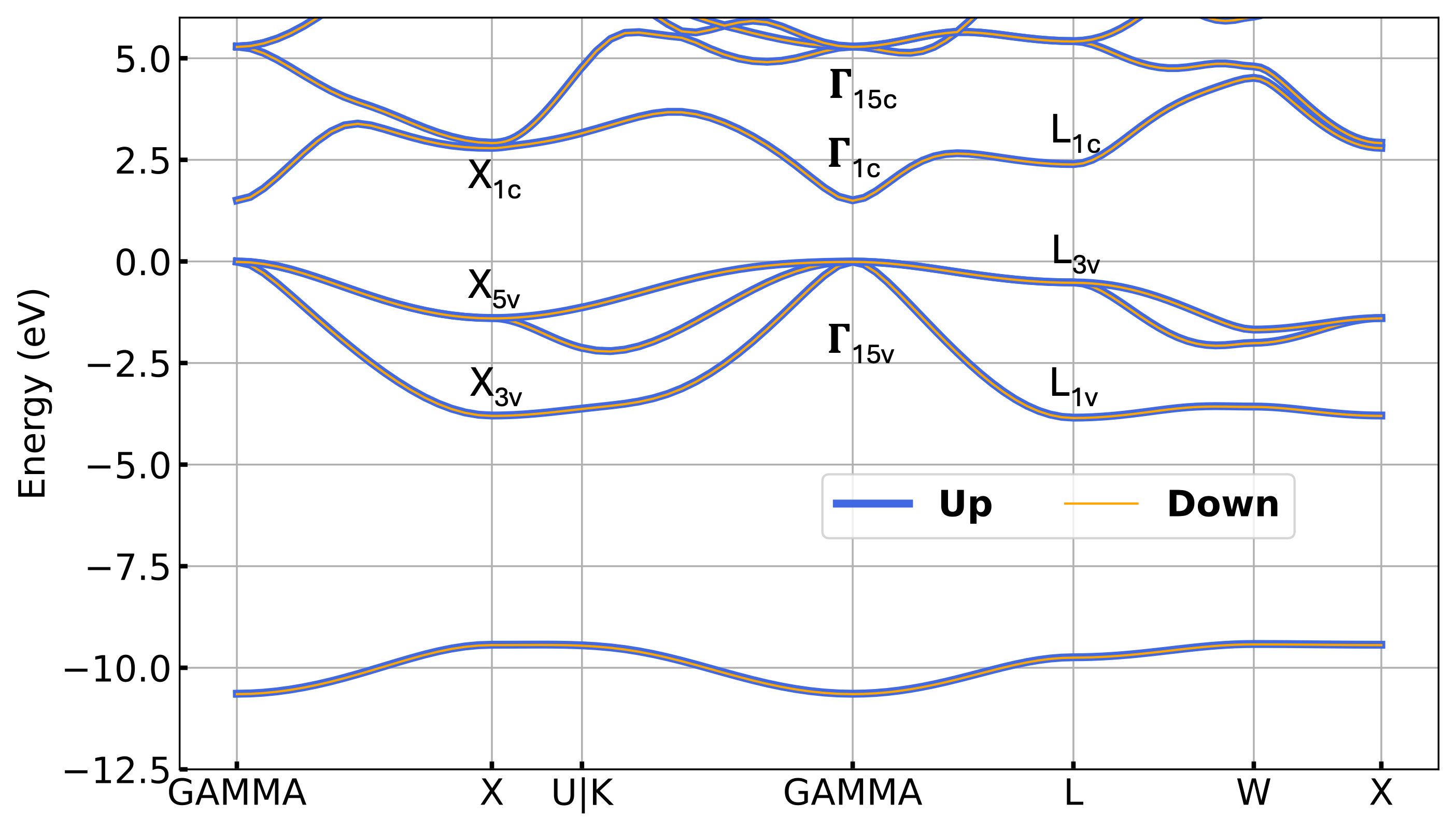}}
\caption{CdTe DOS comparison between GGA (a) and GGA+U (b). The spin-up and spin-down DOS are depicted in (b). The difference between spin-up and spin-down is minimal, resulting in almost overlapping DOS curves.}
\label{fig:DOS}
\end{figure}

To provide more evidence that the problems of DFT arise from the local effects that can be described by the U parameter, we compare the density of states (DOS) of CdTe using GGA, GGA+U in Fig~\ref{fig:DOS}. Additionally, interband transition energies at high symmetry points across different functionals are compared with experimental results in Table~\ref{tab:DOS_compare}. It can be seen that our applied U value effectively corrects the DOS discrepancies between GGA and experimental results, achieving comparable outcomes to HSE and HSE+SOC.

\begin{table*}[!t]
\centering
\caption{\small Comparison of Interband Transition Energies Across Different Functionals and Experimental Data.}
\begin{adjustbox}{max width=0.8\textwidth}
\begin{tabular}{|c|c|c|c|c|c|}
\hline
Interband Transition & GGA & GGA+U (this work) & HSE06\cite{wu2015lda+} & HSE+SOC\cite{kavanagh2021rapid} & Experiment\cite{madelung2004semiconductors} \\
\hline
$\Gamma$ (15v-1c) (eV) & 0.8 & 1.5 & 1.5 & 1.5 & 1.6 (77K), 1.5 (300K) \\
\hline
$\Gamma$ (15v-15c) (eV) & 4.6 & 5.2 & 5.4 & 5.4 & 5.2 \\
\hline
$L$ (3v-1c) (eV) & 2.6 & 3.2 & 3.4 & 3.4 & 3.5 \\
\hline
$X$ (5v-1c) (eV) & 4.5 & 4.8 & 5.2 & 5.4 & 5.1 \\
\hline
\end{tabular}
\end{adjustbox}
\label{tab:DOS_compare}
\end{table*}

\section{\label{app:cutoff}Energy Cutoff Validation}
\begin{figure}[!t]
\centering
\captionsetup{justification=centering}
{\includegraphics[width=1.0\linewidth]{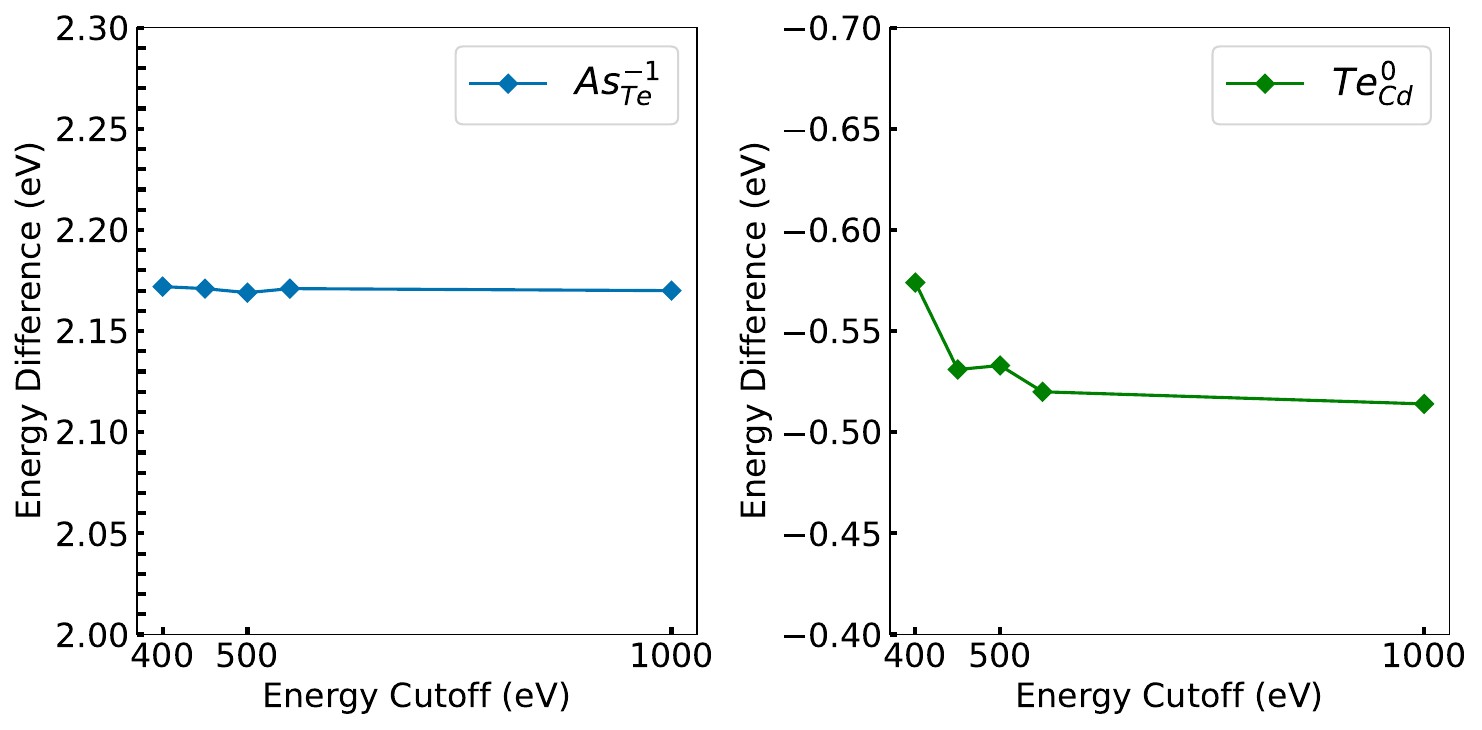}}
\caption{Energy difference ($E_{bulk} - E_{defect}$) of defects ($As_{Te}^{-1}$ and $Te_{Cd}^{0}$) without correction with energy cutoffs ranging from 400 eV to 1000 eV.}
\label{fig:Encut}
\end{figure}

The energy cutoff for the plane-wave basis set in this work is set to be 450 eV. To ensure that this energy cutoff is reasonable, we compared the energy difference between the bulk CdTe supercell and the CdTe supercell with defects (such as $As_{Te}^{-1}$ and $Te_{Cd}^{0}$) without correction across a range of energy cutoffs from 400 eV to 1000 eV. These comparison plots shows that the error between large energy cutoff 1000 eV and 450 eV is within 17 meV (in $Te_{Cd}^{0}$), confirming (Fig~\ref{fig:Encut}) that our chosen energy cutoff is appropriate.\\

\section{\label{app:Se_dep}Se Distribution Dependence in CdSeTe Alloy}
Utilizing a supercell program \cite{okhotnikov2016supercell}, we generated 30 configurations for each alloy composition, prioritizing those with the highest frequency of occurrence. In Table~\ref{tab:alloy_std}, we list the average values and standard deviations of bulk energy and lattice constant of 25\% and 50\% CdSeTe alloy (30 configurations for each). In Table~\ref{tab:Se neighbor model}, we use a linear regression model to explore the relationship between the Se local arrangement and bulk energy. The data consists of 66 points, with Se ratios ranging from 0\% to 50\%. The small standard deviation values and coefficients in Table~\ref{tab:Se neighbor model} imply that there is minimal Se-Se interaction and negligible dependence of lattice constants and bulk energy on Se arrangement.

\begin{table}[htbp]
\centering
\caption{\small Mean Values and Standard Deviation ($\sigma$) of CdSeTe Alloy Bulk Energy $E_{bulk}$ and Lattice Constants $a_0$.}
\begin{adjustbox}{max width=0.47\textwidth}
\begin{tabular}{|c|c|c|c|c|}
\hline
Alloy & $E_{bulk}$ (eV) & $\sigma_{bulk}$ (eV) & $a_0$ (Å) & $\sigma_{lat}$ (Å) \\
\hline
CdSe\textsubscript{0.25}Te\textsubscript{0.75} & -44.25 & 0.048 & 6.34 & 0.003 \\
\hline
CdSe\textsubscript{0.50}Te\textsubscript{0.50} & -47.84 & 0.039 & 6.21 & 0.013 \\
\hline
\end{tabular}
\end{adjustbox}
\label{tab:alloy_std}
\end{table}

\begin{table}[htbp]
\centering
\caption{\small CdSeTe Alloy Linear Regression Model: coefficients $\Delta E$ for each feature. All features are standardized before training. For nearest neighbor counting features, the count is specific to Se atoms. The training error is 0.1 meV while the testing error is 0.6 meV.}
\begin{adjustbox}{max width=0.47\textwidth}
\begin{tabular}{|c|c|c|c|}
\hline
Feature & $\Delta E$ (meV) & Feature & $\Delta E$ (meV) \\
\hline
Se ratio & $57$ & 1NN Se & $8$ \\
\hline
(Se ratio)$^2$ & $24$ & 2NN Se & $0.4$ \\
\hline
3NN Se & $2$ & 4NN Se & $6$ \\
\hline
\end{tabular}
\end{adjustbox}
\label{tab:Se neighbor model}
\end{table}

\section{\label{app:coordinates}Atomic Coordinates of Selected Defects}
\begin{figure}[!t]
\centering
\captionsetup{justification=centering}
{\includegraphics[width=0.5\linewidth]{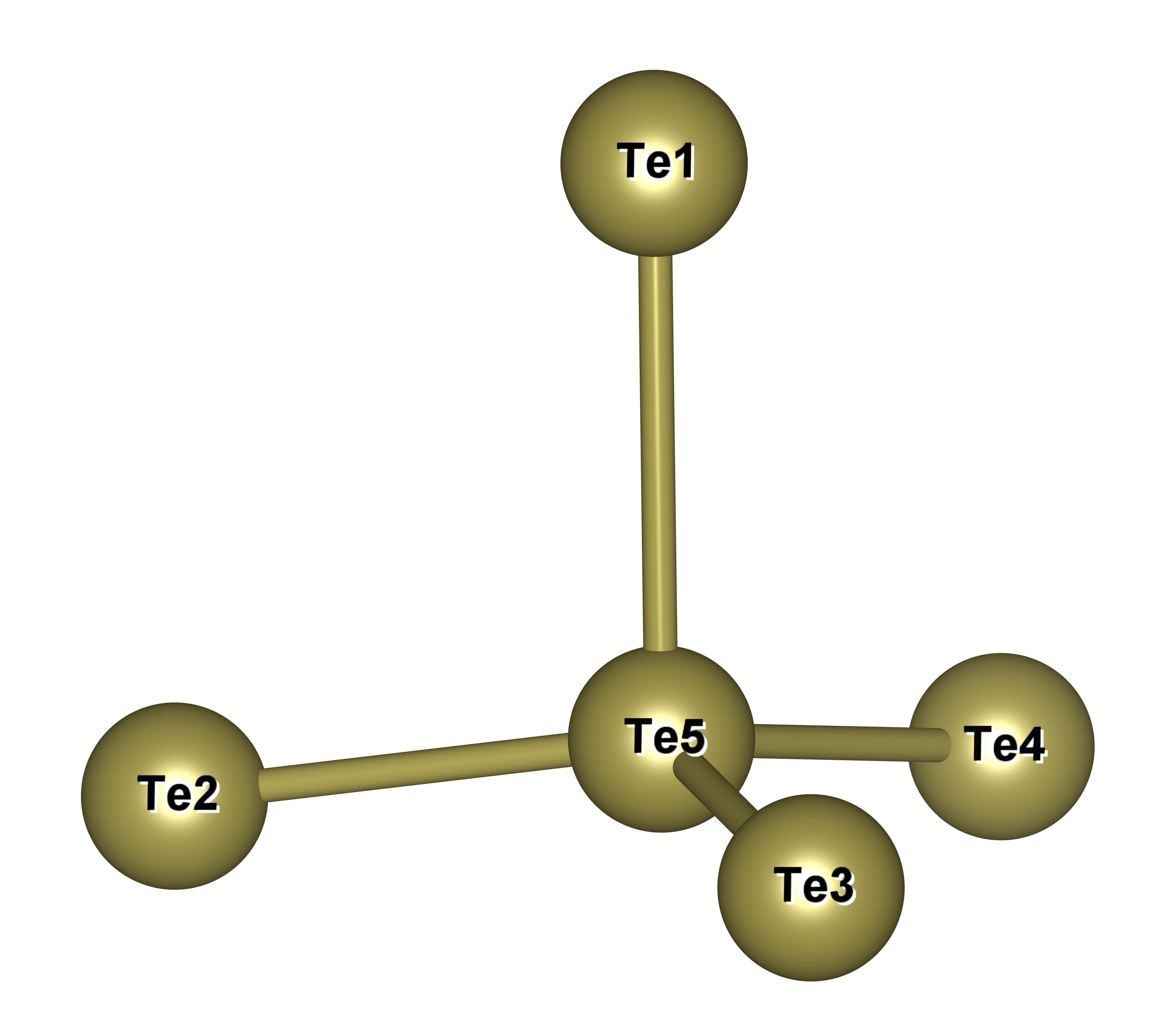}}
\caption{Schematic of $Te_{Cd}^0$ \(C_{3v}\) defect configuration. $Te1-Te5$ is the long bond. In ideal lattice structure, the length of $Te-Cd$ bond is 2.78 \AA. In $Te_{Cd}^0$, when one Te occupies Cd site, the length of $Te2-Te5$ short bond is 2.95 \AA, while the length of $Te1-Te5$ long bond is 3.76 \AA.}
\label{fig:Te_cd_c3v_config}
\end{figure}

\begin{figure}[!t]
\centering
\captionsetup{justification=centering}
{\includegraphics[width=0.6\linewidth]{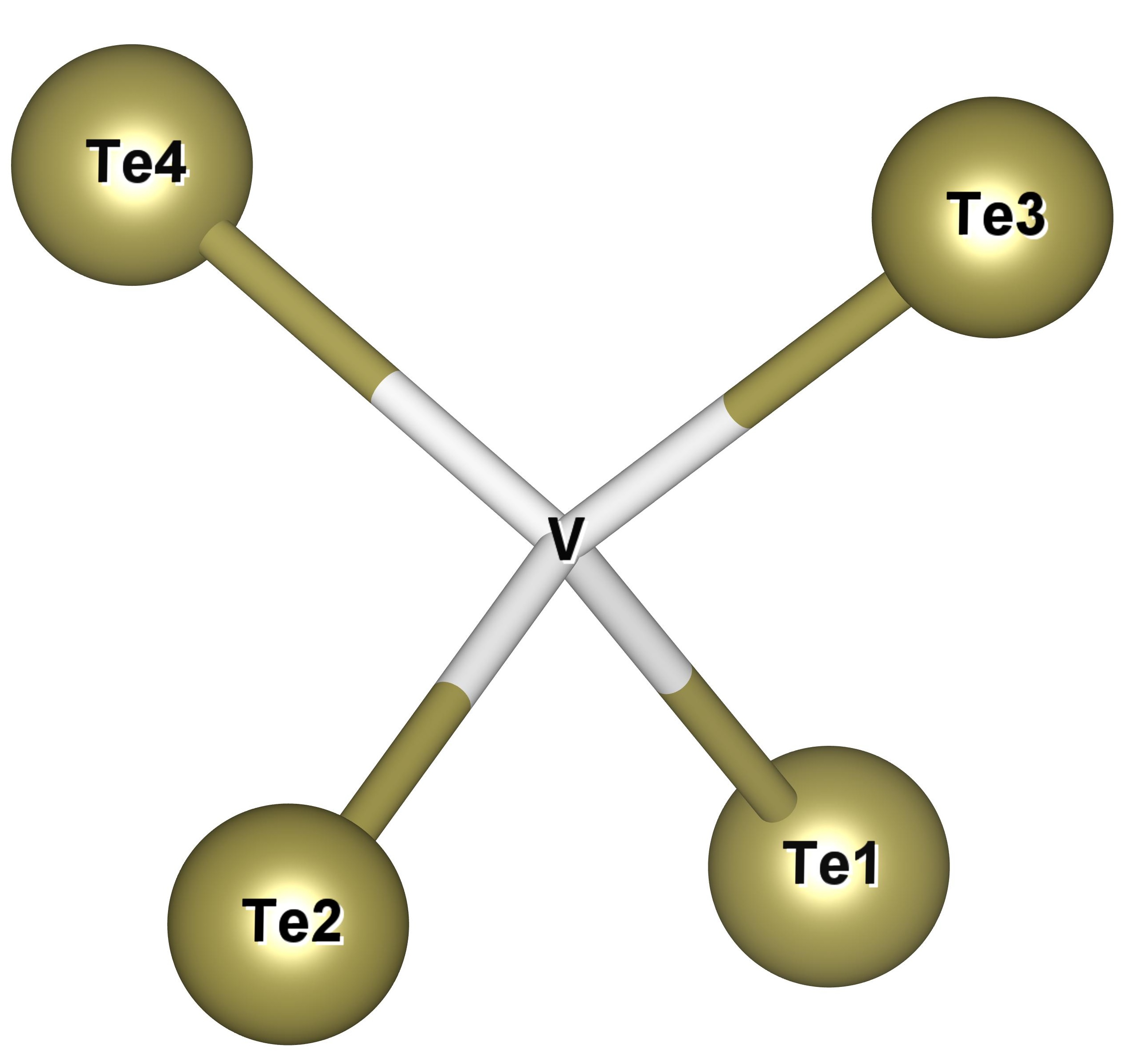}}
\caption{Schematic of $V_{Cd}^0$ \(D_{2d}\) dimer defect configuration. White region represents vacancy site. $Te3-Vacancy$ and $Te4-Vacancy$ are the long bonds. In ideal lattice structure, the length of $Te-Cd$ bond is 2.78 \AA. In $V_{Cd}^0$, the length of $Te3-Vacancy$ and $Te4-Vacancy$ are 2.81 \AA, while the length of other two bonds are 2.78 \AA$\,$ and 2.79 \AA$\,$, respectively.}
\label{fig:V_cd_dimer_config}
\end{figure}

The atomic coordinates and structures for some defects with significant local distortion in CdTe are provided, including $Te_{Cd}^0$ (Fig.~\ref{fig:Te_cd_c3v_config}, Table~\ref{tab:atom_Tecd}), $V_{Cd}^0$ (Fig.~\ref{fig:V_cd_dimer_config}, Table~\ref{tab:atom_Vcd}), $As_{Te}^{+1}$ AX (Fig.~\ref{fig:AX_config}, Table~\ref{tab:atom_AX}) and $(Cd_{int}+As_{Te})^+$ (Fig.~\ref{fig:complex_config}, Table~\ref{tab:atom_As_comp}).
\begin{table}[!t]
\centering
\caption{Fractional atomic coordinates of $Te_{Cd}^0$. In ideal lattice structure, where $Cd$ is not occupied by $Te5$, the coordinate of nearest $Te$ to that $Cd$ site is (0.250, 0.250, 0.250). $Cd$ site in ideal lattice structure is chosen as reference (0, 0, 0).}
\begin{adjustbox}{max width=1.0\textwidth}
\begin{tabular}{|c|c|c|c|}
\hline
Te1 & (-0.253, 0.253, -0.253) & Te2 & (0.271, -0.271, -0.289) \\
\hline
Te3 & (-0.289, -0.271, 0.271) & Te4 & (0.271, 0.289, 0.271) \\
\hline
Te5 & (0.083, -0.083, 0.083) \\
\cline{1-2}
\end{tabular}
\end{adjustbox}
\label{tab:atom_Tecd}
\end{table}

\begin{table}[!t]
\centering
\caption{Fractional atomic coordinates of $V_{Cd}^0$ dimer.}
\begin{adjustbox}{max width=1.0\textwidth}
\begin{tabular}{|c|c|c|c|}
\hline
Te1 & (0.248, 0.248, 0.248) & Te2 & (0.246, -0.252, -0.252) \\
\hline
Te3 & (-0.249, -0.258, 0.249) & Te4 & (-0.249, 0.249, -0.258) \\
\hline
Vacancy & (0, 0, 0) \\
\cline{1-2}
\end{tabular}
\end{adjustbox}
\label{tab:atom_Vcd}
\end{table}
\begin{figure}[!t]
\centering
\captionsetup{justification=centering}
{\includegraphics[width=0.8\linewidth]{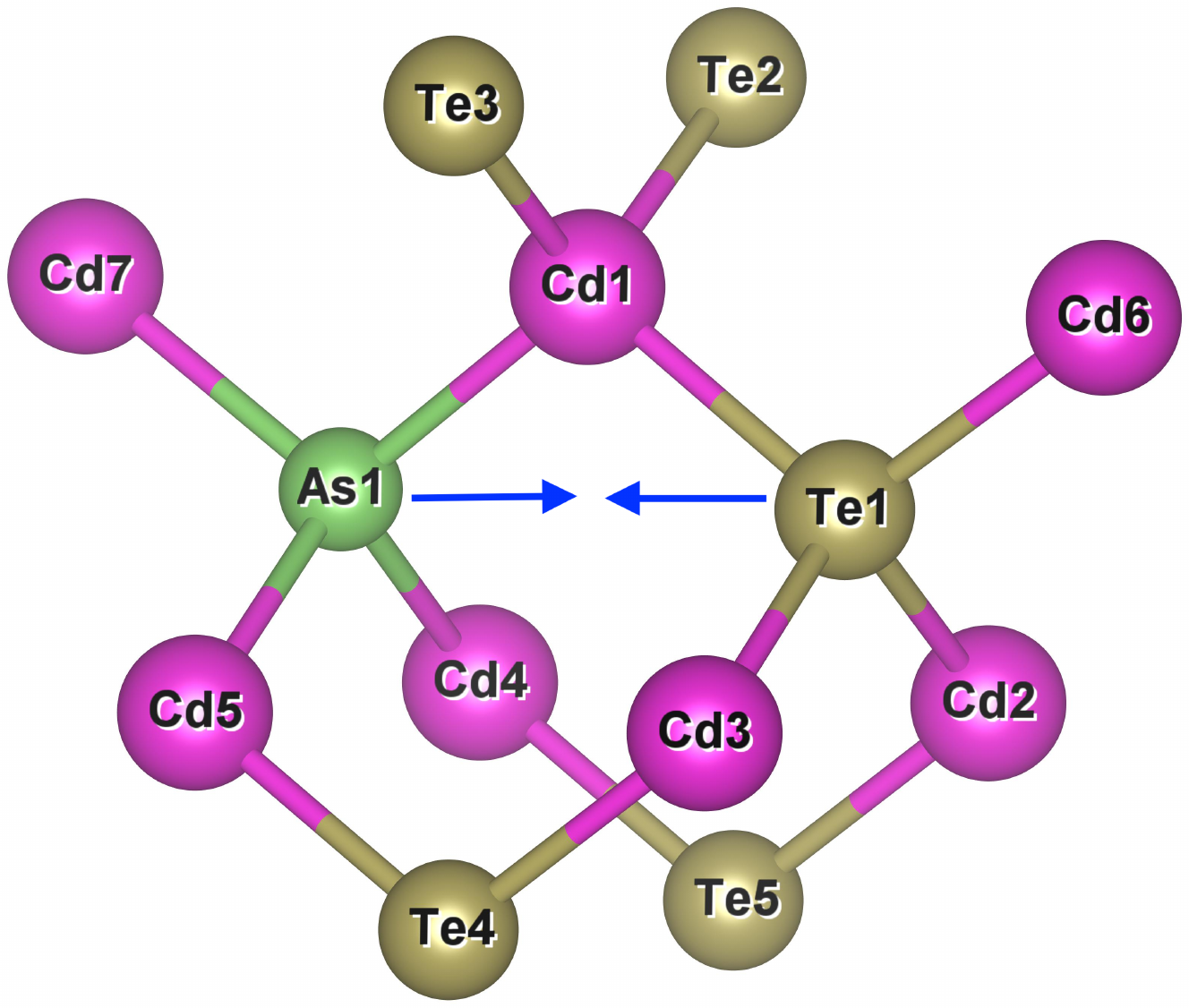}}
\caption{Schematic of $As_{Te}^{+1}$ $T_d$ defect configuration. Blue arrows indicate distortion direction when forming AX defect.}
\label{fig:Td_config}
\end{figure}
\begin{figure}[!t]
\centering
\captionsetup{justification=centering}
{\includegraphics[width=0.6\linewidth]{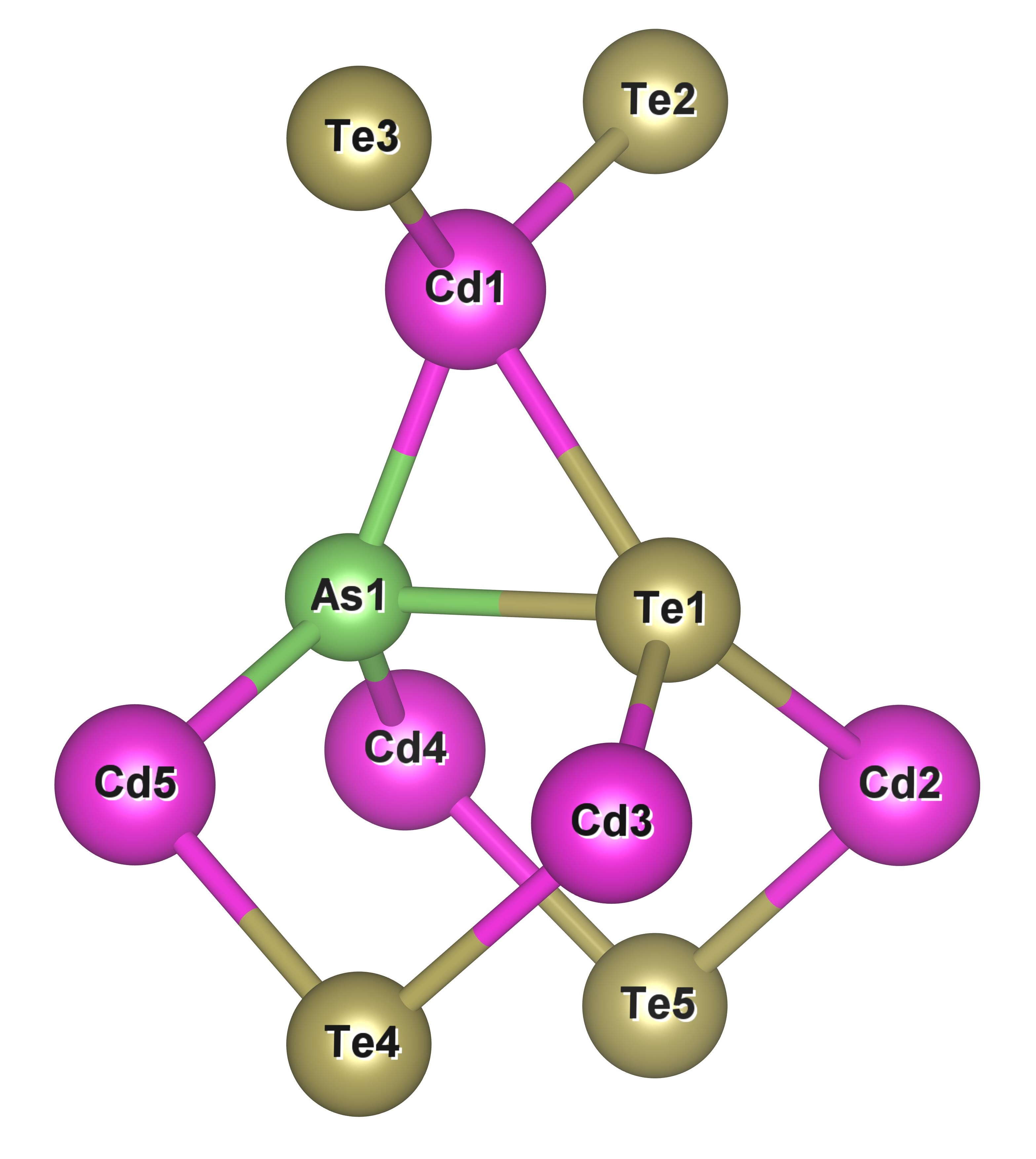}}
\caption{Schematic of $As_{Te}^{+1}$ AX defect configuration. In ideal lattice structure, the length of Te-Te bond is 4.48 \AA. In $As_{Te}^{+1}$ AX, the length of $Te1-As1$ becomes 2.73 \AA. The comparison between Fig.~\ref{fig:Td_config} and Fig.~\ref{fig:AX_config} demonstrates the distortion caused by AX defect formation.}
\label{fig:AX_config}
\end{figure}
\begin{table}[!t]
\centering
\caption{Fractional atomic coordinates of $As_{Te}^{+1}$ AX. $As_{Te}$ \(T_d\) symmetry site in Fig.~\ref{fig:Td_config} is chosen as reference (0, 0, 0).}
\begin{adjustbox}{max width=1.0\textwidth}
\begin{tabular}{|c|c|c|c|}
\hline
Te1 & (0.409, -0.079, -0.409) & Te2 & (-0.003, 0.488, -0.519) \\
\hline
Te3 & (0.519, 0.490, 0.003) & Te4 & (0.518, -0.572, 0.003) \\
\hline
Te5 & (-0.004, -0.572, -0.518) & Cd1 & (0.220, 0.289, -0.219) \\
\hline
Cd2 & (0.237, -0.302, -0.744) & Cd3 & (0.745, -0.303, -0.237) \\
\hline
Cd4 & (-0.202, -0.280, 0.202) & Cd5 & (0.274, -0.280, 0.202) \\
\hline
As1 & (0.110, -0.077, -0.110) \\
\cline{1-2}
\end{tabular}
\end{adjustbox}
\label{tab:atom_AX}
\end{table}

\begin{figure}[!t]
\centering
\captionsetup{justification=centering}
{\includegraphics[width=0.6\linewidth]{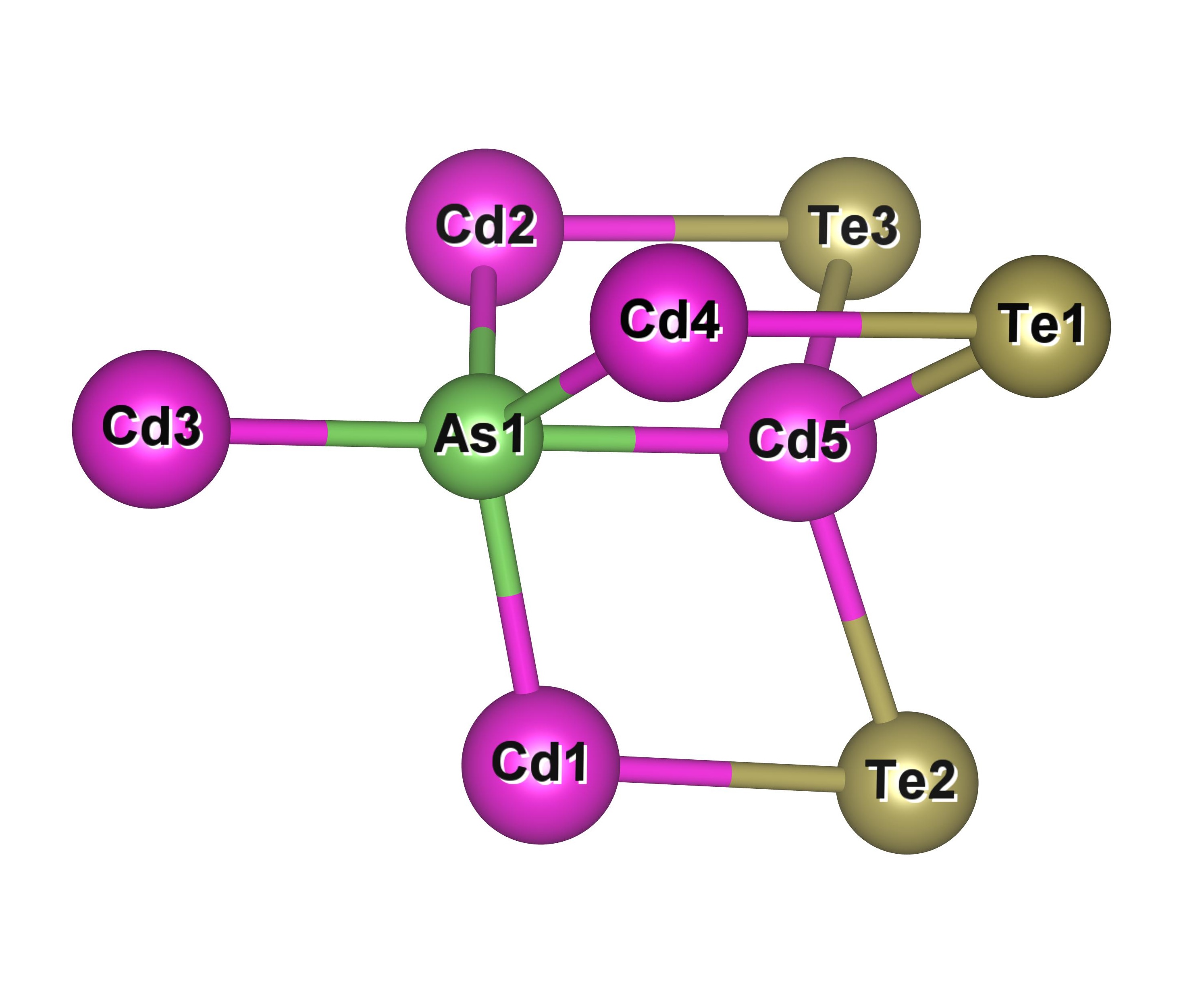}}
\caption{Schematic of $(Cd_{int}+As_{Te})^+$ complex defect configuration. In ideal $Cd_{int}^{2+}$ interstitial structure, where $Cd$ occupies tetrahedral vacant site, the bond length between $Cd_{int}$ and nearby $Te$ is 2.79 \AA. In ideal $As_{Te}^{-}$ structure, where $As$ occupies $Te$ site, the bond length between $As_{Te}$ and nearby $Cd$ is 2.54 \AA. For $(Cd_{int}+As_{Te})^+$ complex defect, the length between $Cd_{int}$ and nearby $As_{Te}$ becomes 2.57 \AA$\,$ and the length between $Cd_{int}$ and nearby $Te$ becomes 2.84 \AA. The length between $As_{Te}$ and nearby three $Cd$ ($Cd1$, $Cd2$ and $Cd4$) becomes 2.63 \AA$\,$, while the length between $As_{Te}$ and $Cd3$ becomes 2.67 \AA. $As_{Te}$ \(T_d\) symmetry site in Fig.~\ref{fig:Td_config} is chosen as reference (0, 0, 0).}
\label{fig:complex_config}
\end{figure}
\begin{table}[!t]
\centering
\caption{Fractional atomic coordinates of $(Cd_{int}+As_{Te})^+$.}
\begin{adjustbox}{max width=1.0\textwidth}
\begin{tabular}{|c|c|c|c|}
\hline
Te1 & (0.009, 0.508, 0.507) & Te2 & (0.507, 0.018, 0.507) \\
\hline
Te3 & (0.507, 0.508, 0.009) & Cd1 & (0.236, -0.245, 0.236) \\
\hline
Cd2 & (0.236, 0.236, -0.247) & Cd3 & (-0.225, -0.225, -0.225) \\
\hline
Cd4 & (-0.247, 0.236, 0.236) & Cd5 & (0.243, 0.246, 0.243) \\
\hline
As1 & (0.014, 0.014, 0.014)\\
\cline{1-2}
\end{tabular}
\end{adjustbox}
\label{tab:atom_As_comp}
\end{table}

\section{\label{app:def_trans}Defect Transition Levels}
\begin{figure}[!t]
\centering
\captionsetup{justification=centering}
\subfloat[]{\label{fig:defects_trans_CdTe}\includegraphics[width=1.0\linewidth]{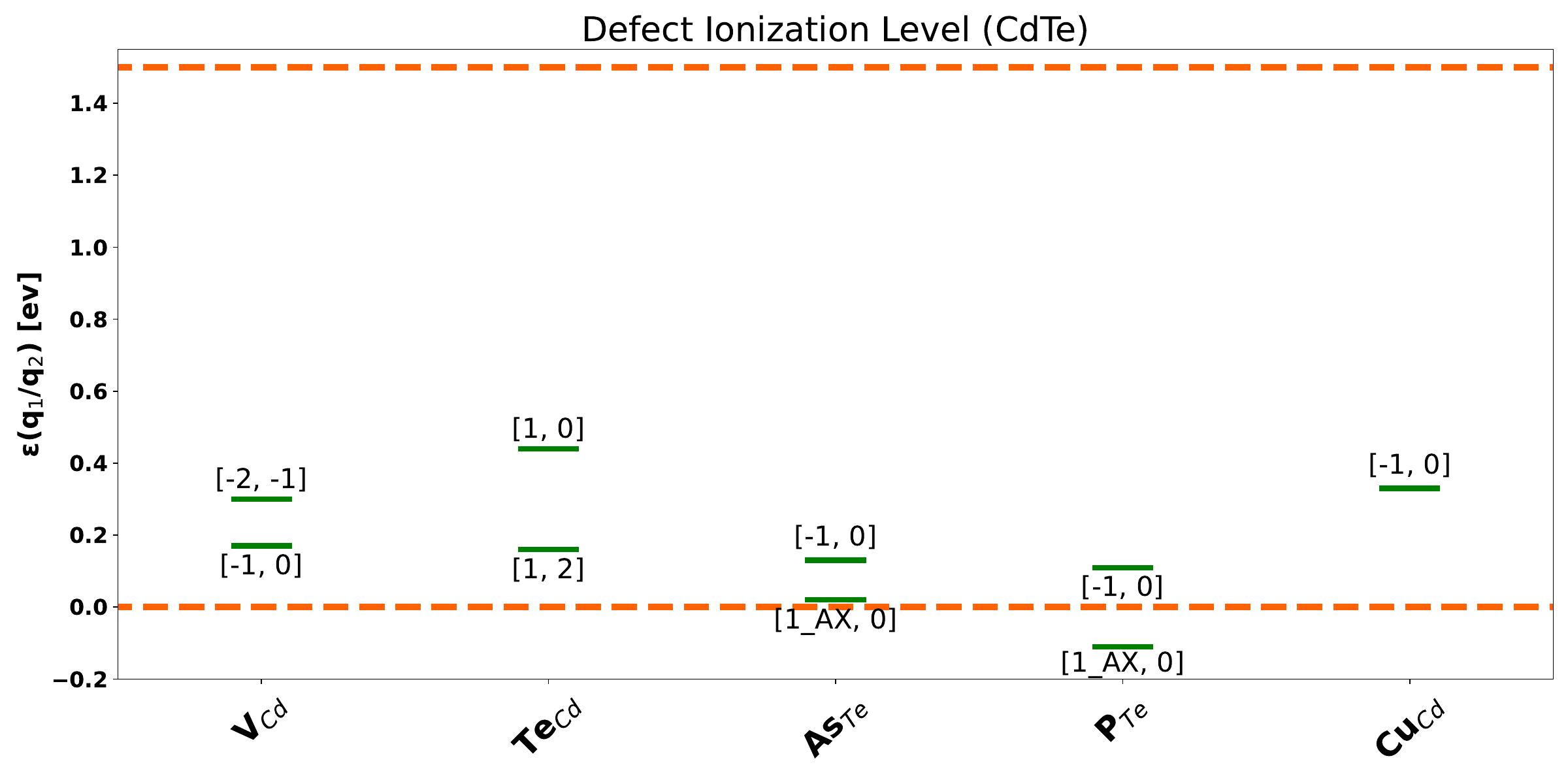}}\\
\subfloat[]{\label{fig:defects_trans_CdSe0.25}\includegraphics[width=1.0\linewidth]{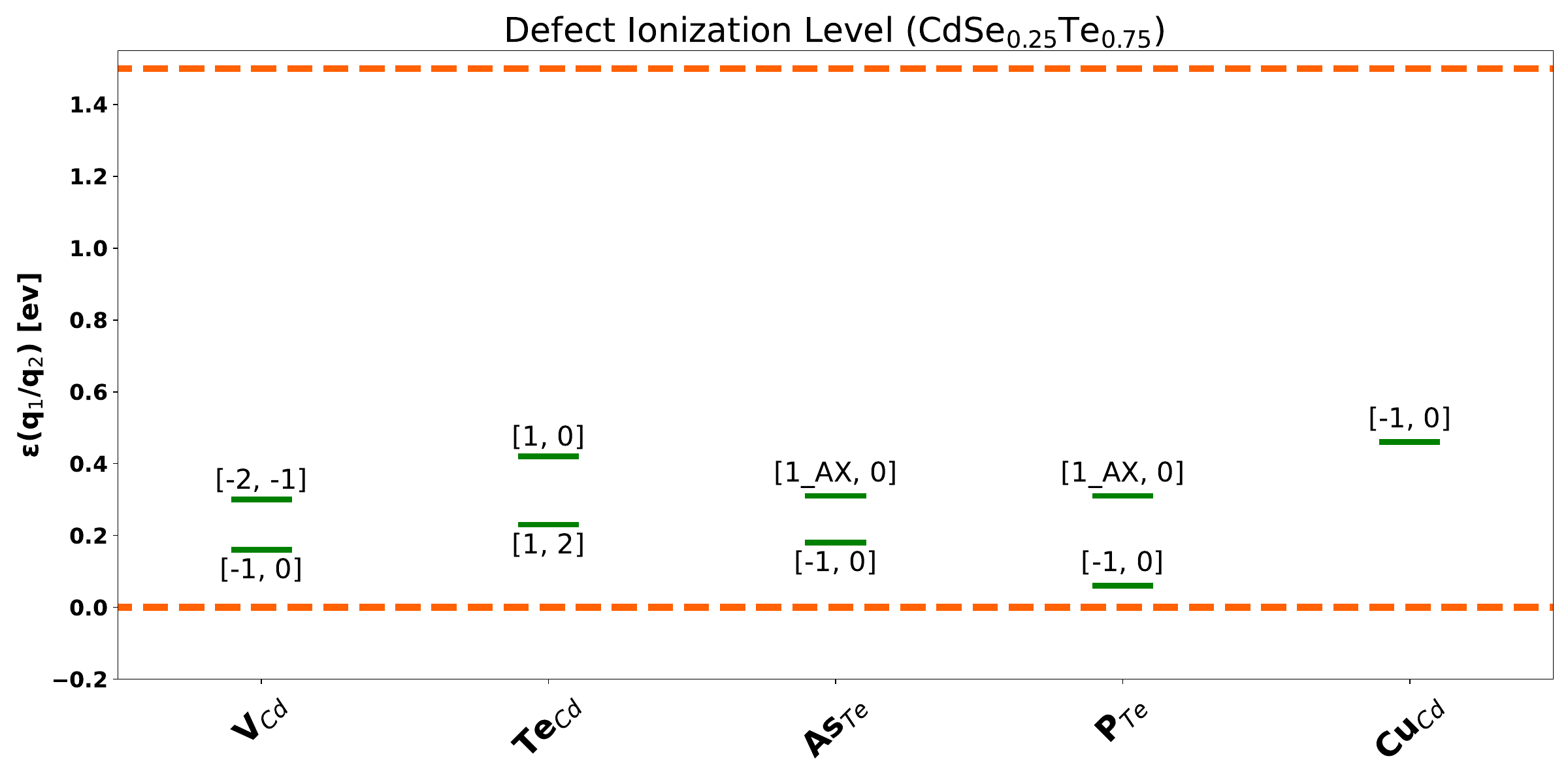}}\\
\subfloat[]{\label{fig:defects_trans_CdSe0.50}\includegraphics[width=1.0\linewidth]{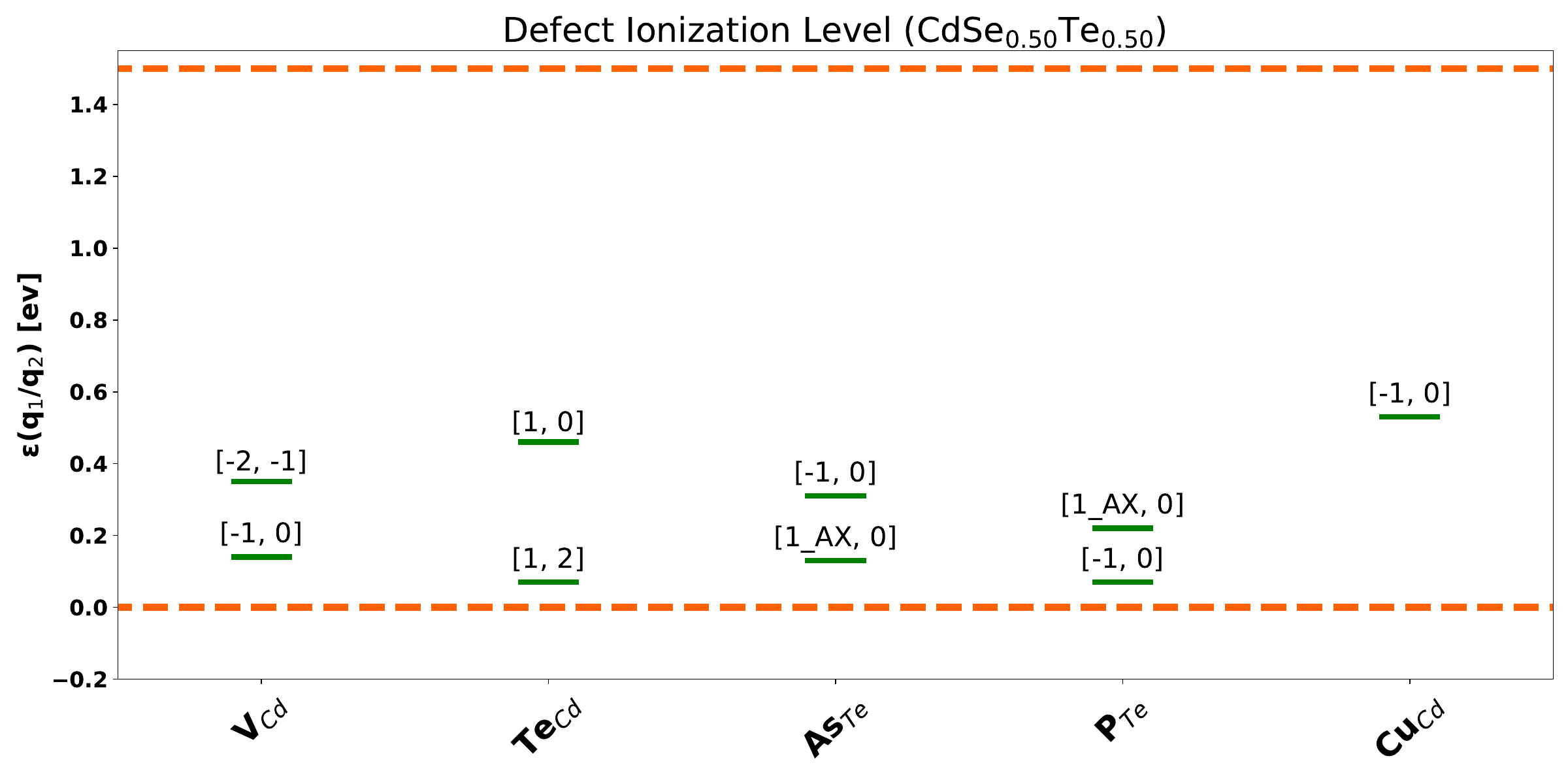}}
%\hspace{-0.5in}
\caption{Selected defect transition levels in 
CdTe, CdSe\textsubscript{0.25}Te\textsubscript{0.75}, and CdSe\textsubscript{0.50}Te\textsubscript{0.50}. Dashed orange lines indicate CBM (above) and VBM (below).}
\label{fig:Defect_trans_level}
\end{figure}

Some of the defects transition levels are depicted in Fig.~\ref{fig:Defect_trans_level}, including dominant deep level such as $Te_{Cd}$ and $V_{Cd}$, and the primary defects associated with As, P, and Cu doping, including $As_{Te}$, $P_{Te}$, and $Cu_{Cd}$. The formation energies of defects in the alloy are calculated using a Boltzmann distribution at 873K as mentioned in Sec~\ref{sec:level3}.
\newpage
\bibliography{aipsamp}% Produces the bibliography via BibTeX.

\begin{thebibliography}{72}%
\makeatletter
\providecommand \@ifxundefined [1]{%
 \@ifx{#1\undefined}
}%
\providecommand \@ifnum [1]{%
 \ifnum #1\expandafter \@firstoftwo
 \else \expandafter \@secondoftwo
 \fi
}%
\providecommand \@ifx [1]{%
 \ifx #1\expandafter \@firstoftwo
 \else \expandafter \@secondoftwo
 \fi
}%
\providecommand \natexlab [1]{#1}%
\providecommand \enquote  [1]{``#1''}%
\providecommand \bibnamefont  [1]{#1}%
\providecommand \bibfnamefont [1]{#1}%
\providecommand \citenamefont [1]{#1}%
\providecommand \href@noop [0]{\@secondoftwo}%
\providecommand \href [0]{\begingroup \@sanitize@url \@href}%
\providecommand \@href[1]{\@@startlink{#1}\@@href}%
\providecommand \@@href[1]{\endgroup#1\@@endlink}%
\providecommand \@sanitize@url [0]{\catcode `\\12\catcode `\$12\catcode `\&12\catcode `\#12\catcode `\^12\catcode `\_12\catcode `\%12\relax}%
\providecommand \@@startlink[1]{}%
\providecommand \@@endlink[0]{}%
\providecommand \url  [0]{\begingroup\@sanitize@url \@url }%
\providecommand \@url [1]{\endgroup\@href {#1}{\urlprefix }}%
\providecommand \urlprefix  [0]{URL }%
\providecommand \Eprint [0]{\href }%
\providecommand \doibase [0]{http://dx.doi.org/}%
\providecommand \selectlanguage [0]{\@gobble}%
\providecommand \bibinfo  [0]{\@secondoftwo}%
\providecommand \bibfield  [0]{\@secondoftwo}%
\providecommand \translation [1]{[#1]}%
\providecommand \BibitemOpen [0]{}%
\providecommand \bibitemStop [0]{}%
\providecommand \bibitemNoStop [0]{.\EOS\space}%
\providecommand \EOS [0]{\spacefactor3000\relax}%
\providecommand \BibitemShut  [1]{\csname bibitem#1\endcsname}%
\let\auto@bib@innerbib\@empty
%</preamble>
\bibitem [{\citenamefont {Fiducia}\ \emph {et~al.}(2019)\citenamefont {Fiducia}, \citenamefont {Mendis}, \citenamefont {Li}, \citenamefont {Grovenor}, \citenamefont {Munshi}, \citenamefont {Barth}, \citenamefont {Sampath}, \citenamefont {Wright}, \citenamefont {Abbas}, \citenamefont {Bowers} \emph {et~al.}}]{fiducia2019understanding}%
  \BibitemOpen
  \bibfield  {author} {\bibinfo {author} {\bibfnamefont {T.~A.}\ \bibnamefont {Fiducia}}, \bibinfo {author} {\bibfnamefont {B.~G.}\ \bibnamefont {Mendis}}, \bibinfo {author} {\bibfnamefont {K.}~\bibnamefont {Li}}, \bibinfo {author} {\bibfnamefont {C.~R.}\ \bibnamefont {Grovenor}}, \bibinfo {author} {\bibfnamefont {A.~H.}\ \bibnamefont {Munshi}}, \bibinfo {author} {\bibfnamefont {K.}~\bibnamefont {Barth}}, \bibinfo {author} {\bibfnamefont {W.~S.}\ \bibnamefont {Sampath}}, \bibinfo {author} {\bibfnamefont {L.~D.}\ \bibnamefont {Wright}}, \bibinfo {author} {\bibfnamefont {A.}~\bibnamefont {Abbas}}, \bibinfo {author} {\bibfnamefont {J.~W.}\ \bibnamefont {Bowers}},  \emph {et~al.},\ }\href@noop {} {\bibfield  {journal} {\bibinfo  {journal} {Nature Energy}\ }\textbf {\bibinfo {volume} {4}},\ \bibinfo {pages} {504} (\bibinfo {year} {2019})}\BibitemShut {NoStop}%
\bibitem [{\citenamefont {Yang}\ \emph {et~al.}(2016{\natexlab{a}})\citenamefont {Yang}, \citenamefont {Yin}, \citenamefont {Park}, \citenamefont {Ma},\ and\ \citenamefont {Wei}}]{yang2016review}%
  \BibitemOpen
  \bibfield  {author} {\bibinfo {author} {\bibfnamefont {J.-H.}\ \bibnamefont {Yang}}, \bibinfo {author} {\bibfnamefont {W.-J.}\ \bibnamefont {Yin}}, \bibinfo {author} {\bibfnamefont {J.-S.}\ \bibnamefont {Park}}, \bibinfo {author} {\bibfnamefont {J.}~\bibnamefont {Ma}}, \ and\ \bibinfo {author} {\bibfnamefont {S.-H.}\ \bibnamefont {Wei}},\ }\href@noop {} {\bibfield  {journal} {\bibinfo  {journal} {Semiconductor Science and Technology}\ }\textbf {\bibinfo {volume} {31}},\ \bibinfo {pages} {083002} (\bibinfo {year} {2016}{\natexlab{a}})}\BibitemShut {NoStop}%
\bibitem [{\citenamefont {Krasikov}\ \emph {et~al.}(2016)\citenamefont {Krasikov}, \citenamefont {Scherbinin}, \citenamefont {Knizhnik}, \citenamefont {Vasiliev}, \citenamefont {Potapkin},\ and\ \citenamefont {Sommerer}}]{krasikov2016theoretical}%
  \BibitemOpen
  \bibfield  {author} {\bibinfo {author} {\bibfnamefont {D.}~\bibnamefont {Krasikov}}, \bibinfo {author} {\bibfnamefont {A.}~\bibnamefont {Scherbinin}}, \bibinfo {author} {\bibfnamefont {A.}~\bibnamefont {Knizhnik}}, \bibinfo {author} {\bibfnamefont {A.}~\bibnamefont {Vasiliev}}, \bibinfo {author} {\bibfnamefont {B.}~\bibnamefont {Potapkin}}, \ and\ \bibinfo {author} {\bibfnamefont {T.}~\bibnamefont {Sommerer}},\ }\href@noop {} {\bibfield  {journal} {\bibinfo  {journal} {Journal of Applied Physics}\ }\textbf {\bibinfo {volume} {119}} (\bibinfo {year} {2016})}\BibitemShut {NoStop}%
\bibitem [{\citenamefont {Kavanagh}\ \emph {et~al.}(2021)\citenamefont {Kavanagh}, \citenamefont {Walsh},\ and\ \citenamefont {Scanlon}}]{kavanagh2021rapid}%
  \BibitemOpen
  \bibfield  {author} {\bibinfo {author} {\bibfnamefont {S.~R.}\ \bibnamefont {Kavanagh}}, \bibinfo {author} {\bibfnamefont {A.}~\bibnamefont {Walsh}}, \ and\ \bibinfo {author} {\bibfnamefont {D.~O.}\ \bibnamefont {Scanlon}},\ }\href@noop {} {\bibfield  {journal} {\bibinfo  {journal} {ACS Energy Letters}\ }\textbf {\bibinfo {volume} {6}},\ \bibinfo {pages} {1392} (\bibinfo {year} {2021})}\BibitemShut {NoStop}%
\bibitem [{\citenamefont {Shepidchenko}\ \emph {et~al.}(2015)\citenamefont {Shepidchenko}, \citenamefont {Sanyal}, \citenamefont {Klintenberg},\ and\ \citenamefont {Mirbt}}]{shepidchenko2015small}%
  \BibitemOpen
  \bibfield  {author} {\bibinfo {author} {\bibfnamefont {A.}~\bibnamefont {Shepidchenko}}, \bibinfo {author} {\bibfnamefont {B.}~\bibnamefont {Sanyal}}, \bibinfo {author} {\bibfnamefont {M.}~\bibnamefont {Klintenberg}}, \ and\ \bibinfo {author} {\bibfnamefont {S.}~\bibnamefont {Mirbt}},\ }\href@noop {} {\bibfield  {journal} {\bibinfo  {journal} {Scientific reports}\ }\textbf {\bibinfo {volume} {5}},\ \bibinfo {pages} {14509} (\bibinfo {year} {2015})}\BibitemShut {NoStop}%
\bibitem [{\citenamefont {Biswas}\ and\ \citenamefont {Du}(2012)}]{biswas2012causes}%
  \BibitemOpen
  \bibfield  {author} {\bibinfo {author} {\bibfnamefont {K.}~\bibnamefont {Biswas}}\ and\ \bibinfo {author} {\bibfnamefont {M.-H.}\ \bibnamefont {Du}},\ }\href@noop {} {\bibfield  {journal} {\bibinfo  {journal} {New Journal of Physics}\ }\textbf {\bibinfo {volume} {14}},\ \bibinfo {pages} {063020} (\bibinfo {year} {2012})}\BibitemShut {NoStop}%
\bibitem [{\citenamefont {Yang}\ \emph {et~al.}(2014)\citenamefont {Yang}, \citenamefont {Park}, \citenamefont {Kang}, \citenamefont {Metzger}, \citenamefont {Barnes},\ and\ \citenamefont {Wei}}]{yang2014tuning}%
  \BibitemOpen
  \bibfield  {author} {\bibinfo {author} {\bibfnamefont {J.-H.}\ \bibnamefont {Yang}}, \bibinfo {author} {\bibfnamefont {J.-S.}\ \bibnamefont {Park}}, \bibinfo {author} {\bibfnamefont {J.}~\bibnamefont {Kang}}, \bibinfo {author} {\bibfnamefont {W.}~\bibnamefont {Metzger}}, \bibinfo {author} {\bibfnamefont {T.}~\bibnamefont {Barnes}}, \ and\ \bibinfo {author} {\bibfnamefont {S.-H.}\ \bibnamefont {Wei}},\ }\href@noop {} {\bibfield  {journal} {\bibinfo  {journal} {Physical Review B}\ }\textbf {\bibinfo {volume} {90}},\ \bibinfo {pages} {245202} (\bibinfo {year} {2014})}\BibitemShut {NoStop}%
\bibitem [{\citenamefont {Orellana}\ \emph {et~al.}(2019)\citenamefont {Orellana}, \citenamefont {Men{\'e}ndez-Proupin},\ and\ \citenamefont {Flores}}]{orellana2019self}%
  \BibitemOpen
  \bibfield  {author} {\bibinfo {author} {\bibfnamefont {W.}~\bibnamefont {Orellana}}, \bibinfo {author} {\bibfnamefont {E.}~\bibnamefont {Men{\'e}ndez-Proupin}}, \ and\ \bibinfo {author} {\bibfnamefont {M.~A.}\ \bibnamefont {Flores}},\ }\href@noop {} {\bibfield  {journal} {\bibinfo  {journal} {Scientific reports}\ }\textbf {\bibinfo {volume} {9}},\ \bibinfo {pages} {9194} (\bibinfo {year} {2019})}\BibitemShut {NoStop}%
\bibitem [{\citenamefont {Du}\ \emph {et~al.}(2008)\citenamefont {Du}, \citenamefont {Takenaka},\ and\ \citenamefont {Singh}}]{du2008native}%
  \BibitemOpen
  \bibfield  {author} {\bibinfo {author} {\bibfnamefont {M.-H.}\ \bibnamefont {Du}}, \bibinfo {author} {\bibfnamefont {H.}~\bibnamefont {Takenaka}}, \ and\ \bibinfo {author} {\bibfnamefont {D.~J.}\ \bibnamefont {Singh}},\ }\href {\doibase 10.1063/1.3000562} {\bibfield  {journal} {\bibinfo  {journal} {Journal of Applied Physics}\ }\textbf {\bibinfo {volume} {104}},\ \bibinfo {pages} {093521} (\bibinfo {year} {2008})},\ \Eprint {http://arxiv.org/abs/https://pubs.aip.org/aip/jap/article-pdf/doi/10.1063/1.3000562/6596855/093521\_1\_online.pdf} {https://pubs.aip.org/aip/jap/article-pdf/doi/10.1063/1.3000562/6596855/093521\_1\_online.pdf} \BibitemShut {NoStop}%
\bibitem [{\citenamefont {Pan}\ \emph {et~al.}(2018)\citenamefont {Pan}, \citenamefont {Metzger},\ and\ \citenamefont {Lany}}]{pan2018spin}%
  \BibitemOpen
  \bibfield  {author} {\bibinfo {author} {\bibfnamefont {J.}~\bibnamefont {Pan}}, \bibinfo {author} {\bibfnamefont {W.~K.}\ \bibnamefont {Metzger}}, \ and\ \bibinfo {author} {\bibfnamefont {S.}~\bibnamefont {Lany}},\ }\href@noop {} {\bibfield  {journal} {\bibinfo  {journal} {Physical Review B}\ }\textbf {\bibinfo {volume} {98}},\ \bibinfo {pages} {054108} (\bibinfo {year} {2018})}\BibitemShut {NoStop}%
\bibitem [{\citenamefont {Men{\'e}ndez-Proupin}\ \emph {et~al.}(2019)\citenamefont {Men{\'e}ndez-Proupin}, \citenamefont {Casanova-P{\'a}ez}, \citenamefont {Montero-Alejo}, \citenamefont {Flores},\ and\ \citenamefont {Orellana}}]{menendez2019symmetry}%
  \BibitemOpen
  \bibfield  {author} {\bibinfo {author} {\bibfnamefont {E.}~\bibnamefont {Men{\'e}ndez-Proupin}}, \bibinfo {author} {\bibfnamefont {M.}~\bibnamefont {Casanova-P{\'a}ez}}, \bibinfo {author} {\bibfnamefont {A.}~\bibnamefont {Montero-Alejo}}, \bibinfo {author} {\bibfnamefont {M.~A.}\ \bibnamefont {Flores}}, \ and\ \bibinfo {author} {\bibfnamefont {W.}~\bibnamefont {Orellana}},\ }\href@noop {} {\bibfield  {journal} {\bibinfo  {journal} {Physica B: Condensed Matter}\ }\textbf {\bibinfo {volume} {568}},\ \bibinfo {pages} {81} (\bibinfo {year} {2019})}\BibitemShut {NoStop}%
\bibitem [{\citenamefont {Krasikov}\ and\ \citenamefont {Sankin}(2018)}]{krasikov2018beyond}%
  \BibitemOpen
  \bibfield  {author} {\bibinfo {author} {\bibfnamefont {D.}~\bibnamefont {Krasikov}}\ and\ \bibinfo {author} {\bibfnamefont {I.}~\bibnamefont {Sankin}},\ }\href@noop {} {\bibfield  {journal} {\bibinfo  {journal} {Physical Review Materials}\ }\textbf {\bibinfo {volume} {2}},\ \bibinfo {pages} {103803} (\bibinfo {year} {2018})}\BibitemShut {NoStop}%
\bibitem [{\citenamefont {Selvaraj}\ \emph {et~al.}(2021)\citenamefont {Selvaraj}, \citenamefont {Gupta}, \citenamefont {Caliste},\ and\ \citenamefont {Pochet}}]{selvaraj2021passivation}%
  \BibitemOpen
  \bibfield  {author} {\bibinfo {author} {\bibfnamefont {S.~C.}\ \bibnamefont {Selvaraj}}, \bibinfo {author} {\bibfnamefont {S.}~\bibnamefont {Gupta}}, \bibinfo {author} {\bibfnamefont {D.}~\bibnamefont {Caliste}}, \ and\ \bibinfo {author} {\bibfnamefont {P.}~\bibnamefont {Pochet}},\ }\href@noop {} {\bibfield  {journal} {\bibinfo  {journal} {Applied Physics Letters}\ }\textbf {\bibinfo {volume} {119}} (\bibinfo {year} {2021})}\BibitemShut {NoStop}%
\bibitem [{\citenamefont {Krasikov}\ and\ \citenamefont {Sankin}(2017)}]{krasikov2017defect}%
  \BibitemOpen
  \bibfield  {author} {\bibinfo {author} {\bibfnamefont {D.}~\bibnamefont {Krasikov}}\ and\ \bibinfo {author} {\bibfnamefont {I.}~\bibnamefont {Sankin}},\ }\href@noop {} {\bibfield  {journal} {\bibinfo  {journal} {Journal of Materials Chemistry A}\ }\textbf {\bibinfo {volume} {5}},\ \bibinfo {pages} {3503} (\bibinfo {year} {2017})}\BibitemShut {NoStop}%
\bibitem [{\citenamefont {Yang}\ \emph {et~al.}(2016{\natexlab{b}})\citenamefont {Yang}, \citenamefont {Yin}, \citenamefont {Park}, \citenamefont {Metzger},\ and\ \citenamefont {Wei}}]{yang2016first}%
  \BibitemOpen
  \bibfield  {author} {\bibinfo {author} {\bibfnamefont {J.-H.}\ \bibnamefont {Yang}}, \bibinfo {author} {\bibfnamefont {W.-J.}\ \bibnamefont {Yin}}, \bibinfo {author} {\bibfnamefont {J.-S.}\ \bibnamefont {Park}}, \bibinfo {author} {\bibfnamefont {W.}~\bibnamefont {Metzger}}, \ and\ \bibinfo {author} {\bibfnamefont {S.-H.}\ \bibnamefont {Wei}},\ }\href {\doibase 10.1063/1.4940722} {\bibfield  {journal} {\bibinfo  {journal} {Journal of Applied Physics}\ }\textbf {\bibinfo {volume} {119}},\ \bibinfo {pages} {045104} (\bibinfo {year} {2016}{\natexlab{b}})},\ \Eprint {http://arxiv.org/abs/https://pubs.aip.org/aip/jap/article-pdf/doi/10.1063/1.4940722/15176609/045104\_1\_online.pdf} {https://pubs.aip.org/aip/jap/article-pdf/doi/10.1063/1.4940722/15176609/045104\_1\_online.pdf} \BibitemShut {NoStop}%
\bibitem [{\citenamefont {Dou}\ \emph {et~al.}(2021)\citenamefont {Dou}, \citenamefont {Sun},\ and\ \citenamefont {Wei}}]{dou2021effects}%
  \BibitemOpen
  \bibfield  {author} {\bibinfo {author} {\bibfnamefont {B.}~\bibnamefont {Dou}}, \bibinfo {author} {\bibfnamefont {Q.}~\bibnamefont {Sun}}, \ and\ \bibinfo {author} {\bibfnamefont {S.-H.}\ \bibnamefont {Wei}},\ }\href@noop {} {\bibfield  {journal} {\bibinfo  {journal} {Physical Review B}\ }\textbf {\bibinfo {volume} {104}},\ \bibinfo {pages} {245202} (\bibinfo {year} {2021})}\BibitemShut {NoStop}%
\bibitem [{\citenamefont {Yang}\ \emph {et~al.}(2015)\citenamefont {Yang}, \citenamefont {Yin}, \citenamefont {Park}, \citenamefont {Burst}, \citenamefont {Metzger}, \citenamefont {Gessert}, \citenamefont {Barnes},\ and\ \citenamefont {Wei}}]{yang2015enhanced}%
  \BibitemOpen
  \bibfield  {author} {\bibinfo {author} {\bibfnamefont {J.-H.}\ \bibnamefont {Yang}}, \bibinfo {author} {\bibfnamefont {W.-J.}\ \bibnamefont {Yin}}, \bibinfo {author} {\bibfnamefont {J.-S.}\ \bibnamefont {Park}}, \bibinfo {author} {\bibfnamefont {J.}~\bibnamefont {Burst}}, \bibinfo {author} {\bibfnamefont {W.~K.}\ \bibnamefont {Metzger}}, \bibinfo {author} {\bibfnamefont {T.}~\bibnamefont {Gessert}}, \bibinfo {author} {\bibfnamefont {T.}~\bibnamefont {Barnes}}, \ and\ \bibinfo {author} {\bibfnamefont {S.-H.}\ \bibnamefont {Wei}},\ }\href@noop {} {\bibfield  {journal} {\bibinfo  {journal} {Journal of Applied Physics}\ }\textbf {\bibinfo {volume} {118}},\ \bibinfo {pages} {025102} (\bibinfo {year} {2015})}\BibitemShut {NoStop}%
\bibitem [{\citenamefont {Wu}\ \emph {et~al.}(2015)\citenamefont {Wu}, \citenamefont {Chen}, \citenamefont {Zhu}, \citenamefont {Yin}, \citenamefont {Yan}, \citenamefont {Al-Jassim},\ and\ \citenamefont {Pennycook}}]{wu2015lda+}%
  \BibitemOpen
  \bibfield  {author} {\bibinfo {author} {\bibfnamefont {Y.}~\bibnamefont {Wu}}, \bibinfo {author} {\bibfnamefont {G.}~\bibnamefont {Chen}}, \bibinfo {author} {\bibfnamefont {Y.}~\bibnamefont {Zhu}}, \bibinfo {author} {\bibfnamefont {W.-J.}\ \bibnamefont {Yin}}, \bibinfo {author} {\bibfnamefont {Y.}~\bibnamefont {Yan}}, \bibinfo {author} {\bibfnamefont {M.}~\bibnamefont {Al-Jassim}}, \ and\ \bibinfo {author} {\bibfnamefont {S.~J.}\ \bibnamefont {Pennycook}},\ }\href@noop {} {\bibfield  {journal} {\bibinfo  {journal} {Computational Materials Science}\ }\textbf {\bibinfo {volume} {98}},\ \bibinfo {pages} {18} (\bibinfo {year} {2015})}\BibitemShut {NoStop}%
\bibitem [{\citenamefont {Men{\'e}ndez-Proupin}\ \emph {et~al.}(2014)\citenamefont {Men{\'e}ndez-Proupin}, \citenamefont {Am{\'e}zaga},\ and\ \citenamefont {Hern{\'a}ndez}}]{menendez2014electronic}%
  \BibitemOpen
  \bibfield  {author} {\bibinfo {author} {\bibfnamefont {E.}~\bibnamefont {Men{\'e}ndez-Proupin}}, \bibinfo {author} {\bibfnamefont {A.}~\bibnamefont {Am{\'e}zaga}}, \ and\ \bibinfo {author} {\bibfnamefont {N.~C.}\ \bibnamefont {Hern{\'a}ndez}},\ }\href@noop {} {\bibfield  {journal} {\bibinfo  {journal} {Physica B: Condensed Matter}\ }\textbf {\bibinfo {volume} {452}},\ \bibinfo {pages} {119} (\bibinfo {year} {2014})}\BibitemShut {NoStop}%
\bibitem [{\citenamefont {Xiang}\ \emph {et~al.}(2023)\citenamefont {Xiang}, \citenamefont {Gehrke},\ and\ \citenamefont {Dunham}}]{xiang2023understanding}%
  \BibitemOpen
  \bibfield  {author} {\bibinfo {author} {\bibfnamefont {X.}~\bibnamefont {Xiang}}, \bibinfo {author} {\bibfnamefont {A.}~\bibnamefont {Gehrke}}, \ and\ \bibinfo {author} {\bibfnamefont {S.}~\bibnamefont {Dunham}},\ }in\ \href@noop {} {\emph {\bibinfo {booktitle} {2023 IEEE 50th Photovoltaic Specialists Conference (PVSC)}}}\ (\bibinfo  {publisher} {IEEE},\ \bibinfo {year} {2023})\ pp.\ \bibinfo {pages} {1--3}\BibitemShut {NoStop}%
\bibitem [{\citenamefont {Castleton}\ and\ \citenamefont {Mirbt}(2004)}]{castleton2004finite}%
  \BibitemOpen
  \bibfield  {author} {\bibinfo {author} {\bibfnamefont {C.}~\bibnamefont {Castleton}}\ and\ \bibinfo {author} {\bibfnamefont {S.}~\bibnamefont {Mirbt}},\ }\href@noop {} {\bibfield  {journal} {\bibinfo  {journal} {Physical Review B}\ }\textbf {\bibinfo {volume} {70}},\ \bibinfo {pages} {195202} (\bibinfo {year} {2004})}\BibitemShut {NoStop}%
\bibitem [{\citenamefont {Schenk}\ and\ \citenamefont {Silber}(1998)}]{schenk1998lattice}%
  \BibitemOpen
  \bibfield  {author} {\bibinfo {author} {\bibfnamefont {M.}~\bibnamefont {Schenk}}\ and\ \bibinfo {author} {\bibfnamefont {C.}~\bibnamefont {Silber}},\ }\href@noop {} {\bibfield  {journal} {\bibinfo  {journal} {Journal of Materials Science: Materials in Electronics}\ }\textbf {\bibinfo {volume} {9}},\ \bibinfo {pages} {295} (\bibinfo {year} {1998})}\BibitemShut {NoStop}%
\bibitem [{\citenamefont {Poplawsky}\ \emph {et~al.}(2016)\citenamefont {Poplawsky}, \citenamefont {Guo}, \citenamefont {Paudel}, \citenamefont {Ng}, \citenamefont {More}, \citenamefont {Leonard},\ and\ \citenamefont {Yan}}]{poplawsky2016structural}%
  \BibitemOpen
  \bibfield  {author} {\bibinfo {author} {\bibfnamefont {J.~D.}\ \bibnamefont {Poplawsky}}, \bibinfo {author} {\bibfnamefont {W.}~\bibnamefont {Guo}}, \bibinfo {author} {\bibfnamefont {N.}~\bibnamefont {Paudel}}, \bibinfo {author} {\bibfnamefont {A.}~\bibnamefont {Ng}}, \bibinfo {author} {\bibfnamefont {K.}~\bibnamefont {More}}, \bibinfo {author} {\bibfnamefont {D.}~\bibnamefont {Leonard}}, \ and\ \bibinfo {author} {\bibfnamefont {Y.}~\bibnamefont {Yan}},\ }\href@noop {} {\bibfield  {journal} {\bibinfo  {journal} {Nature communications}\ }\textbf {\bibinfo {volume} {7}},\ \bibinfo {pages} {12537} (\bibinfo {year} {2016})}\BibitemShut {NoStop}%
\bibitem [{\citenamefont {Li}\ \emph {et~al.}(2022)\citenamefont {Li}, \citenamefont {Bista}, \citenamefont {Awni}, \citenamefont {Neupane}, \citenamefont {Abudulimu}, \citenamefont {Wang}, \citenamefont {Subedi}, \citenamefont {Jamarkattel}, \citenamefont {Phillips}, \citenamefont {Heben} \emph {et~al.}}]{li202220}%
  \BibitemOpen
  \bibfield  {author} {\bibinfo {author} {\bibfnamefont {D.-B.}\ \bibnamefont {Li}}, \bibinfo {author} {\bibfnamefont {S.~S.}\ \bibnamefont {Bista}}, \bibinfo {author} {\bibfnamefont {R.~A.}\ \bibnamefont {Awni}}, \bibinfo {author} {\bibfnamefont {S.}~\bibnamefont {Neupane}}, \bibinfo {author} {\bibfnamefont {A.}~\bibnamefont {Abudulimu}}, \bibinfo {author} {\bibfnamefont {X.}~\bibnamefont {Wang}}, \bibinfo {author} {\bibfnamefont {K.~K.}\ \bibnamefont {Subedi}}, \bibinfo {author} {\bibfnamefont {M.~K.}\ \bibnamefont {Jamarkattel}}, \bibinfo {author} {\bibfnamefont {A.~B.}\ \bibnamefont {Phillips}}, \bibinfo {author} {\bibfnamefont {M.~J.}\ \bibnamefont {Heben}},  \emph {et~al.},\ }\href@noop {} {\bibfield  {journal} {\bibinfo  {journal} {Nature Communications}\ }\textbf {\bibinfo {volume} {13}},\ \bibinfo {pages} {7849} (\bibinfo {year} {2022})}\BibitemShut {NoStop}%
\bibitem [{\citenamefont {Kuciauskas}\ \emph {et~al.}(2023)\citenamefont {Kuciauskas}, \citenamefont {Nardone}, \citenamefont {Bothwell}, \citenamefont {Albin}, \citenamefont {Reich}, \citenamefont {Lee},\ and\ \citenamefont {Colegrove}}]{kuciauskas2023increased}%
  \BibitemOpen
  \bibfield  {author} {\bibinfo {author} {\bibfnamefont {D.}~\bibnamefont {Kuciauskas}}, \bibinfo {author} {\bibfnamefont {M.}~\bibnamefont {Nardone}}, \bibinfo {author} {\bibfnamefont {A.}~\bibnamefont {Bothwell}}, \bibinfo {author} {\bibfnamefont {D.}~\bibnamefont {Albin}}, \bibinfo {author} {\bibfnamefont {C.}~\bibnamefont {Reich}}, \bibinfo {author} {\bibfnamefont {C.}~\bibnamefont {Lee}}, \ and\ \bibinfo {author} {\bibfnamefont {E.}~\bibnamefont {Colegrove}},\ }\href@noop {} {\bibfield  {journal} {\bibinfo  {journal} {Advanced Energy Materials}\ }\textbf {\bibinfo {volume} {13}},\ \bibinfo {pages} {2301784} (\bibinfo {year} {2023})}\BibitemShut {NoStop}%
\bibitem [{\citenamefont {Okhotnikov}\ \emph {et~al.}(2016)\citenamefont {Okhotnikov}, \citenamefont {Charpentier},\ and\ \citenamefont {Cadars}}]{okhotnikov2016supercell}%
  \BibitemOpen
  \bibfield  {author} {\bibinfo {author} {\bibfnamefont {K.}~\bibnamefont {Okhotnikov}}, \bibinfo {author} {\bibfnamefont {T.}~\bibnamefont {Charpentier}}, \ and\ \bibinfo {author} {\bibfnamefont {S.}~\bibnamefont {Cadars}},\ }\href@noop {} {\bibfield  {journal} {\bibinfo  {journal} {Journal of cheminformatics}\ }\textbf {\bibinfo {volume} {8}},\ \bibinfo {pages} {1} (\bibinfo {year} {2016})}\BibitemShut {NoStop}%
\bibitem [{\citenamefont {Scarpulla}\ \emph {et~al.}(2023)\citenamefont {Scarpulla}, \citenamefont {McCandless}, \citenamefont {Phillips}, \citenamefont {Yan}, \citenamefont {Heben}, \citenamefont {Wolden}, \citenamefont {Xiong}, \citenamefont {Metzger}, \citenamefont {Mao}, \citenamefont {Krasikov} \emph {et~al.}}]{scarpulla2023cdte}%
  \BibitemOpen
  \bibfield  {author} {\bibinfo {author} {\bibfnamefont {M.~A.}\ \bibnamefont {Scarpulla}}, \bibinfo {author} {\bibfnamefont {B.}~\bibnamefont {McCandless}}, \bibinfo {author} {\bibfnamefont {A.~B.}\ \bibnamefont {Phillips}}, \bibinfo {author} {\bibfnamefont {Y.}~\bibnamefont {Yan}}, \bibinfo {author} {\bibfnamefont {M.~J.}\ \bibnamefont {Heben}}, \bibinfo {author} {\bibfnamefont {C.}~\bibnamefont {Wolden}}, \bibinfo {author} {\bibfnamefont {G.}~\bibnamefont {Xiong}}, \bibinfo {author} {\bibfnamefont {W.~K.}\ \bibnamefont {Metzger}}, \bibinfo {author} {\bibfnamefont {D.}~\bibnamefont {Mao}}, \bibinfo {author} {\bibfnamefont {D.}~\bibnamefont {Krasikov}},  \emph {et~al.},\ }\href@noop {} {\bibfield  {journal} {\bibinfo  {journal} {Solar Energy Materials and Solar Cells}\ }\textbf {\bibinfo {volume} {255}},\ \bibinfo {pages} {112289} (\bibinfo {year} {2023})}\BibitemShut {NoStop}%
\bibitem [{\citenamefont {Albin}\ \emph {et~al.}(2002)\citenamefont {Albin}, \citenamefont {Yan},\ and\ \citenamefont {Al-Jassim}}]{albin2002effect}%
  \BibitemOpen
  \bibfield  {author} {\bibinfo {author} {\bibfnamefont {D.}~\bibnamefont {Albin}}, \bibinfo {author} {\bibfnamefont {Y.}~\bibnamefont {Yan}}, \ and\ \bibinfo {author} {\bibfnamefont {M.}~\bibnamefont {Al-Jassim}},\ }\href@noop {} {\bibfield  {journal} {\bibinfo  {journal} {Progress in Photovoltaics: Research and Applications}\ }\textbf {\bibinfo {volume} {10}},\ \bibinfo {pages} {309} (\bibinfo {year} {2002})}\BibitemShut {NoStop}%
\bibitem [{\citenamefont {Landolt}\ \emph {et~al.}(1961)\citenamefont {Landolt}, \citenamefont {B{\"o}rnstein},\ and\ \citenamefont {Hellwege}}]{landolt1966numerical}%
  \BibitemOpen
  \bibfield  {author} {\bibinfo {author} {\bibfnamefont {H.}~\bibnamefont {Landolt}}, \bibinfo {author} {\bibfnamefont {R.}~\bibnamefont {B{\"o}rnstein}}, \ and\ \bibinfo {author} {\bibfnamefont {H.}~\bibnamefont {Hellwege}},\ }\href@noop {} {\emph {\bibinfo {title} {Numerical Data and Functional Relationships in Science and Technology}}}\ (\bibinfo  {publisher} {Springer},\ \bibinfo {address} {Berlin},\ \bibinfo {year} {1961})\BibitemShut {NoStop}%
\bibitem [{\citenamefont {Freysoldt}\ \emph {et~al.}(2014)\citenamefont {Freysoldt}, \citenamefont {Grabowski}, \citenamefont {Hickel}, \citenamefont {Neugebauer}, \citenamefont {Kresse}, \citenamefont {Janotti},\ and\ \citenamefont {Van~de Walle}}]{freysoldt2014first}%
  \BibitemOpen
  \bibfield  {author} {\bibinfo {author} {\bibfnamefont {C.}~\bibnamefont {Freysoldt}}, \bibinfo {author} {\bibfnamefont {B.}~\bibnamefont {Grabowski}}, \bibinfo {author} {\bibfnamefont {T.}~\bibnamefont {Hickel}}, \bibinfo {author} {\bibfnamefont {J.}~\bibnamefont {Neugebauer}}, \bibinfo {author} {\bibfnamefont {G.}~\bibnamefont {Kresse}}, \bibinfo {author} {\bibfnamefont {A.}~\bibnamefont {Janotti}}, \ and\ \bibinfo {author} {\bibfnamefont {C.~G.}\ \bibnamefont {Van~de Walle}},\ }\href@noop {} {\bibfield  {journal} {\bibinfo  {journal} {Reviews of modern physics}\ }\textbf {\bibinfo {volume} {86}},\ \bibinfo {pages} {253} (\bibinfo {year} {2014})}\BibitemShut {NoStop}%
\bibitem [{\citenamefont {Freysoldt}\ \emph {et~al.}(2009)\citenamefont {Freysoldt}, \citenamefont {Neugebauer},\ and\ \citenamefont {Van~de Walle}}]{freysoldt2009fully}%
  \BibitemOpen
  \bibfield  {author} {\bibinfo {author} {\bibfnamefont {C.}~\bibnamefont {Freysoldt}}, \bibinfo {author} {\bibfnamefont {J.}~\bibnamefont {Neugebauer}}, \ and\ \bibinfo {author} {\bibfnamefont {C.~G.}\ \bibnamefont {Van~de Walle}},\ }\href@noop {} {\bibfield  {journal} {\bibinfo  {journal} {Physical review letters}\ }\textbf {\bibinfo {volume} {102}},\ \bibinfo {pages} {016402} (\bibinfo {year} {2009})}\BibitemShut {NoStop}%
\bibitem [{\citenamefont {Komsa}\ \emph {et~al.}(2012)\citenamefont {Komsa}, \citenamefont {Rantala},\ and\ \citenamefont {Pasquarello}}]{komsa2012finite}%
  \BibitemOpen
  \bibfield  {author} {\bibinfo {author} {\bibfnamefont {H.-P.}\ \bibnamefont {Komsa}}, \bibinfo {author} {\bibfnamefont {T.~T.}\ \bibnamefont {Rantala}}, \ and\ \bibinfo {author} {\bibfnamefont {A.}~\bibnamefont {Pasquarello}},\ }\href@noop {} {\bibfield  {journal} {\bibinfo  {journal} {Physical Review B}\ }\textbf {\bibinfo {volume} {86}},\ \bibinfo {pages} {045112} (\bibinfo {year} {2012})}\BibitemShut {NoStop}%
\bibitem [{\citenamefont {Tong}\ \emph {et~al.}()\citenamefont {Tong}, \citenamefont {Xiang},\ and\ \citenamefont {Dunham}}]{tong2024pervoskite}%
  \BibitemOpen
  \bibfield  {author} {\bibinfo {author} {\bibfnamefont {Y.}~\bibnamefont {Tong}}, \bibinfo {author} {\bibfnamefont {X.}~\bibnamefont {Xiang}}, \ and\ \bibinfo {author} {\bibfnamefont {S.~T.}\ \bibnamefont {Dunham}},\ }\href@noop {} {\bibinfo  {journal} {(unpublished)}\ }\BibitemShut {NoStop}%
\bibitem [{\citenamefont {Limas}\ and\ \citenamefont {Manz}(2016)}]{limas2016introducing}%
  \BibitemOpen
\bibfield  {journal} {  }\bibfield  {author} {\bibinfo {author} {\bibfnamefont {N.~G.}\ \bibnamefont {Limas}}\ and\ \bibinfo {author} {\bibfnamefont {T.~A.}\ \bibnamefont {Manz}},\ }\href@noop {} {\bibfield  {journal} {\bibinfo  {journal} {RSC advances}\ }\textbf {\bibinfo {volume} {6}},\ \bibinfo {pages} {45727} (\bibinfo {year} {2016})}\BibitemShut {NoStop}%
\bibitem [{\citenamefont {Manz}\ and\ \citenamefont {Limas}(2016)}]{manz2016introducing}%
  \BibitemOpen
  \bibfield  {author} {\bibinfo {author} {\bibfnamefont {T.~A.}\ \bibnamefont {Manz}}\ and\ \bibinfo {author} {\bibfnamefont {N.~G.}\ \bibnamefont {Limas}},\ }\href@noop {} {\bibfield  {journal} {\bibinfo  {journal} {RSC advances}\ }\textbf {\bibinfo {volume} {6}},\ \bibinfo {pages} {47771} (\bibinfo {year} {2016})}\BibitemShut {NoStop}%
\bibitem [{\citenamefont {Oba}\ \emph {et~al.}(2008)\citenamefont {Oba}, \citenamefont {Togo}, \citenamefont {Tanaka}, \citenamefont {Paier},\ and\ \citenamefont {Kresse}}]{oba2008defect}%
  \BibitemOpen
  \bibfield  {author} {\bibinfo {author} {\bibfnamefont {F.}~\bibnamefont {Oba}}, \bibinfo {author} {\bibfnamefont {A.}~\bibnamefont {Togo}}, \bibinfo {author} {\bibfnamefont {I.}~\bibnamefont {Tanaka}}, \bibinfo {author} {\bibfnamefont {J.}~\bibnamefont {Paier}}, \ and\ \bibinfo {author} {\bibfnamefont {G.}~\bibnamefont {Kresse}},\ }\href@noop {} {\bibfield  {journal} {\bibinfo  {journal} {Physical Review B}\ }\textbf {\bibinfo {volume} {77}},\ \bibinfo {pages} {245202} (\bibinfo {year} {2008})}\BibitemShut {NoStop}%
\bibitem [{\citenamefont {Nagaoka}\ \emph {et~al.}(2020)\citenamefont {Nagaoka}, \citenamefont {Nishioka}, \citenamefont {Yoshino}, \citenamefont {Katsube}, \citenamefont {Nose}, \citenamefont {Masuda},\ and\ \citenamefont {Scarpulla}}]{nagaoka2020comparison}%
  \BibitemOpen
  \bibfield  {author} {\bibinfo {author} {\bibfnamefont {A.}~\bibnamefont {Nagaoka}}, \bibinfo {author} {\bibfnamefont {K.}~\bibnamefont {Nishioka}}, \bibinfo {author} {\bibfnamefont {K.}~\bibnamefont {Yoshino}}, \bibinfo {author} {\bibfnamefont {R.}~\bibnamefont {Katsube}}, \bibinfo {author} {\bibfnamefont {Y.}~\bibnamefont {Nose}}, \bibinfo {author} {\bibfnamefont {T.}~\bibnamefont {Masuda}}, \ and\ \bibinfo {author} {\bibfnamefont {M.~A.}\ \bibnamefont {Scarpulla}},\ }\href@noop {} {\bibfield  {journal} {\bibinfo  {journal} {Applied Physics Letters}\ }\textbf {\bibinfo {volume} {116}} (\bibinfo {year} {2020})}\BibitemShut {NoStop}%
\bibitem [{\citenamefont {Seeger}(2013)}]{seeger2013semiconductor}%
  \BibitemOpen
  \bibfield  {author} {\bibinfo {author} {\bibfnamefont {K.}~\bibnamefont {Seeger}},\ }\href {https://books.google.com/books?id=Hp36CAAAQBAJ} {\emph {\bibinfo {title} {Semiconductor Physics}}},\ Springer Study Edition\ (\bibinfo  {publisher} {Springer Vienna},\ \bibinfo {year} {2013})\BibitemShut {NoStop}%
\bibitem [{\citenamefont {Emanuelsson}\ \emph {et~al.}(1993)\citenamefont {Emanuelsson}, \citenamefont {Omling}, \citenamefont {Meyer}, \citenamefont {Wienecke},\ and\ \citenamefont {Schenk}}]{emanuelsson1993identification}%
  \BibitemOpen
  \bibfield  {author} {\bibinfo {author} {\bibfnamefont {P.}~\bibnamefont {Emanuelsson}}, \bibinfo {author} {\bibfnamefont {P.}~\bibnamefont {Omling}}, \bibinfo {author} {\bibfnamefont {B.}~\bibnamefont {Meyer}}, \bibinfo {author} {\bibfnamefont {M.}~\bibnamefont {Wienecke}}, \ and\ \bibinfo {author} {\bibfnamefont {M.}~\bibnamefont {Schenk}},\ }\href@noop {} {\bibfield  {journal} {\bibinfo  {journal} {Physical Review B}\ }\textbf {\bibinfo {volume} {47}},\ \bibinfo {pages} {15578} (\bibinfo {year} {1993})}\BibitemShut {NoStop}%
\bibitem [{\citenamefont {Berding}(1999)}]{berding1999native}%
  \BibitemOpen
  \bibfield  {author} {\bibinfo {author} {\bibfnamefont {M.}~\bibnamefont {Berding}},\ }\href@noop {} {\bibfield  {journal} {\bibinfo  {journal} {Physical Review B}\ }\textbf {\bibinfo {volume} {60}},\ \bibinfo {pages} {8943} (\bibinfo {year} {1999})}\BibitemShut {NoStop}%
\bibitem [{\citenamefont {Wei}\ and\ \citenamefont {Zhang}(2002)}]{wei2002chemical}%
  \BibitemOpen
  \bibfield  {author} {\bibinfo {author} {\bibfnamefont {S.-H.}\ \bibnamefont {Wei}}\ and\ \bibinfo {author} {\bibfnamefont {S.}~\bibnamefont {Zhang}},\ }\href@noop {} {\bibfield  {journal} {\bibinfo  {journal} {Physical Review B}\ }\textbf {\bibinfo {volume} {66}},\ \bibinfo {pages} {155211} (\bibinfo {year} {2002})}\BibitemShut {NoStop}%
\bibitem [{\citenamefont {Chang}\ \emph {et~al.}(2006)\citenamefont {Chang}, \citenamefont {James},\ and\ \citenamefont {Davenport}}]{chang2006symmetrized}%
  \BibitemOpen
  \bibfield  {author} {\bibinfo {author} {\bibfnamefont {Y.-C.}\ \bibnamefont {Chang}}, \bibinfo {author} {\bibfnamefont {R.}~\bibnamefont {James}}, \ and\ \bibinfo {author} {\bibfnamefont {J.}~\bibnamefont {Davenport}},\ }\href@noop {} {\bibfield  {journal} {\bibinfo  {journal} {Physical Review B}\ }\textbf {\bibinfo {volume} {73}},\ \bibinfo {pages} {035211} (\bibinfo {year} {2006})}\BibitemShut {NoStop}%
\bibitem [{\citenamefont {Castaldini}\ \emph {et~al.}(1998)\citenamefont {Castaldini}, \citenamefont {Cavallini}, \citenamefont {Fraboni}, \citenamefont {Fernandez},\ and\ \citenamefont {Piqueras}}]{castaldini1998deep}%
  \BibitemOpen
  \bibfield  {author} {\bibinfo {author} {\bibfnamefont {A.}~\bibnamefont {Castaldini}}, \bibinfo {author} {\bibfnamefont {A.}~\bibnamefont {Cavallini}}, \bibinfo {author} {\bibfnamefont {B.}~\bibnamefont {Fraboni}}, \bibinfo {author} {\bibfnamefont {P.}~\bibnamefont {Fernandez}}, \ and\ \bibinfo {author} {\bibfnamefont {J.}~\bibnamefont {Piqueras}},\ }\href@noop {} {\bibfield  {journal} {\bibinfo  {journal} {Journal of applied physics}\ }\textbf {\bibinfo {volume} {83}},\ \bibinfo {pages} {2121} (\bibinfo {year} {1998})}\BibitemShut {NoStop}%
\bibitem [{\citenamefont {Reisl{\"o}hner}\ \emph {et~al.}(1998)\citenamefont {Reisl{\"o}hner}, \citenamefont {Grillenberger},\ and\ \citenamefont {Witthuhn}}]{reislohner1998band}%
  \BibitemOpen
  \bibfield  {author} {\bibinfo {author} {\bibfnamefont {U.}~\bibnamefont {Reisl{\"o}hner}}, \bibinfo {author} {\bibfnamefont {J.}~\bibnamefont {Grillenberger}}, \ and\ \bibinfo {author} {\bibfnamefont {W.}~\bibnamefont {Witthuhn}},\ }\href@noop {} {\bibfield  {journal} {\bibinfo  {journal} {Journal of crystal growth}\ }\textbf {\bibinfo {volume} {184}},\ \bibinfo {pages} {1160} (\bibinfo {year} {1998})}\BibitemShut {NoStop}%
\bibitem [{\citenamefont {Park}\ \emph {et~al.}(2018)\citenamefont {Park}, \citenamefont {Kim}, \citenamefont {Xie},\ and\ \citenamefont {Walsh}}]{park2018point}%
  \BibitemOpen
  \bibfield  {author} {\bibinfo {author} {\bibfnamefont {J.~S.}\ \bibnamefont {Park}}, \bibinfo {author} {\bibfnamefont {S.}~\bibnamefont {Kim}}, \bibinfo {author} {\bibfnamefont {Z.}~\bibnamefont {Xie}}, \ and\ \bibinfo {author} {\bibfnamefont {A.}~\bibnamefont {Walsh}},\ }\href {\doibase 10.1038/s41578-018-0026-7} {\bibfield  {journal} {\bibinfo  {journal} {Nat. Rev. Mater.}\ }\textbf {\bibinfo {volume} {3}},\ \bibinfo {pages} {194} (\bibinfo {year} {2018})}\BibitemShut {NoStop}%
\bibitem [{\citenamefont {Stoneham}(1981)}]{stoneham1981non}%
  \BibitemOpen
  \bibfield  {author} {\bibinfo {author} {\bibfnamefont {A.}~\bibnamefont {Stoneham}},\ }\href {\doibase 10.1088/0034-4885/44/12/001} {\bibfield  {journal} {\bibinfo  {journal} {Rep. Prog. Phys.}\ }\textbf {\bibinfo {volume} {44}},\ \bibinfo {pages} {1251} (\bibinfo {year} {1981})}\BibitemShut {NoStop}%
\bibitem [{\citenamefont {Alkauskas}\ \emph {et~al.}(2014)\citenamefont {Alkauskas}, \citenamefont {Yan},\ and\ \citenamefont {Van~de Walle}}]{alkauskas2014first}%
  \BibitemOpen
  \bibfield  {author} {\bibinfo {author} {\bibfnamefont {A.}~\bibnamefont {Alkauskas}}, \bibinfo {author} {\bibfnamefont {Q.}~\bibnamefont {Yan}}, \ and\ \bibinfo {author} {\bibfnamefont {C.~G.}\ \bibnamefont {Van~de Walle}},\ }\href {\doibase 10.1103/PhysRevB.90.075202} {\bibfield  {journal} {\bibinfo  {journal} {Phys. Rev. B}\ }\textbf {\bibinfo {volume} {90}},\ \bibinfo {pages} {075202} (\bibinfo {year} {2014})}\BibitemShut {NoStop}%
\bibitem [{\citenamefont {Kim}\ \emph {et~al.}(2019{\natexlab{a}})\citenamefont {Kim}, \citenamefont {Park}, \citenamefont {Hood},\ and\ \citenamefont {Walsh}}]{Kim2019kesterite}%
  \BibitemOpen
  \bibfield  {author} {\bibinfo {author} {\bibfnamefont {S.}~\bibnamefont {Kim}}, \bibinfo {author} {\bibfnamefont {J.-S.}\ \bibnamefont {Park}}, \bibinfo {author} {\bibfnamefont {S.~N.}\ \bibnamefont {Hood}}, \ and\ \bibinfo {author} {\bibfnamefont {A.}~\bibnamefont {Walsh}},\ }\href {\doibase 10.1039/C8TA10130B} {\bibfield  {journal} {\bibinfo  {journal} {J. Mater. Chem. A}\ }\textbf {\bibinfo {volume} {7}},\ \bibinfo {pages} {2686} (\bibinfo {year} {2019}{\natexlab{a}})}\BibitemShut {NoStop}%
\bibitem [{\citenamefont {Kim}\ \emph {et~al.}(2019{\natexlab{b}})\citenamefont {Kim}, \citenamefont {Hood},\ and\ \citenamefont {Walsh}}]{kim2019anham}%
  \BibitemOpen
  \bibfield  {author} {\bibinfo {author} {\bibfnamefont {S.}~\bibnamefont {Kim}}, \bibinfo {author} {\bibfnamefont {S.~N.}\ \bibnamefont {Hood}}, \ and\ \bibinfo {author} {\bibfnamefont {A.}~\bibnamefont {Walsh}},\ }\href {\doibase 10.1103/PhysRevB.100.041202} {\bibfield  {journal} {\bibinfo  {journal} {Phys. Rev. B}\ }\textbf {\bibinfo {volume} {100}},\ \bibinfo {pages} {041202} (\bibinfo {year} {2019}{\natexlab{b}})}\BibitemShut {NoStop}%
\bibitem [{\citenamefont {Efron}\ and\ \citenamefont {Tibshirani}(1993)}]{tibshirani1993introduction}%
  \BibitemOpen
  \bibfield  {author} {\bibinfo {author} {\bibfnamefont {B.}~\bibnamefont {Efron}}\ and\ \bibinfo {author} {\bibfnamefont {R.}~\bibnamefont {Tibshirani}},\ }\href@noop {} {\emph {\bibinfo {title} {An introduction to the bootstrap}}},\ Monographs on statistics and applied probability ; 57\ (\bibinfo  {publisher} {Chapman \& Hall},\ \bibinfo {address} {New York},\ \bibinfo {year} {1993})\BibitemShut {NoStop}%
\bibitem [{\citenamefont {Ma}\ \emph {et~al.}(2013)\citenamefont {Ma}, \citenamefont {Kuciauskas}, \citenamefont {Albin}, \citenamefont {Bhattacharya}, \citenamefont {Reese}, \citenamefont {Barnes}, \citenamefont {Li}, \citenamefont {Gessert},\ and\ \citenamefont {Wei}}]{ma2013dependence}%
  \BibitemOpen
  \bibfield  {author} {\bibinfo {author} {\bibfnamefont {J.}~\bibnamefont {Ma}}, \bibinfo {author} {\bibfnamefont {D.}~\bibnamefont {Kuciauskas}}, \bibinfo {author} {\bibfnamefont {D.}~\bibnamefont {Albin}}, \bibinfo {author} {\bibfnamefont {R.}~\bibnamefont {Bhattacharya}}, \bibinfo {author} {\bibfnamefont {M.}~\bibnamefont {Reese}}, \bibinfo {author} {\bibfnamefont {T.}~\bibnamefont {Barnes}}, \bibinfo {author} {\bibfnamefont {J.~V.}\ \bibnamefont {Li}}, \bibinfo {author} {\bibfnamefont {T.}~\bibnamefont {Gessert}}, \ and\ \bibinfo {author} {\bibfnamefont {S.-H.}\ \bibnamefont {Wei}},\ }\href@noop {} {\bibfield  {journal} {\bibinfo  {journal} {Physical review letters}\ }\textbf {\bibinfo {volume} {111}},\ \bibinfo {pages} {067402} (\bibinfo {year} {2013})}\BibitemShut {NoStop}%
\bibitem [{\citenamefont {Ma}\ \emph {et~al.}(2014)\citenamefont {Ma}, \citenamefont {Yang}, \citenamefont {Wei},\ and\ \citenamefont {Da~Silva}}]{ma2014correlation}%
  \BibitemOpen
  \bibfield  {author} {\bibinfo {author} {\bibfnamefont {J.}~\bibnamefont {Ma}}, \bibinfo {author} {\bibfnamefont {J.}~\bibnamefont {Yang}}, \bibinfo {author} {\bibfnamefont {S.-H.}\ \bibnamefont {Wei}}, \ and\ \bibinfo {author} {\bibfnamefont {J.~L.}\ \bibnamefont {Da~Silva}},\ }\href@noop {} {\bibfield  {journal} {\bibinfo  {journal} {Physical Review B}\ }\textbf {\bibinfo {volume} {90}},\ \bibinfo {pages} {155208} (\bibinfo {year} {2014})}\BibitemShut {NoStop}%
\bibitem [{\citenamefont {Mutter}\ and\ \citenamefont {Dunham}(2015)}]{mutter2015calculation}%
  \BibitemOpen
  \bibfield  {author} {\bibinfo {author} {\bibfnamefont {D.}~\bibnamefont {Mutter}}\ and\ \bibinfo {author} {\bibfnamefont {S.~T.}\ \bibnamefont {Dunham}},\ }\href@noop {} {\bibfield  {journal} {\bibinfo  {journal} {IEEE Journal of Photovoltaics}\ }\textbf {\bibinfo {volume} {5}},\ \bibinfo {pages} {1188} (\bibinfo {year} {2015})}\BibitemShut {NoStop}%
\bibitem [{\citenamefont {Balcioglu}\ \emph {et~al.}(2000)\citenamefont {Balcioglu}, \citenamefont {Ahrenkiel},\ and\ \citenamefont {Hasoon}}]{balcioglu2000deep}%
  \BibitemOpen
  \bibfield  {author} {\bibinfo {author} {\bibfnamefont {A.}~\bibnamefont {Balcioglu}}, \bibinfo {author} {\bibfnamefont {R.}~\bibnamefont {Ahrenkiel}}, \ and\ \bibinfo {author} {\bibfnamefont {F.}~\bibnamefont {Hasoon}},\ }\href@noop {} {\bibfield  {journal} {\bibinfo  {journal} {Journal of Applied Physics}\ }\textbf {\bibinfo {volume} {88}},\ \bibinfo {pages} {7175} (\bibinfo {year} {2000})}\BibitemShut {NoStop}%
\bibitem [{\citenamefont {Krasikov}(2019)}]{krasikov2019selenium}%
  \BibitemOpen
  \bibfield  {author} {\bibinfo {author} {\bibfnamefont {D.}~\bibnamefont {Krasikov}},\ }\href@noop {} {\bibfield  {journal} {\bibinfo  {journal} {Nature Energy}\ }\textbf {\bibinfo {volume} {4}},\ \bibinfo {pages} {442} (\bibinfo {year} {2019})}\BibitemShut {NoStop}%
\bibitem [{\citenamefont {Xiang}\ \emph {et~al.}(2021)\citenamefont {Xiang}, \citenamefont {Sommer}, \citenamefont {Gehrke},\ and\ \citenamefont {Dunham}}]{xiang2021coupled}%
  \BibitemOpen
  \bibfield  {author} {\bibinfo {author} {\bibfnamefont {X.}~\bibnamefont {Xiang}}, \bibinfo {author} {\bibfnamefont {D.~E.}\ \bibnamefont {Sommer}}, \bibinfo {author} {\bibfnamefont {A.}~\bibnamefont {Gehrke}}, \ and\ \bibinfo {author} {\bibfnamefont {S.~T.}\ \bibnamefont {Dunham}},\ }in\ \href@noop {} {\emph {\bibinfo {booktitle} {2021 IEEE 48th Photovoltaic Specialists Conference (PVSC)}}}\ (\bibinfo  {publisher} {IEEE},\ \bibinfo {year} {2021})\ pp.\ \bibinfo {pages} {1707--1711}\BibitemShut {NoStop}%
\bibitem [{\citenamefont {Gehrke}\ \emph {et~al.}(2023)\citenamefont {Gehrke}, \citenamefont {Sommer},\ and\ \citenamefont {Dunham}}]{gehrke2023atomistic}%
  \BibitemOpen
  \bibfield  {author} {\bibinfo {author} {\bibfnamefont {A.~S.}\ \bibnamefont {Gehrke}}, \bibinfo {author} {\bibfnamefont {D.~E.}\ \bibnamefont {Sommer}}, \ and\ \bibinfo {author} {\bibfnamefont {S.~T.}\ \bibnamefont {Dunham}},\ }\href@noop {} {\bibfield  {journal} {\bibinfo  {journal} {Journal of Applied Physics}\ }\textbf {\bibinfo {volume} {134}} (\bibinfo {year} {2023})}\BibitemShut {NoStop}%
\bibitem [{\citenamefont {Xiang}\ \emph {et~al.}(2024)\citenamefont {Xiang}, \citenamefont {Sommer}, \citenamefont {Gehrke},\ and\ \citenamefont {Dunham}}]{xiang2024coupled}%
  \BibitemOpen
  \bibfield  {author} {\bibinfo {author} {\bibfnamefont {X.}~\bibnamefont {Xiang}}, \bibinfo {author} {\bibfnamefont {D.~E.}\ \bibnamefont {Sommer}}, \bibinfo {author} {\bibfnamefont {A.}~\bibnamefont {Gehrke}}, \ and\ \bibinfo {author} {\bibfnamefont {S.~T.}\ \bibnamefont {Dunham}},\ }\href {\doibase 10.1109/JPHOTOV.2024.3366652} {\bibfield  {journal} {\bibinfo  {journal} {IEEE Journal of Photovoltaics}\ }\textbf {\bibinfo {volume} {14}},\ \bibinfo {pages} {422} (\bibinfo {year} {2024})}\BibitemShut {NoStop}%
\bibitem [{\citenamefont {Chatratin}\ \emph {et~al.}(2023)\citenamefont {Chatratin}, \citenamefont {Dou}, \citenamefont {Wei},\ and\ \citenamefont {Janotti}}]{chatratin2023doping}%
  \BibitemOpen
  \bibfield  {author} {\bibinfo {author} {\bibfnamefont {I.}~\bibnamefont {Chatratin}}, \bibinfo {author} {\bibfnamefont {B.}~\bibnamefont {Dou}}, \bibinfo {author} {\bibfnamefont {S.-H.}\ \bibnamefont {Wei}}, \ and\ \bibinfo {author} {\bibfnamefont {A.}~\bibnamefont {Janotti}},\ }\href@noop {} {\bibfield  {journal} {\bibinfo  {journal} {The Journal of Physical Chemistry Letters}\ }\textbf {\bibinfo {volume} {14}},\ \bibinfo {pages} {273} (\bibinfo {year} {2023})}\BibitemShut {NoStop}%
\bibitem [{\citenamefont {Dreyer}\ \emph {et~al.}(2020)\citenamefont {Dreyer}, \citenamefont {Alkauskas}, \citenamefont {Lyons},\ and\ \citenamefont {Van~de Walle}}]{dreyer2020radiative}%
  \BibitemOpen
  \bibfield  {author} {\bibinfo {author} {\bibfnamefont {C.~E.}\ \bibnamefont {Dreyer}}, \bibinfo {author} {\bibfnamefont {A.}~\bibnamefont {Alkauskas}}, \bibinfo {author} {\bibfnamefont {J.~L.}\ \bibnamefont {Lyons}}, \ and\ \bibinfo {author} {\bibfnamefont {C.~G.}\ \bibnamefont {Van~de Walle}},\ }\href@noop {} {\bibfield  {journal} {\bibinfo  {journal} {Physical Review B}\ }\textbf {\bibinfo {volume} {102}},\ \bibinfo {pages} {085305} (\bibinfo {year} {2020})}\BibitemShut {NoStop}%
\bibitem [{\citenamefont {Ullrich}\ and\ \citenamefont {Yang}(2016)}]{ullrich2016excitons}%
  \BibitemOpen
  \bibfield  {author} {\bibinfo {author} {\bibfnamefont {C.~A.}\ \bibnamefont {Ullrich}}\ and\ \bibinfo {author} {\bibfnamefont {Z.-h.}\ \bibnamefont {Yang}},\ }\href {\doibase 10.1007/128_2014_610} {\emph {\bibinfo {title} {Density-Functional Methods for Excited States}}},\ edited by\ \bibinfo {editor} {\bibfnamefont {N.}~\bibnamefont {Ferr{\'e}}}, \bibinfo {editor} {\bibfnamefont {M.}~\bibnamefont {Filatov}}, \ and\ \bibinfo {editor} {\bibfnamefont {M.}~\bibnamefont {Huix-Rotllant}}\ (\bibinfo  {publisher} {Springer International Publishing},\ \bibinfo {address} {Cham},\ \bibinfo {year} {2016})\ pp.\ \bibinfo {pages} {185--217}\BibitemShut {NoStop}%
\bibitem [{\citenamefont {Li}\ \emph {et~al.}(2021)\citenamefont {Li}, \citenamefont {Yao}, \citenamefont {Vijayaraghavan}, \citenamefont {Awni}, \citenamefont {Subedi}, \citenamefont {Ellingson}, \citenamefont {Li}, \citenamefont {Yan},\ and\ \citenamefont {Yan}}]{li2021low}%
  \BibitemOpen
  \bibfield  {author} {\bibinfo {author} {\bibfnamefont {D.-B.}\ \bibnamefont {Li}}, \bibinfo {author} {\bibfnamefont {C.}~\bibnamefont {Yao}}, \bibinfo {author} {\bibfnamefont {S.}~\bibnamefont {Vijayaraghavan}}, \bibinfo {author} {\bibfnamefont {R.~A.}\ \bibnamefont {Awni}}, \bibinfo {author} {\bibfnamefont {K.~K.}\ \bibnamefont {Subedi}}, \bibinfo {author} {\bibfnamefont {R.~J.}\ \bibnamefont {Ellingson}}, \bibinfo {author} {\bibfnamefont {L.}~\bibnamefont {Li}}, \bibinfo {author} {\bibfnamefont {Y.}~\bibnamefont {Yan}}, \ and\ \bibinfo {author} {\bibfnamefont {F.}~\bibnamefont {Yan}},\ }\href@noop {} {\bibfield  {journal} {\bibinfo  {journal} {Nature Energy}\ }\textbf {\bibinfo {volume} {6}},\ \bibinfo {pages} {715} (\bibinfo {year} {2021})}\BibitemShut {NoStop}%
\bibitem [{\citenamefont {Artegiani}\ \emph {et~al.}(2020)\citenamefont {Artegiani}, \citenamefont {Major}, \citenamefont {Shiel}, \citenamefont {Dhanak}, \citenamefont {Ferrari},\ and\ \citenamefont {Romeo}}]{artegiani2020amount}%
  \BibitemOpen
  \bibfield  {author} {\bibinfo {author} {\bibfnamefont {E.}~\bibnamefont {Artegiani}}, \bibinfo {author} {\bibfnamefont {J.~D.}\ \bibnamefont {Major}}, \bibinfo {author} {\bibfnamefont {H.}~\bibnamefont {Shiel}}, \bibinfo {author} {\bibfnamefont {V.}~\bibnamefont {Dhanak}}, \bibinfo {author} {\bibfnamefont {C.}~\bibnamefont {Ferrari}}, \ and\ \bibinfo {author} {\bibfnamefont {A.}~\bibnamefont {Romeo}},\ }\href@noop {} {\bibfield  {journal} {\bibinfo  {journal} {Solar Energy Materials and Solar Cells}\ }\textbf {\bibinfo {volume} {204}},\ \bibinfo {pages} {110228} (\bibinfo {year} {2020})}\BibitemShut {NoStop}%
\bibitem [{\citenamefont {Krasikov}\ \emph {et~al.}(2021)\citenamefont {Krasikov}, \citenamefont {Guo}, \citenamefont {Demtsu},\ and\ \citenamefont {Sankin}}]{krasikov2021comparative}%
  \BibitemOpen
  \bibfield  {author} {\bibinfo {author} {\bibfnamefont {D.}~\bibnamefont {Krasikov}}, \bibinfo {author} {\bibfnamefont {D.}~\bibnamefont {Guo}}, \bibinfo {author} {\bibfnamefont {S.}~\bibnamefont {Demtsu}}, \ and\ \bibinfo {author} {\bibfnamefont {I.}~\bibnamefont {Sankin}},\ }\href@noop {} {\bibfield  {journal} {\bibinfo  {journal} {Solar Energy Materials and Solar Cells}\ }\textbf {\bibinfo {volume} {224}},\ \bibinfo {pages} {111012} (\bibinfo {year} {2021})}\BibitemShut {NoStop}%
\bibitem [{\citenamefont {Chou}\ \emph {et~al.}(1996)\citenamefont {Chou}, \citenamefont {Rohatgi}, \citenamefont {Jokerst}, \citenamefont {Thomas},\ and\ \citenamefont {Kamra}}]{chou1996copper}%
  \BibitemOpen
  \bibfield  {author} {\bibinfo {author} {\bibfnamefont {H.}~\bibnamefont {Chou}}, \bibinfo {author} {\bibfnamefont {A.}~\bibnamefont {Rohatgi}}, \bibinfo {author} {\bibfnamefont {N.}~\bibnamefont {Jokerst}}, \bibinfo {author} {\bibfnamefont {E.}~\bibnamefont {Thomas}}, \ and\ \bibinfo {author} {\bibfnamefont {S.}~\bibnamefont {Kamra}},\ }\href@noop {} {\bibfield  {journal} {\bibinfo  {journal} {Journal of electronic materials}\ }\textbf {\bibinfo {volume} {25}},\ \bibinfo {pages} {1093} (\bibinfo {year} {1996})}\BibitemShut {NoStop}%
\bibitem [{\citenamefont {Gippius}\ \emph {et~al.}(1974)\citenamefont {Gippius}, \citenamefont {Panossian},\ and\ \citenamefont {Chapnin}}]{gippius1974deep}%
  \BibitemOpen
  \bibfield  {author} {\bibinfo {author} {\bibfnamefont {A.}~\bibnamefont {Gippius}}, \bibinfo {author} {\bibfnamefont {J.}~\bibnamefont {Panossian}}, \ and\ \bibinfo {author} {\bibfnamefont {V.}~\bibnamefont {Chapnin}},\ }\href@noop {} {\bibfield  {journal} {\bibinfo  {journal} {physica status solidi (a)}\ }\textbf {\bibinfo {volume} {21}},\ \bibinfo {pages} {753} (\bibinfo {year} {1974})}\BibitemShut {NoStop}%
\bibitem [{\citenamefont {Chamonal}\ \emph {et~al.}(1982)\citenamefont {Chamonal}, \citenamefont {Molva},\ and\ \citenamefont {Pautrat}}]{chamonal1982identification}%
  \BibitemOpen
  \bibfield  {author} {\bibinfo {author} {\bibfnamefont {J.}~\bibnamefont {Chamonal}}, \bibinfo {author} {\bibfnamefont {E.}~\bibnamefont {Molva}}, \ and\ \bibinfo {author} {\bibfnamefont {J.}~\bibnamefont {Pautrat}},\ }\href@noop {} {\bibfield  {journal} {\bibinfo  {journal} {Solid State Communications}\ }\textbf {\bibinfo {volume} {43}},\ \bibinfo {pages} {801} (\bibinfo {year} {1982})}\BibitemShut {NoStop}%
\bibitem [{\citenamefont {Said}\ and\ \citenamefont {Kanehisa}(1990)}]{said1990excited}%
  \BibitemOpen
  \bibfield  {author} {\bibinfo {author} {\bibfnamefont {M.}~\bibnamefont {Said}}\ and\ \bibinfo {author} {\bibfnamefont {M.}~\bibnamefont {Kanehisa}},\ }\href@noop {} {\bibfield  {journal} {\bibinfo  {journal} {Journal of crystal growth}\ }\textbf {\bibinfo {volume} {101}},\ \bibinfo {pages} {488} (\bibinfo {year} {1990})}\BibitemShut {NoStop}%
\bibitem [{\citenamefont {Zhang}\ \emph {et~al.}(2008)\citenamefont {Zhang}, \citenamefont {Da~Silva}, \citenamefont {Li}, \citenamefont {Yan}, \citenamefont {Gessert},\ and\ \citenamefont {Wei}}]{zhang2008effect}%
  \BibitemOpen
  \bibfield  {author} {\bibinfo {author} {\bibfnamefont {L.}~\bibnamefont {Zhang}}, \bibinfo {author} {\bibfnamefont {J.~L.}\ \bibnamefont {Da~Silva}}, \bibinfo {author} {\bibfnamefont {J.}~\bibnamefont {Li}}, \bibinfo {author} {\bibfnamefont {Y.}~\bibnamefont {Yan}}, \bibinfo {author} {\bibfnamefont {T.}~\bibnamefont {Gessert}}, \ and\ \bibinfo {author} {\bibfnamefont {S.-H.}\ \bibnamefont {Wei}},\ }\href@noop {} {\bibfield  {journal} {\bibinfo  {journal} {Physical review letters}\ }\textbf {\bibinfo {volume} {101}},\ \bibinfo {pages} {155501} (\bibinfo {year} {2008})}\BibitemShut {NoStop}%
\bibitem [{\citenamefont {Nagaoka}\ \emph {et~al.}(2019)\citenamefont {Nagaoka}, \citenamefont {Nishioka}, \citenamefont {Yoshino}, \citenamefont {Kuciauskas},\ and\ \citenamefont {Scarpulla}}]{nagaoka2019arsenic}%
  \BibitemOpen
  \bibfield  {author} {\bibinfo {author} {\bibfnamefont {A.}~\bibnamefont {Nagaoka}}, \bibinfo {author} {\bibfnamefont {K.}~\bibnamefont {Nishioka}}, \bibinfo {author} {\bibfnamefont {K.}~\bibnamefont {Yoshino}}, \bibinfo {author} {\bibfnamefont {D.}~\bibnamefont {Kuciauskas}}, \ and\ \bibinfo {author} {\bibfnamefont {M.~A.}\ \bibnamefont {Scarpulla}},\ }\href@noop {} {\bibfield  {journal} {\bibinfo  {journal} {Applied Physics Express}\ }\textbf {\bibinfo {volume} {12}},\ \bibinfo {pages} {081002} (\bibinfo {year} {2019})}\BibitemShut {NoStop}%
\bibitem [{\citenamefont {Brebrick}(2010)}]{brebrick2010high}%
  \BibitemOpen
  \bibfield  {author} {\bibinfo {author} {\bibfnamefont {R.~F.}\ \bibnamefont {Brebrick}},\ }\href@noop {} {\bibfield  {journal} {\bibinfo  {journal} {Journal of phase equilibria and diffusion}\ }\textbf {\bibinfo {volume} {31}},\ \bibinfo {pages} {260} (\bibinfo {year} {2010})}\BibitemShut {NoStop}%
\bibitem [{\citenamefont {Madelung}(2004)}]{madelung2004semiconductors}%
  \BibitemOpen
  \bibfield  {author} {\bibinfo {author} {\bibfnamefont {O.}~\bibnamefont {Madelung}},\ }\href {https://books.google.com/books?id=v_8sMfNAcA4C} {\emph {\bibinfo {title} {Semiconductors: Data Handbook}}}\ (\bibinfo  {publisher} {Cleaver-Hume},\ \bibinfo {year} {2004})\BibitemShut {NoStop}%
\end{thebibliography}%

\end{document}